\documentclass[twocolumn,pra,aps,superscriptaddress,amsmath,amssymb,floatfix,10pt]{revtex4-2}

\usepackage{cmap} 
\usepackage[utf8]{inputenc}
\usepackage[english]{babel}
\usepackage[T1]{fontenc}
\usepackage{appendix}
\usepackage{braket}
\usepackage{url, amsfonts, tikz, physics, listings, xcolor, graphicx}
\usepackage[shortlabels]{enumitem}
\usepackage{dsfont}
\usepackage{amsmath, amssymb, amstext, amscd, amsthm, makeidx, graphicx, url, mathrsfs, mathtools, longdivision, polynom, bbm, complexity}
\usepackage{nicefrac}
\usepackage[linesnumbered,ruled,vlined]{algorithm2e}
\usepackage{xcolor}
\usepackage[normalem]{ulem}
\usepackage{booktabs}

\definecolor{blueviolet}{rgb}{0.2, 0.2, 0.6}
\definecolor{webgreen}{rgb}{0,.5,0}
\definecolor{webbrown}{rgb}{.6,0,0}
\usepackage[
  colorlinks=false, 
  urlcolor=webbrown,
  linkcolor=blueviolet, 
  citecolor=webgreen,
  ]{hyperref}
\usepackage{cleveref}
\usepackage{quantikz}

\newcommand{\bigo}{O}
\renewcommand{\E}{\mathbb{E}}
\renewcommand{\Pr}{\mathbb{P}}

\numberwithin{equation}{section}

\newtheorem{theorem}{Theorem}

\newcommand{\stkout}[1]{\ifmmode\text{\sout{\ensuremath{#1}}}\else\sout{#1}\fi}
\newif\ifverbose
\verbosefalse
\newcommand{\ins}[1]{\ifverbose\textcolor{blue}{#1}\else#1\fi}
\newcommand{\edit}[2]{\ifverbose\textcolor{red}{\stkout{#1} #2}\else#2\fi}

\begin{document}

\title{Entanglement-induced provable and robust quantum learning advantages}

\author{Haimeng Zhao}
\email{haimengzhao@icloud.com}
\affiliation{Center for Quantum Information, IIIS, Tsinghua University, Beijing 100084, China}
\affiliation{Institute for Quantum Information and Matter, California Institute of Technology, Pasadena, CA 91125, USA}

 \author{Dong-Ling Deng}
\email[]{dldeng@tsinghua.edu.cn}
\affiliation{Center for Quantum Information, IIIS, Tsinghua University, Beijing 100084, China}
\affiliation{Hefei National Laboratory, Hefei 230088, China}
\affiliation{Shanghai Qi Zhi Institute, Shanghai 200232, China}


\begin{abstract}
\textbf{Abstract.}
Quantum computing holds unparalleled potentials to enhance machine learning. However, a demonstration of quantum learning advantage has not been achieved so far. We make a step forward by rigorously establishing a noise-robust, unconditional quantum learning advantage in expressivity, inference speed, and training efficiency, compared to commonly-used classical models. Our proof is information-theoretic and pinpoints the origin of this advantage: entanglement can be used to reduce the communication required by non-local tasks. In particular, we design a task that can be solved with certainty by quantum models with a constant number of parameters using entanglement, whereas commonly-used classical models must scale linearly to achieve a larger-than-exponentially-small accuracy. We show that the quantum model is trainable with constant resources and robust against constant noise. Through numerical and trapped-ion experiments on IonQ Aria, we demonstrate the desired advantage. Our results provide valuable guidance for demonstrating quantum learning advantages with current noisy intermediate-scale devices.
\end{abstract}

\maketitle

\section{Introduction}

Demonstrating quantum advantages in real-world applications is a central goal of quantum computing \cite{preskill2012quantum,preskill2018quantum,nielsen2010quantum}.
An intriguing and promising approach concerns quantum machine learning (ML) \cite{biamonte2017quantum,cerezo2022challenges,ciliberto2018quantum,carleo2019machine}, owing to the great success of its classical counterpart \cite{goodfellow2016deep,lecun2015deep,jordan2015machine}.
In recent years, a number of quantum ML algorithms have been developed, which offer potential polynomial or even exponential speedups in ML tasks \cite{dalzell2023quantum,harrow2009quantum,lloyd2014quantum,rebentrost2014quantum,dunjko2016quantum,gao2018quantum,liu2021rigorous,gyurik2023exponential,lloyd2018quantum,schuld2019quantum,cerezo2021variational,huang2022quantum,aharonov2022quantum,oh2024entanglement,zhu2022generative}.
However, whether they can lead to practically useful advantage on classical tasks remains largely unclear to date.
This is because existing proposals are either heuristic \cite{cerezo2021variational}, depend on unproven complexity-theoretic conjectures \cite{gao2018quantum,liu2021rigorous,gyurik2023exponential}, specific to particular quantum tasks \cite{huang2022quantum,aharonov2022quantum,molteni2024exponential,oh2024entanglement}, require inefficient data loading or extracting procedures \cite{aaronson2015read,ciliberto2018quantum}, or can be dequantized into efficient classical algorithms \cite{tang2023quantum,tang2022dequantizing,tang2019quantum,tang2021quantum}.
In fact, it is still under active debate whether practical quantum advantage is even possible in ML tasks \cite{schuld2022quantum,cerezo2023does,gil2024relation}.

A few recent works have studied \emph{unconditional} quantum advantage in ML based on unique features of quantum mechanics.
By invoking quantum non-locality, a logarithmic quantum advantage with shallow quantum circuits has been proved \cite{bravyi2018quantum,bravyi2020quantum} and extended into advantage in ML tasks \cite{zhang2024quantum}.
But this logarithmic advantage requires a huge system size to be manifested, which is beyond the scope of current noisy intermediate-scale quantum devices \cite{preskill2018quantum}.
On the other hand, polynomial advantage against commonly-used classical models has been proven via quantum contextuality \cite{gao2022enhancing,anschuetz2023interpretable,anschuetz2024arbitrary}, which, however, consumes polynomially many qubits and relies on loss functions being worst-case as well as assumptions on the analytical properties of the corresponding classical models.
Whether this advantage survives under the effect of noises on near-term quantum devices is also unclear.
Moreover, these results focus on expressivity, i.e., the number of parameters used to solve the ML task, whereas potential advantages in training remains to be elucidated.

\begin{figure*}[t]
    \centering
    \includegraphics[width=\linewidth]{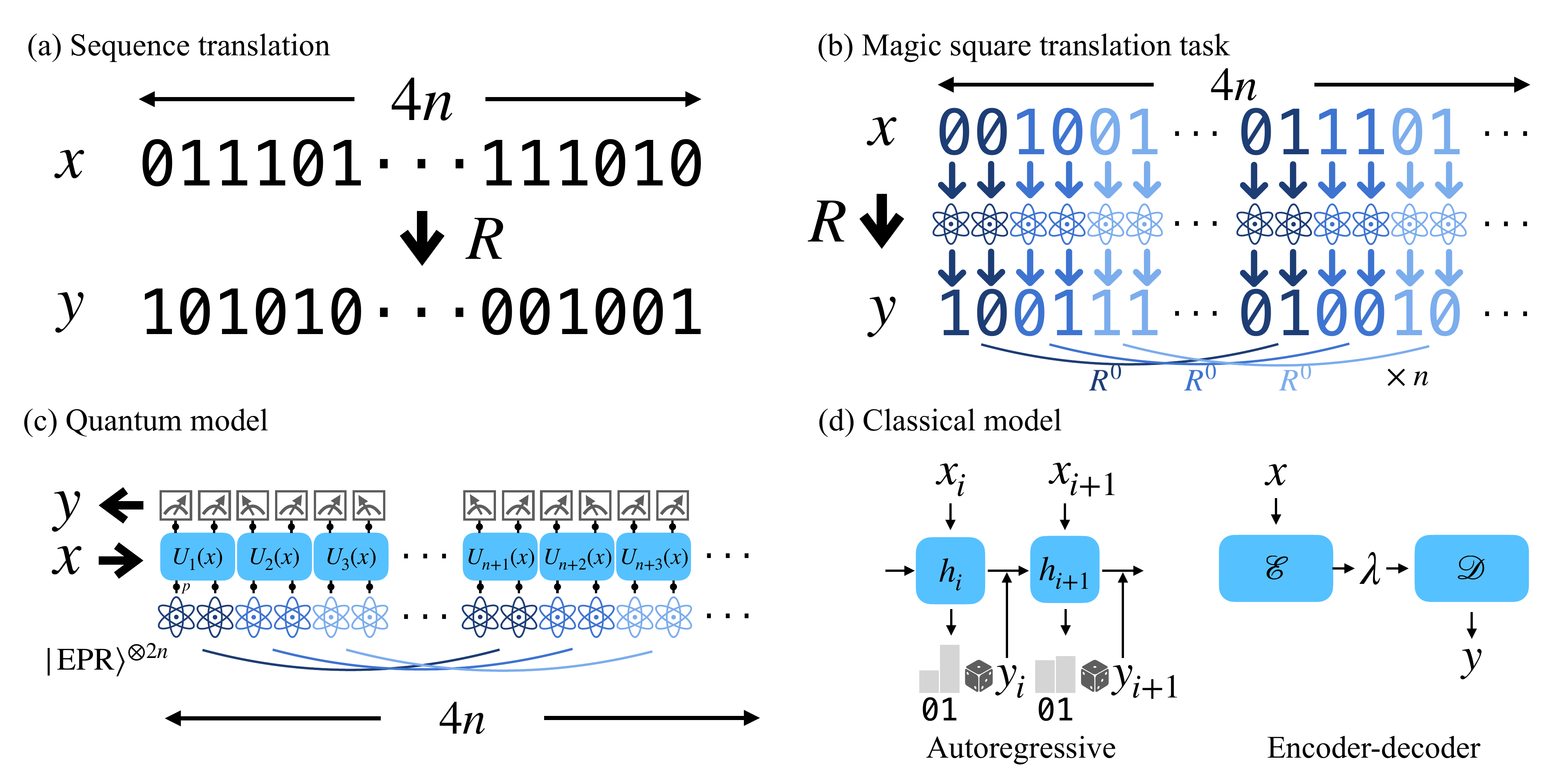}
    \caption{\textbf{Schematic overview of the machine learning tasks and models.}
    \textbf{(a)} Sequence translation task $R: \{0, 1\}^{4n}\times \{0, 1\}^{4n}\to \{0, 1\}$ where $R(x, y)=1$ indicates valid translation.
    \textbf{(b)} The magic square translation task $R$ constructed from $n$ parallel repetition of the sub-task $R^0$ adopted from the Mermin-Peres magic square game.
    \textbf{(c)} The constant-parameter-size quantum learning model with entanglement supply.
    Here, $\ket{\mathrm{EPR}}=(\ket{00}+\ket{11})/\sqrt{2}$ is the Einstein–Podolsky–Rosen (EPR) state, and qubits with the same color are entangled.
    \textbf{(d)} The two commonly-used classical machine learning models: autoregressive and encoder-decoder models.
    }
    \label{fig:task-model}
\end{figure*}

In this work, we establish a noise-robust, unconditional, constant-vs-linear quantum advantage in ML in terms of expressivity, inference speed, and training efficiency, compared to commonly-used classical models.
We design a ML task with $n$-bit classical inputs and outputs that can be solved by a quantum model with maximum score, using a constant number of variational parameters and linearly-scaling entanglement resources.
Meanwhile, we rigorously prove that commonly-used classical ML models must scale at least linearly with the size of the task to achieve a larger-than-exponentially-small accuracy.
In terms of training, we show that constant time and $\bigo(1/n)$ samples suffice to train the quantum model, while the training of classical models \ins{is expected to} require quadratic sample size and training time.
This decreasing sample complexity of the quantum model comes from the increasing amount of information contained in each sample.
We further prove that this advantage still persists under constant-strength depolarization noise.
We conduct numerical simulations and trapped-ion experiments on IonQ Aria \cite{ionq} to validate its noise-robustness, and find that this advantage is already evident on about $25$ qubits. 

Our results improve upon previous works \cite{gao2022enhancing,anschuetz2023interpretable,anschuetz2024arbitrary} in several ways.
First, ours prove an average-case separation and holds for most inputs, whereas previous ones are worst-case separations.
Second, our quantum algorithm is shallow and noise-robust, while previous ones require deep circuits to implement and the advantage may disappear under noise.
Third, our $O(1)$-vs-$\Omega(n)$ separation exponentially improves over previous logarithmic separations \cite{bravyi2018quantum,bravyi2020quantum,zhang2024quantum} and matches the arbitrary polynomial separation $O(1)$-vs-$\Omega(n^k)$  in the tokenized setting \cite{anschuetz2024arbitrary}, where the inputs and outputs are $n$ tokens each containing $n^{k-1}$ bits, resulting in a total problem size of $n^k$.
These improvements make our results more suitable for near-term demonstration.

Our proof makes use of information-theoretic tools and centers around the communication capacity of classical ML models, which is limited by their size.
For quantum models, the use of entanglement and its non-local effect can reduce the classical communication needed to solve certain distributed computational tasks, a fact often referred to as quantum pseudo-telepathy \cite{yao1993quantum,brassard2003quantum,brassard2005quantum}.
This reduction in classical communication allows us to go beyond inherently quantum problems (e.g., with quantum states as inputs) and trade pre-computable \cite{huggins2024accelerating} entanglement for advantage in a fully classical setting.

\section{Results}

\subsection{Framework and models}
We consider sequence translation tasks (illustrated in \Cref{fig:task-model}(a)) that are ubiquitous in the study of natural language processing \cite{lecun2015deep,manning1999foundations,sutskever2014sequence}. 
For simplicity, we assume that the sequences are bitstrings with length $4n$.
Such a translation task is specified by a translation rule $R: \{0, 1\}^{4n}\times \{0, 1\}^{4n} \to \{0, 1\}$, where $R(x, y)=1$ means that $y$ is a valid translation of $x$.
The goal is to design a ML model $\mathcal{M}$ that takes $x$ as input and produce a (possibly random) translation result $y=\mathcal{M}(x)$ with an expected success rate as high as possible.
The performance can be measured by a score function $S(\mathcal{M}) = \E[R(x, \mathcal{M}(x))]=\Pr[R(x, \mathcal{M}(x))=1]\in [0, 1]$.
We note that this is a fully classical ML task with no particular quantum feature, in contrast to the inherently quantum tasks of learning properties of quantum states or processes considered in Refs. \cite{huang2022quantum,aharonov2022quantum,molteni2024exponential,oh2024entanglement}.

\ins{
As a simple example of a translation task and its corresponding rule $R$, consider the identity translation where every input sequence $x$ is mapped to itself $y=x$.
In this case, the translation rule is the Kronecker delta $R(x, y) = \delta_{y, x}$ which has value one if and only if $y=x$.
For a random model $\mathcal{M}$ that generates a random translation result $y$ uniformly sampled from $\{0, 1\}^{4n}$ regardless of the input $x$, its score is the probability of randomly guessing $S(\mathcal{M}) = 1/2^{4n}$.
}

Our quantum model $\mathcal{M}_Q$ for such translation tasks is a simple, parameterized shallow quantum circuit shown in \Cref{fig:task-model}(c).
It is a quantum circuit on $4n$ qubits where a two-qubit unitary gate $U_i(x_{2i-1}, x_{2i})$ acts on each consecutive pair of qubits $(2i-1, 2i)$, $\forall 1\leq i\leq 2n$.
After a single layer of gates, we measure the qubits in the computational basis, and output the resulting bitstring $y$ as the translation result.
The initial state is independent of $x$, but can be prepared strategically to supply entanglement resources.
Additional layers and pre-/post-processing can be added for more general purposes, but this most basic setting is enough here and straightforward for experimental realization.
To accommodate variable sequence length, we apply weight sharing.
That is, a common set of unitaries are used at different places of the sequence.
Therefore, only a constant number of parameters are used in this quantum model.

For the classical model $\mathcal{M}_C$, we consider the commonly-used autoregressive and encoder-decoder models (\Cref{fig:task-model}(d)), also with weight sharing.
In general, their communication capacity $c$ (i.e., the amount of information \ins{about the input sequence} that the model can carry and spread to other parts of the sequence) is limited by the number of parameters $d$ of the model \cite{collins2022capacity}.
\ins{
Here, we define the number of parameters more broadly as the number of bits used to store the intermediate state at each time step during the forward computation of the model.
}
For example, the communication capacity of an autoregressive (encoder-decoder) model is proportional to the dimension of its hidden state (latent space).
Based on this observation, we focus on this general family of classical models whose communication capacity is bounded by its parameter size.
We call them \emph{communication-bounded} classical models, which include commonly-used ones such as recurrent neural networks or Transformers \cite{vaswani2017attention}.
\ins{
We also allow the classical models to have shared randomness distributed across different parts of the model, as the classical analog of the shared entanglement in the quantum model. 
This gives commonly-used classical model extra power so that we have a fair comparison between classical and quantum.
Note that in defining the communication complexity, we only count the transmitted information that depends on the input sequences.
Therefore, neither shared entanglement nor shared randomness count as they are precomputed and do not depend on the inputs.
More rigorous definitions and details of these classical models can be found in the Supplementary Materials Sec. I.
}

\subsection{Quantum advantage in inference}

Communication is a key bottleneck in solving natural language processing tasks.
To deal with long-range context in texts, the model must be able to memorize features of previous parts of the sequence and spread the information to other parts.
To overcome this bottleneck, we note that entanglement can be used to reduce the classical communication needed to solve certain distributed computation tasks \cite{brassard2005quantum}. 
This is achieved by allowing quantum protocols that are more general than classical ones, without direct transmission of classical information.
This gives us a way to trade entanglement for advantage in sequence translation tasks.

To establish such a quantum advantage, we design a ML task that is easy for quantum computers while hard for classical ones.
In particular, we design a translation rule $R$ based on quantum pseudo-telepathy, the most extreme form of the above phenomenon, where the task can be solved by an entangled protocol with no classical communication at all.
We adopt a variant of the Mermin-Peres magic square game \cite{mermin1990simple,peres1990incompatible,bravyi2020quantum,cabello2001all,cabello2001bell} as the building block of our translation task $R:\{0, 1\}^{4n}\times \{0, 1\}^{4n}\to \{0, 1\}$.
In such a game, two players $A, B$ each receive two random bits $x^A, x^B\in \{0, 1\}^2$ and output two bits $y^A, y^B\in \{0, 1\}^2$.
Let $I(x)=2x_1+x_2\in \{0, 1, 2, 3\}$ and $y_3^A=y_1^A\oplus y_2^A\oplus 1, y_3^B=y_1^B\oplus y_2^B$, where $\oplus$ is the addition in the field $\mathbb{Z}_2$.
The game is won if and only if at least one of the following is satisfied: (1) either $x^A$ or $x^B$ is $00$; (2) $y^B_{I(x^A)}=y^A_{I(x^B)}$.
This winning condition defines a translation rule $R^0: \{0, 1\}^4\times \{0, 1\}^4\to \{0, 1\}$.
\ins{
For clarity, we include a transparent illustration of this translation rule $R^0$ in Supplementary Materials Table S2.
}
We show that this translation task $R^0$ cannot be solved with probability more than $\omega=15/16$ by any classical non-communicating strategies.
Meanwhile, a quantum strategy using two Bell pairs shared between $A$ and $B$ can solve this task with certainty.

To boost the separation in score, we combine $n$ parallel repetition of $R^0$ to form the translation task $R$.
For any $x, y\in \{0, 1\}^{4n}$ and $1\leq i\leq n$, we set the two players of the $i$th game to receive $x_{2i-1}x_{2i}, x_{2n+2i-1}x_{2n+2i}$ and output the corresponding bits of $y$.
The translation $(x, y)$ is valid ($R(x, y)=1$) when all these $n$ sub-tasks $R^0$ are simultaneously solved.
We call this task $R$ the \emph{magic square translation task}.
It can also be viewed as a non-local game with two players $A, B$, where they each corresponds to the first and last $2n$ bits of $x, y$.
In this way, the score of any classical non-communicating strategy is reduced to $2^{-\Omega(n)}$ \cite{rao2008parallel}, while that of the quantum protocol using $2n$ Bell pairs remains one.
This allows us to prove the following theorem, which exponentially improves the logarithmic advantage in Ref. \cite{zhang2024quantum}.

\begin{theorem}[Quantum advantage without noise]
\label{thm:noiseless-inf}
    For the magic square translation task $R: \{0, 1\}^{4n}\times \{0, 1\}^{4n}\to \{0, 1\}$, there exists an $\bigo(1)$-parameter-size quantum model $\mathcal{M}_Q$ that can achieve a score $S(\mathcal{M}_Q)=1$ using $2n$ Bell pairs.
    Meanwhile, any communication-bounded classical model $\mathcal{M}_C$ that can achieve a score $S(\mathcal{M}_C)\geq 2^{-o(n)}$ must have $\Omega(n)$ parameter size.
\end{theorem}

\Cref{thm:noiseless-inf} gives an \textit{unconditional} constant-versus-linear separation between quantum and commonly-used classical ML models, in terms of expressivity and inference speed.
Here, higher expressivity means fewer parameters needed to express the target translation rule, and faster inference speed refers to less time needed to translate a input sequence.
We remark that this can be straightforwardly boosted to arbitrary polynomial separation by supplying polynomially many qubits, similar to the approach exploited in Ref. \cite{anschuetz2024arbitrary}.
We prove \Cref{thm:noiseless-inf} by dividing the sequence in two halves and bounding the communication between the two parts required to solve the magic square translation task.
We derive the score of the classical model from bounds on the success probability of non-communicating classical strategies of the magic square game \cite{rao2008parallel,jain2022direct}, and relate it to the communication capacity and the size of the classical model.
To show the quantum easiness, we embed the quantum strategy of the game into the quantum model, which can win the game with certainty.
See the Supplementary Materials Sec. II for technique details.

To demonstrate this quantum advantage on near-term noisy quantum devices, we show that it persists under constant-strength noise.
In particular, we consider single-qubit depolarization noises, with strength $p$ following each gate that corrupts the state $\rho$ into $(1-p)\rho + \frac{p}{3}(X\rho X+Y\rho Y+Z\rho Z)$ \cite{nielsen2010quantum}.
We prove the following theorem.

\begin{theorem}[Quantum advantage with noise]
\label{thm:noisy-inf}
    For any noise strength $p\leq p^\star\approx 0.0064$, there exists a noisy magic square translation task $R_p: \{0, 1\}^{4n}\times \{0, 1\}^{4n}\to \{0, 1\}$, such that it can be solved by an $\bigo(1)$-parameter-size quantum model $\mathcal{M}_Q$ under depolarization noise of strength $p$ with score $S(\mathcal{M}_Q)\geq 1-2^{-\Omega(n)}$ using $2n$ Bell pairs.
    Meanwhile, any communication-bounded classical model $\mathcal{M}_C$ that can achieve a score $S(\mathcal{M}_C)\geq 2^{-o(n)}$ must have $\Omega(n)$ parameter size.
\end{theorem}

We prove \Cref{thm:noisy-inf} by constructing a family of noisy magic square translation tasks $R_p$ that are a bit easier than the original task (see the Supplementary Materials Sec. II for technical details).
Instead of demanding simultaneous winning of all sub-games as in $R$, the noisy task $R_p$ only requires winning a certain fraction of them.
The fraction is chosen such that the task remains hard for classical models, but can tolerate mistakes caused by constant-strength noise, thereby going beyond the best known $1/\poly(n)$ noise tolerance proved in Ref. \cite{zhang2024quantum}.
This requirement leads to the noise threshold $p\leq p^\star= 1-(15/16)^{0.1}\approx 0.0064$, which can be improved by designing more sophisticated sub-tasks $R^0$.

\subsection{Quantum advantage in training}

We have proven the existence of a constant-depth quantum model $\mathcal{M}_Q$ that can solve the magic square translation task near-perfectly.
Next, we show that this optimal model can be found efficiently with maximum likelihood estimation using constant-size training data $M=\bigo(1)$ and training time $T=\bigo(1)$.
Here, the size of the training data $\{(x^{(i)}, y^{(i)})\}_{i=1}^N$ with $R(x^{(i)}, y^{(i)})=1$ is defined as $M=\Theta(nN)$, since each data point $(x^{(i)}, y^{(i)})$ is of size $\Theta(n)$.
This implies that the number of data samples $N$ required to train the quantum model scales inversely proportional to the size of the task.

\begin{theorem}[Quantum advantage in training]
\label{thm:train}
    There exists a training algorithm that, with probability at least $2/3$, takes $\{(x^{(i)}, y^{(i)})\}_{i=1}^N$ with size $M=\Theta(nN)=\bigo(1)$ as input and outputs the optimal quantum model $\mathcal{M}_Q$ for the task $R$.
    Moreover, the running time of this training algorithm is $T=\bigo(1)$.
\end{theorem}

We prove \Cref{thm:train} by explicitly constructing the training algorithm (in Supplementary Materials Sec. III).
The algorithm performs maximal likelihood estimation by an exhaustive search over all possible sets of parameters in $\mathcal{M}_Q$ and outputs the one with the maximal empirical likelihood calculated from the training data.
We then show that $N=\bigo(1/n)$ number of data samples suffices to estimate the likelihood accurately enough.
Combining with the fact that the quantum model only has $\bigo(1)$ many parameters, this gives the desired space and time complexity to train the quantum model.

In contrast, as shown in \Cref{thm:noiseless-inf,thm:noisy-inf}, any classical model requires at least $\Omega(n)$ parameters.
This makes their training \edit{necessarily}{generally} harder.
In particular, if we assume that the $d$ parameters in the model can only take $K=\Theta(1)$ discrete values (e.g., due to machine precision), then the number of possible hypothesis functions represented by the classical model is $K^d$.
Therefore, the generalization error of the model (i.e., test loss minus training loss) \ins{is expected to be} $\sqrt{d\log K/N}$ \cite{mohri2018foundations}.
This means that only when $N\geq \Omega(d)\geq \Omega(n)$ will the statistical fluctuation be small enough such that accurate training can be guaranteed.
\ins{
Intuitively, this is because that the information needed to specify a model from a set of $K^d$ possible models is $\log(K^d)=d\log K$, so the training algorithm is expected to need roughly these many samples to train well.
}
This requirement corresponds to a training data of size $M\geq \Omega(n^2)$ and training time $T\geq \Omega(n^2)$, quadratically worse than the quantum case.
\ins{Though we cannot rigorously prove that no training algorithm can do better, it gives a general expectation on the amount of training resource needed by classical models.
Below, we conduct numerical experiments to empirically validate this quantum advantage in training.}

\subsection{Numerical and experimental results}
\label{sec:numerics}

\begin{figure*}
    \centering
    \includegraphics[width=\linewidth]{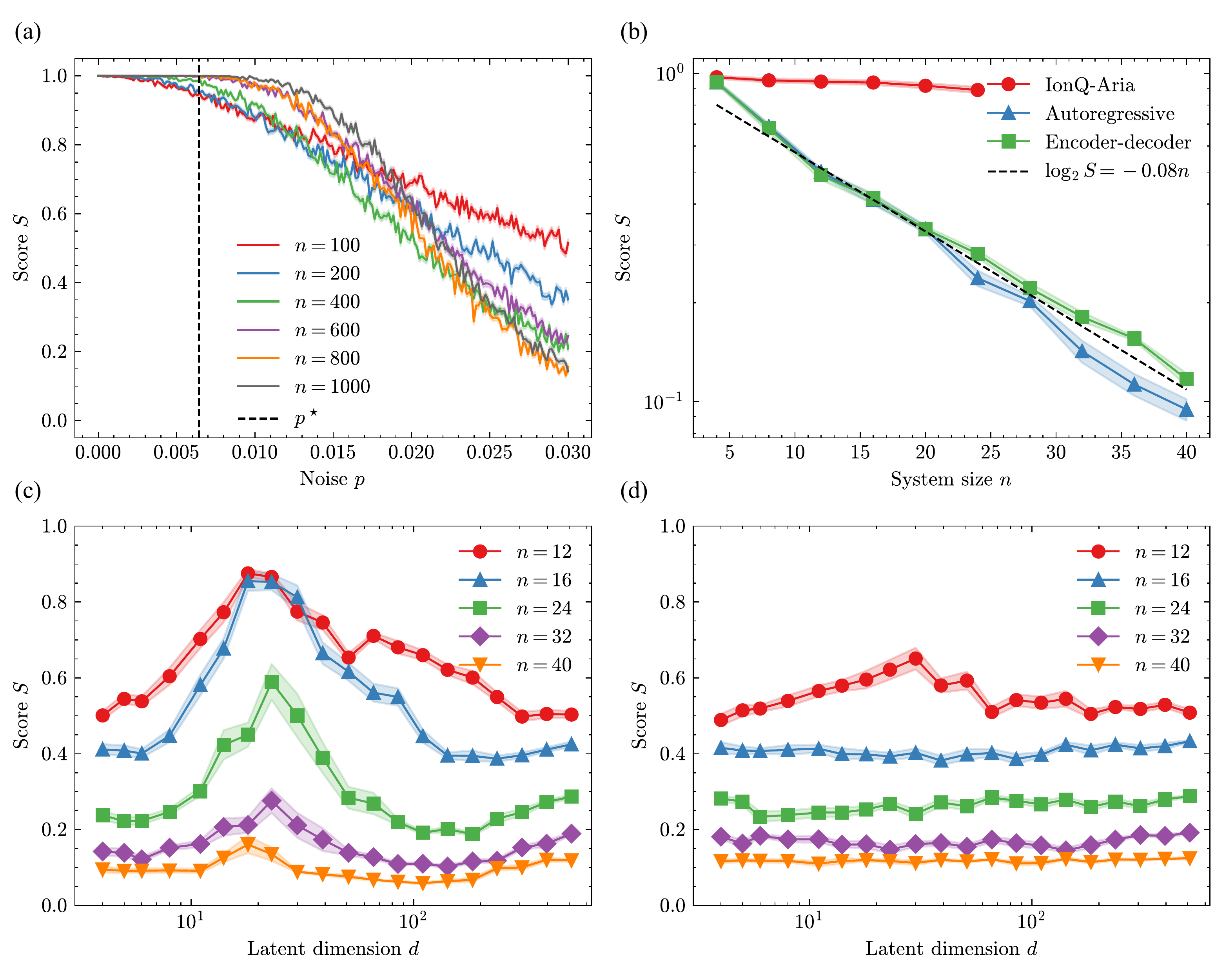}
    \caption{\textbf{Results on numerical simulations and trapped-ion experiments on IonQ Aria.} \textbf{(a)} Performance of the quantum model on the noisy magic square translation task $R_p$ with different problem size $n$ and noisy strength $p$. For each $n$ and $p$, the score of the model is averaged over $10^4$ randomly sampled inputs $x$, and plotted in solid lines. Shaded areas indicate standard deviations. \textbf{(b)} The quantum model executed on IonQ Aria achieves consistent high score, while the performance of classical models with $d=4$ decays exponentially as the system size $n$ grows. \textbf{(c-d)} Performance of classical autoregressive \textbf{(c)} and encoder-decoder \textbf{(d)} models on the magic sqaure translation task $R$ with different problem size $n$ and model size $d$. For each $n$ and $d$, the score of the model is calculated from $10^3$ randomly sampled test inputs $x$ and $10$ random training instances. The average scores are plotted in solid lines and the shaded areas indicate standard deviations. }
    \label{fig:numerics}
\end{figure*}

Since the optimal quantum model $\mathcal{M}_Q$ is composed of classically-controlled Clifford operations, we can simulate it efficiently on classical computers in $\bigo(n^3)$ time \cite{aaronson2004improved}.
Note that this is compatible with the proved $\Omega(n)$ lower bound.
This enables us to study the noise robustness of the quantum model on up to $n=1000$ qubits using PyClifford \cite{pyclifford}.
Note that in this section, we use $n$ to denote the system size, which is previously denoted as $4n$.
For each system size $n$ and noise level $p\in [0, 0.03]$, we calculate the average score on the task $R_p$ with $10^4$ random inputs.
The resulting scores and their standard deviations are plotted in \Cref{fig:numerics}(a) as solid lines and shaded areas.
We observe that as the problem size $n$ grows, a plateau of perfect score below the noise threshold $p^\star\approx 0.0064$ emerges, in accordance with the theoretical prediction $S(\mathcal{M}_Q)\geq 1-2^{-\Omega(n)}$ in \Cref{thm:noisy-inf}.
This validates the robustness of the quantum model against constant-strength noise.

For classical models, we choose the Gated Recurrent Unit (GRU) architecture \cite{cho2014learning} as the common backbone of the autoregressive and encoder-decoder models.
We adopt the standard implementation of GRU from Keras \cite{chollet2015keras} and tune its latent dimension $d$ to control the model size.
For each problem size $n$ up to $40$, we generate training data of size $N=10^4$ by randomly sample the input $x$, run the noiseless optimal quantum model for $R$, and collect the output $y$.
We train the model by minimizing the negative log-likelihood loss using the Adam optimizer with learning rate $10^{-2}$ and batch size $10^3$ for $10^4$ epochs.
If the loss stops to decrease for $500$ epochs, we terminate the training.
Then we calculate the score of the model on $10^3$ newly sampled inputs $x$, and average it over $10$ random training instances.
The results are plotted in \Cref{fig:numerics}(c-d) and details of the implementation can be found in the Supplementary Materials Sec. IV.

We find that for each $n$, the score of each classical model first increases as the latent dimension grows.
Then it reaches a peak at $d\sim 30$ and starts to decrease.
The scores at small and large $d$ are roughly the same, which decreases exponentially with $n$, as shown in \Cref{fig:numerics}(b) to be around $S(\mathcal{M}_C)\approx 2^{-0.08n}$ for fixed $d=4$, confirming the prediction in \Cref{thm:noiseless-inf}.
This phenomenon comes from the delicate dilemma due to insufficient communication capacity at small $d$ and overfitting at large $d$. 
At small $d$, the communication capacity of the model increases with the size $d$, allowing the model to capture contextual information with longer range.
However, when $d$ is large and surpasses the control of $N$, the limited sample size $N$ cannot provide enough information to train the model well.
This overfitting issue causes the performance of the model on newly sampled test data to decrease again.
This explanation is confirmed by the large generalization error observed when $d\gtrsim 100$.
The constant-versus-linear separation together with the overfitting problem makes the quantum advantage already evident at a relative small system size $n\sim 40$.

To better demonstrate this noise-robust quantum advantage, we carry out experiments on IonQ's 25-qubit trapped-ion quantum device Aria via Amazon Web Services (AWS).
We execute the quantum model to generate $10^3$ random translation results on up to $24$ qubits and calculate its average score on the task $R_p$ with fraction threshold set to $0.95$.
The results are plotted in \Cref{fig:numerics}(b) and calibration details are provided in Supplementary Materials Sec. IV.
We find that the scores achieved on current quantum devices are consistently above $0.88$ for up to problem size $n=24$.
When compared to the performance of classical models, we observe a clear exponential quantum advantage in score.
Given that the noise strength in IonQ Aria is not yet below $p^\star$, we expect that the performance of the quantum model will improve with smaller noise and larger system size according to \Cref{thm:noisy-inf}.

We note that the performance of autoregressive models are in general better than encoder-decoder models.
Their peaks of score are sharper and appear at smaller $d$.
This is because that autoregressive models have smaller communication burdens as they do not need to memorize the whole input with their hidden states.
In contrast, for encoder-decoder models, the $d$ required for communication quickly grows and falls into the overfitting regime, causing the peaks to disappear when $n$ is large.

\section{Discussion}
Our results establish a noise-robust, unconditional quantum advantage in ML in terms of expressivity, inference speed, and training efficiency, compared to commonly-used classical models.
Through numerical and trapped-ion experiments on cloud, we show that this advantage is already evident on a relatively small system size.
These features make it a promising candidate to be experimentally demonstrated at scale in the near term, particularly on trapped-ion and neutral atom systems where all-to-all connectivity aids entanglement generation.
We also pinpoint the origin of this advantage: quantum entanglement can be utilized to reduce the classical communication of certain ML tasks.

We remark that our results are based on a surprisingly simple task, parallel repetition of a pseudo-telepathy game, and yet it has many desired merits for near-term demonstration that have not been achieved in the literature before.
This is realized by offloading certain burdens into pre-computable entanglement resources \cite{huggins2024accelerating} and only aiming at constant-versus-linear advantage over commonly-used classical ML models (i.e., autoregressive and encoder-decoder models).
In a way, this points out a drawback in the current design of classical neural networks that leads to quantum advantage: when data sequences with long-range correlations arrive in a streaming fashion, these classical models must memorize a linearly growing amount of data, while quantum models need not.
We leave to future work the problem of finding stronger advantages that are provable against broader families of classical algorithms.

Our work also raises many interesting questions for future studies.
First, we have shown that $\bigo(n)$ Bell pairs can be converted into a constant-versus-linear quantum advantage.
It is natural to ask how does entanglement resources relate to quantum advantage in general.
In particular, what is the maximal efficiency of trading entanglement for advantage?
This would provide a new perspective to recent works studying the connection between entanglement and magic (or non-stabilizerness) \cite{fattal2004entanglement,tirrito2024quantifying,gu2024magic,frau2024nonstabilizerness}.

Second, our ML task is constructed using the parallel repetition of a two-body non-local game.
Our analysis also proceeds by dividing the sequence into two halves.
Within the framework of two-body games, our communication complexity lower bound $c\geq \Omega(n)$ is already optimal, since one can always win the game by transmitting all the $\bigo(n)$ input bits of $A$ to $B$.
A potential way to improve would be to use genuine many-body non-local games instead and look at multiparty communication, which may provide exponential violation of many-body Bell inequalities.
An obstacle to tackle would be to efficiently measure all the terms (possibly exponentially many) in the Bell inequality.
Strategic measurement schemes from recent experimental advances on Bell experiments could be useful \cite{wang2024probing}.

Third, our results may also be improved by designing quantum pseudo-telepathy tasks that are more secure against classical communication.
In particular, cryptographic primitives might be useful in making certain parts of the input information hard to decipher by classical ML models.
Meanwhile, the quantum model must be able to perform the quantum winning strategy knowing only the ciphertext of the inputs.
This may give rise to much larger quantum advantage against a wider range of classical models at the cost of introducing cryptographic assumptions \cite{vidick2023introduction}.

Lastly, the quantum advantage explored in this paper builds upon a specific family of translation tasks that are contrivedly designed to best exemplify the role of entanglement in reducing classical communication.
However, they naturally occurs in ML tasks that aim to model the correspondence between measurement statistics and measurement instructions on entangled quantum systems.
We also expect the information-theoretic origin of the advantage to hold for more generic translation tasks (e.g., those generated from random constant-depth Clifford circuits as in \cite{watts2024quantum}).
This would open up a new direction for finding practical advantage in more general classical ML problems.
For instance, it would be interesting to (empirically) study the communication cost of other natural language processing or computer vision problems and whether quantum entanglement can lead to advantages.

\section{Methods}
In this section, we elaborate on the proof ideas behind our quantum advantage results in inference (\Cref{thm:noiseless-inf,thm:noisy-inf}) and in training (\Cref{thm:train}).

\subsection{Inference advantage}
To prove the quantum advantage on $R$, we first show that the optimal quantum strategy for the non-local game can be converted into a constant-parameter-size quantum ML model $\mathcal{M}_Q$ in \Cref{fig:task-model}(c).
Weight sharing is guaranteed because the winning strategies for all sub-games are the same.
Therefore, the quantum model can solve $R$ with perfect score and only $\bigo(1)$ parameters.

Next, for classical models, we quantify the communication required to solve $R$.
We do so by constructing a non-communicating protocol out of a communicating protocol \cite{jain2022direct}.
In particular, suppose there is a protocol $\mathcal{P}_\mathrm{com}$ that can solve $R$ with probability $p_\mathrm{com}$ using $c$ bits of communication between the two players $A, B$.
Then we can construct a non-communicating protocol $\mathcal{P}_\mathrm{non-com}$ by first randomly guessing the communicated message $m\in \{0, 1\}^c$ and feed it into $\mathcal{P}_\mathrm{com}$.
The success probability $p_\mathrm{non-com}$ of $\mathcal{P}_\mathrm{non-com}$ is thus lower bounded by $(\frac{1}{2})^c p_\mathrm{com}$.
On the other hand, the hardness of $R$ against non-communicating protocols asserts that $p_\mathrm{non-com}\leq 2^{-\Omega(n)}$.
Therefore, we have $p_\mathrm{com}\leq 2^{-\Omega(n)+c}$, which means that at least $c\geq \Omega(n)$ bits of communication are needed to achieve a score $S\geq 2^{-o(n)}$.

\ins{
For general classical ML models $\mathcal{M}_C$, the communication capacity $c$ is proportional to the model size $d$ because of the subadditivity of information.
}
For example, in autoregressive and encoder-decoder models, we have $c\leq \bigo(d)$ where $d$ is the dimension of the latent space.
This gives use the desired classical hardness result as stated in \Cref{thm:noiseless-inf}.

To account for noise, we first show that the winning probability of the quantum model $\mathcal{M}_Q$ on each $R^0$ decays as $\omega^\star_p=(1-p)^{10}$ with the noise strength $p$ in the worst case.
This comes from the constant-depth structure of the quantum model.
When $\omega^\star_p>\omega$ (i.e., $p<p^\star=1-(15/16)^{0.1}\approx 0.0064$), the advantage in winning probability persists.
Then we can again boost the separation by an $n$-fold parallel repetition $R_p$ of $R^0$ that requires a fraction $\eta_p = (\omega^\star_p+\omega)/2$ of the sub-games winning.
By Hoeffding's inequality \cite{vershynin2018high}, we show that the quantum model still has a score at least $1-2^{-\Omega(n)}$ under noise $p$, and the score of the classical model stays $2^{-\Omega(n)}$ unless it has linear size, thus proving \Cref{thm:noisy-inf}.
Note that the noise threshold $p^\star$ can be straightforwardly improved by replacing $R^0$ with non-local games with larger quantum-classical separation.

\subsection{Training Advantage}
To prove the advantage in training, we explicitly construct a training algorithm that can find the optimal quantum model based on randomly sampled training data $\{(x^{(i)}, y^{(i)})\}_{i=1}^N$, where $R(x^{(i)}, y^{(i)})=1, \forall 1\leq i\leq N$.
The training algorithm proceeds by an exhaustive search over all possible parameters $\theta$ of the quantum model $\mathcal{M}_Q$, outputting the one with the maximal empirical likelihood.
Since the optimal quantum strategy consists of only classically-controlled Clifford operations, we restrict our search space to two-qubit Clifford gates $\mathcal{C}(2)$.
Due to weight-sharing, there's only a constant number of gates that need training, and each gate is only controlled by two classical bits.
This gives rise to a search space of size $|\mathcal{C}(2)|^{\bigo(1)}=\bigo(1)$.

For each possible parameter $\theta$, we execute the quantum model to calculate the empirical likelihood $l(\theta) = \frac{1}{N}\sum_{i=1}^N\log p_\theta (y^{(i)}|x^{(i)})$, where $p_\theta (y|x)$ is the condition distribution given by the quantum model $\mathcal{M}_Q$ with parameter $\theta$.
Since each sub-task $R^0$ of $R$ is independent to each other, $p_\theta$ factorizes into the product of $n$ sub-distributions $p_\theta^0$, giving rise to a sample mean approximation of $\mathbb{E}[\log p_\theta^0]$ with $nN$ independent samples. 
Standard concentration inequalities \cite{vershynin2018high} then assert that $nN=\bigo(1/\epsilon^2)$ suffices to estimate the likelihood to $\epsilon$ error with high probability.
Meanwhile, a constant error $\epsilon=\Theta(1)$ is enough for distinguishing between a constant number of parameters on $p^0_\theta$ which is independent of $n$.
Therefore, the training algorithm only needs $N=\bigo(1/n)$ many training data.
This $1/n$ decay reflects the fact that each data entry $(x^{(i)}, y^{(i)})$ already contains $n$ independent samples on the sub-task $R_0$, and the total number of such samples $M=\bigo(nN)=\bigo(1)$ stays constant.

To obtain the likelihood of each sample $p_\theta (y^{(i)}|x^{(i)})$ to constant error, we measure the expectation value of single-qubit observables $\mathcal{O}(y_j^{(i)}) = (1-y_j^{(i)})\ket{0}\bra{0}+y_j^{(i)}\ket{1}\bra{1}$ on qubit $j$ of the output state from $\mathcal{M}_Q$.
One such step costs $\bigo(n)$ time to run the constant-depth circuit, measure, and calculate the product.
Thus, the total running time of the training algorithm is $T=|\mathcal{C}(2)|^{\bigo(1)}\cdot N\cdot \bigo(n)=\bigo(1)$.

\vbox{}

\noindent\textbf{Data availability}\\
The data presented in the figures and that support the other findings of this study are available for download at \url{https://github.com/haimengzhao/qml-advantage}.

\vbox{}

\noindent \textbf{Acknowledgments}\\
We thank Eric R. Anschuetz, Matthias C. Caro, Xun Gao, Weiyuan Gong, Yingfei Gu, Hsin-Yuan Huang, Minghao Liu, Zhide Lu, Mehdi Soleimanifar, Penghui Yao, and Zhihan Zhang for helpful discussions. This work was supported by the National Natural Science Foundation of China (grant nos. T2225008, and  12075128), the Innovation Program for Quantum Science and Technology (grant no. 2021ZD0302203), the Tsinghua University Dushi Program, and the Shanghai Qi Zhi Institute. 

\vbox{}

\noindent \textbf{Author contributions}\\
H.Z. developed the theory and carried out the numerical simulations and experiments on IonQ Aria. D.-L.D. supervised the project.

\vbox{}

\noindent \textbf{Competing interests}\\
The authors declare no competing financial or non-financial interests.

\vbox{}

\noindent \textbf{Additional information}\\
Supplementary Materials are available in the online version of the paper.

\nocite{devlin2018bert,brown2020language,achiam2023gpt,jumper2021highly,merchant2023scaling,akiyama2019first,zhao2022magic,huang2022provably,carleo2017solving,torlai2018neural,zhao2024empirical,zhao2023learning,hochreiter1997long,cho2014learning,sutskever2014sequence,vaswani2017attention,radford2018improving,achiam2023gpt,collins2022capacity,li2022recent,li2021quantum,zhao2023non,zhao2023learning,caro2020pseudo,perez2021one,schuld2021effect,brunner2014bell,bell2004speakable,brassard2005quantum,yao1993quantum,brassard2003quantum,einstein1935can,mermin1990simple,peres1990incompatible,cabello2001all,cabello2001bell,jain2022direct,bharti2023power}
\bibliography{ref}

\providecommand{\noopsort}[1]{}\providecommand{\singleletter}[1]{#1}%
\begin{thebibliography}{51}%
\makeatletter
\providecommand \@ifxundefined [1]{%
 \@ifx{#1\undefined}
}%
\providecommand \@ifnum [1]{%
 \ifnum #1\expandafter \@firstoftwo
 \else \expandafter \@secondoftwo
 \fi
}%
\providecommand \@ifx [1]{%
 \ifx #1\expandafter \@firstoftwo
 \else \expandafter \@secondoftwo
 \fi
}%
\providecommand \natexlab [1]{#1}%
\providecommand \enquote  [1]{``#1''}%
\providecommand \bibnamefont  [1]{#1}%
\providecommand \bibfnamefont [1]{#1}%
\providecommand \citenamefont [1]{#1}%
\providecommand \href@noop [0]{\@secondoftwo}%
\providecommand \href [0]{\begingroup \@sanitize@url \@href}%
\providecommand \@href[1]{\@@startlink{#1}\@@href}%
\providecommand \@@href[1]{\endgroup#1\@@endlink}%
\providecommand \@sanitize@url [0]{\catcode `\\12\catcode `\$12\catcode `\&12\catcode `\#12\catcode `\^12\catcode `\_12\catcode `\%12\relax}%
\providecommand \@@startlink[1]{}%
\providecommand \@@endlink[0]{}%
\providecommand \url  [0]{\begingroup\@sanitize@url \@url }%
\providecommand \@url [1]{\endgroup\@href {#1}{\urlprefix }}%
\providecommand \urlprefix  [0]{URL }%
\providecommand \Eprint [0]{\href }%
\providecommand \doibase [0]{https://doi.org/}%
\providecommand \selectlanguage [0]{\@gobble}%
\providecommand \bibinfo  [0]{\@secondoftwo}%
\providecommand \bibfield  [0]{\@secondoftwo}%
\providecommand \translation [1]{[#1]}%
\providecommand \BibitemOpen [0]{}%
\providecommand \bibitemStop [0]{}%
\providecommand \bibitemNoStop [0]{.\EOS\space}%
\providecommand \EOS [0]{\spacefactor3000\relax}%
\providecommand \BibitemShut  [1]{\csname bibitem#1\endcsname}%
\let\auto@bib@innerbib\@empty
\bibitem [{\citenamefont {Goodfellow}\ \emph {et~al.}(2016)\citenamefont {Goodfellow}, \citenamefont {Bengio},\ and\ \citenamefont {Courville}}]{goodfellow2016deep}%
  \BibitemOpen
  \bibfield  {author} {\bibinfo {author} {\bibfnamefont {I.}~\bibnamefont {Goodfellow}}, \bibinfo {author} {\bibfnamefont {Y.}~\bibnamefont {Bengio}},\ and\ \bibinfo {author} {\bibfnamefont {A.}~\bibnamefont {Courville}},\ }\href@noop {} {\emph {\bibinfo {title} {Deep learning}}}\ (\bibinfo  {publisher} {MIT press},\ \bibinfo {year} {2016})\BibitemShut {NoStop}%
\bibitem [{\citenamefont {Devlin}\ \emph {et~al.}(2019)\citenamefont {Devlin}, \citenamefont {Chang}, \citenamefont {Lee},\ and\ \citenamefont {Toutanova}}]{devlin2018bert}%
  \BibitemOpen
  \bibfield  {author} {\bibinfo {author} {\bibfnamefont {J.}~\bibnamefont {Devlin}}, \bibinfo {author} {\bibfnamefont {M.-W.}\ \bibnamefont {Chang}}, \bibinfo {author} {\bibfnamefont {K.}~\bibnamefont {Lee}},\ and\ \bibinfo {author} {\bibfnamefont {K.}~\bibnamefont {Toutanova}},\ }\bibfield  {title} {\bibinfo {title} {{BERT}: Pre-training of deep bidirectional transformers for language understanding},\ }in\ \href {https://doi.org/10.18653/v1/N19-1423} {\emph {\bibinfo {booktitle} {Proceedings of the 2019 Conference of the North {A}merican Chapter of the Association for Computational Linguistics: Human Language Technologies, Volume 1 (Long and Short Papers)}}},\ \bibinfo {editor} {edited by\ \bibinfo {editor} {\bibfnamefont {J.}~\bibnamefont {Burstein}}, \bibinfo {editor} {\bibfnamefont {C.}~\bibnamefont {Doran}},\ and\ \bibinfo {editor} {\bibfnamefont {T.}~\bibnamefont {Solorio}}}\ (\bibinfo  {publisher} {Association for Computational Linguistics},\ \bibinfo {address} {Minneapolis, Minnesota},\ \bibinfo
  {year} {2019})\ pp.\ \bibinfo {pages} {4171--4186}\BibitemShut {NoStop}%
\bibitem [{\citenamefont {Brown}\ \emph {et~al.}(2020)\citenamefont {Brown}, \citenamefont {Mann}, \citenamefont {Ryder}, \citenamefont {Subbiah}, \citenamefont {Kaplan}, \citenamefont {Dhariwal}, \citenamefont {Neelakantan}, \citenamefont {Shyam}, \citenamefont {Sastry}, \citenamefont {Askell} \emph {et~al.}}]{brown2020language}%
  \BibitemOpen
  \bibfield  {author} {\bibinfo {author} {\bibfnamefont {T.}~\bibnamefont {Brown}}, \bibinfo {author} {\bibfnamefont {B.}~\bibnamefont {Mann}}, \bibinfo {author} {\bibfnamefont {N.}~\bibnamefont {Ryder}}, \bibinfo {author} {\bibfnamefont {M.}~\bibnamefont {Subbiah}}, \bibinfo {author} {\bibfnamefont {J.~D.}\ \bibnamefont {Kaplan}}, \bibinfo {author} {\bibfnamefont {P.}~\bibnamefont {Dhariwal}}, \bibinfo {author} {\bibfnamefont {A.}~\bibnamefont {Neelakantan}}, \bibinfo {author} {\bibfnamefont {P.}~\bibnamefont {Shyam}}, \bibinfo {author} {\bibfnamefont {G.}~\bibnamefont {Sastry}}, \bibinfo {author} {\bibfnamefont {A.}~\bibnamefont {Askell}}, \emph {et~al.},\ }\bibfield  {title} {\bibinfo {title} {Language models are few-shot learners},\ }\href@noop {} {\bibfield  {journal} {\bibinfo  {journal} {Advances in neural information processing systems}\ }\textbf {\bibinfo {volume} {33}},\ \bibinfo {pages} {1877} (\bibinfo {year} {2020})}\BibitemShut {NoStop}%
\bibitem [{\citenamefont {Achiam}\ \emph {et~al.}(2023)\citenamefont {Achiam}, \citenamefont {Adler}, \citenamefont {Agarwal}, \citenamefont {Ahmad}, \citenamefont {Akkaya}, \citenamefont {Aleman}, \citenamefont {Almeida}, \citenamefont {Altenschmidt}, \citenamefont {Altman}, \citenamefont {Anadkat} \emph {et~al.}}]{achiam2023gpt}%
  \BibitemOpen
  \bibfield  {author} {\bibinfo {author} {\bibfnamefont {J.}~\bibnamefont {Achiam}}, \bibinfo {author} {\bibfnamefont {S.}~\bibnamefont {Adler}}, \bibinfo {author} {\bibfnamefont {S.}~\bibnamefont {Agarwal}}, \bibinfo {author} {\bibfnamefont {L.}~\bibnamefont {Ahmad}}, \bibinfo {author} {\bibfnamefont {I.}~\bibnamefont {Akkaya}}, \bibinfo {author} {\bibfnamefont {F.~L.}\ \bibnamefont {Aleman}}, \bibinfo {author} {\bibfnamefont {D.}~\bibnamefont {Almeida}}, \bibinfo {author} {\bibfnamefont {J.}~\bibnamefont {Altenschmidt}}, \bibinfo {author} {\bibfnamefont {S.}~\bibnamefont {Altman}}, \bibinfo {author} {\bibfnamefont {S.}~\bibnamefont {Anadkat}}, \emph {et~al.},\ }\bibfield  {title} {\bibinfo {title} {Gpt-4 technical report},\ }\href@noop {} {\bibfield  {journal} {\bibinfo  {journal} {arXiv preprint arXiv:2303.08774}\ } (\bibinfo {year} {2023})}\BibitemShut {NoStop}%
\bibitem [{\citenamefont {Jumper}\ \emph {et~al.}(2021)\citenamefont {Jumper}, \citenamefont {Evans}, \citenamefont {Pritzel}, \citenamefont {Green}, \citenamefont {Figurnov}, \citenamefont {Ronneberger}, \citenamefont {Tunyasuvunakool}, \citenamefont {Bates}, \citenamefont {{\v{Z}}{\'\i}dek}, \citenamefont {Potapenko} \emph {et~al.}}]{jumper2021highly}%
  \BibitemOpen
  \bibfield  {author} {\bibinfo {author} {\bibfnamefont {J.}~\bibnamefont {Jumper}}, \bibinfo {author} {\bibfnamefont {R.}~\bibnamefont {Evans}}, \bibinfo {author} {\bibfnamefont {A.}~\bibnamefont {Pritzel}}, \bibinfo {author} {\bibfnamefont {T.}~\bibnamefont {Green}}, \bibinfo {author} {\bibfnamefont {M.}~\bibnamefont {Figurnov}}, \bibinfo {author} {\bibfnamefont {O.}~\bibnamefont {Ronneberger}}, \bibinfo {author} {\bibfnamefont {K.}~\bibnamefont {Tunyasuvunakool}}, \bibinfo {author} {\bibfnamefont {R.}~\bibnamefont {Bates}}, \bibinfo {author} {\bibfnamefont {A.}~\bibnamefont {{\v{Z}}{\'\i}dek}}, \bibinfo {author} {\bibfnamefont {A.}~\bibnamefont {Potapenko}}, \emph {et~al.},\ }\bibfield  {title} {\bibinfo {title} {Highly accurate protein structure prediction with alphafold},\ }\href@noop {} {\bibfield  {journal} {\bibinfo  {journal} {Nature}\ }\textbf {\bibinfo {volume} {596}},\ \bibinfo {pages} {583} (\bibinfo {year} {2021})}\BibitemShut {NoStop}%
\bibitem [{\citenamefont {Merchant}\ \emph {et~al.}(2023)\citenamefont {Merchant}, \citenamefont {Batzner}, \citenamefont {Schoenholz}, \citenamefont {Aykol}, \citenamefont {Cheon},\ and\ \citenamefont {Cubuk}}]{merchant2023scaling}%
  \BibitemOpen
  \bibfield  {author} {\bibinfo {author} {\bibfnamefont {A.}~\bibnamefont {Merchant}}, \bibinfo {author} {\bibfnamefont {S.}~\bibnamefont {Batzner}}, \bibinfo {author} {\bibfnamefont {S.~S.}\ \bibnamefont {Schoenholz}}, \bibinfo {author} {\bibfnamefont {M.}~\bibnamefont {Aykol}}, \bibinfo {author} {\bibfnamefont {G.}~\bibnamefont {Cheon}},\ and\ \bibinfo {author} {\bibfnamefont {E.~D.}\ \bibnamefont {Cubuk}},\ }\bibfield  {title} {\bibinfo {title} {Scaling deep learning for materials discovery},\ }\href@noop {} {\bibfield  {journal} {\bibinfo  {journal} {Nature}\ }\textbf {\bibinfo {volume} {624}},\ \bibinfo {pages} {80} (\bibinfo {year} {2023})}\BibitemShut {NoStop}%
\bibitem [{\citenamefont {Akiyama}\ \emph {et~al.}(2019)\citenamefont {Akiyama}, \citenamefont {Alberdi}, \citenamefont {Alef}, \citenamefont {Asada}, \citenamefont {Azulay}, \citenamefont {Baczko}, \citenamefont {Ball}, \citenamefont {Balokovi{\'c}}, \citenamefont {Barrett}, \citenamefont {Bintley} \emph {et~al.}}]{akiyama2019first}%
  \BibitemOpen
  \bibfield  {author} {\bibinfo {author} {\bibfnamefont {K.}~\bibnamefont {Akiyama}}, \bibinfo {author} {\bibfnamefont {A.}~\bibnamefont {Alberdi}}, \bibinfo {author} {\bibfnamefont {W.}~\bibnamefont {Alef}}, \bibinfo {author} {\bibfnamefont {K.}~\bibnamefont {Asada}}, \bibinfo {author} {\bibfnamefont {R.}~\bibnamefont {Azulay}}, \bibinfo {author} {\bibfnamefont {A.-K.}\ \bibnamefont {Baczko}}, \bibinfo {author} {\bibfnamefont {D.}~\bibnamefont {Ball}}, \bibinfo {author} {\bibfnamefont {M.}~\bibnamefont {Balokovi{\'c}}}, \bibinfo {author} {\bibfnamefont {J.}~\bibnamefont {Barrett}}, \bibinfo {author} {\bibfnamefont {D.}~\bibnamefont {Bintley}}, \emph {et~al.},\ }\bibfield  {title} {\bibinfo {title} {First m87 event horizon telescope results. iv. imaging the central supermassive black hole},\ }\href@noop {} {\bibfield  {journal} {\bibinfo  {journal} {The Astrophysical Journal Letters}\ }\textbf {\bibinfo {volume} {875}},\ \bibinfo {pages} {L4} (\bibinfo {year} {2019})}\BibitemShut {NoStop}%
\bibitem [{\citenamefont {Zhao}\ and\ \citenamefont {Zhu}(2022)}]{zhao2022magic}%
  \BibitemOpen
  \bibfield  {author} {\bibinfo {author} {\bibfnamefont {H.}~\bibnamefont {Zhao}}\ and\ \bibinfo {author} {\bibfnamefont {W.}~\bibnamefont {Zhu}},\ }\bibfield  {title} {\bibinfo {title} {Magic: Microlensing analysis guided by intelligent computation},\ }\href@noop {} {\bibfield  {journal} {\bibinfo  {journal} {The Astronomical Journal}\ }\textbf {\bibinfo {volume} {164}},\ \bibinfo {pages} {192} (\bibinfo {year} {2022})}\BibitemShut {NoStop}%
\bibitem [{\citenamefont {Huang}\ \emph {et~al.}(2022{\natexlab{a}})\citenamefont {Huang}, \citenamefont {Kueng}, \citenamefont {Torlai}, \citenamefont {Albert},\ and\ \citenamefont {Preskill}}]{huang2022provably}%
  \BibitemOpen
  \bibfield  {author} {\bibinfo {author} {\bibfnamefont {H.-Y.}\ \bibnamefont {Huang}}, \bibinfo {author} {\bibfnamefont {R.}~\bibnamefont {Kueng}}, \bibinfo {author} {\bibfnamefont {G.}~\bibnamefont {Torlai}}, \bibinfo {author} {\bibfnamefont {V.~V.}\ \bibnamefont {Albert}},\ and\ \bibinfo {author} {\bibfnamefont {J.}~\bibnamefont {Preskill}},\ }\bibfield  {title} {\bibinfo {title} {Provably efficient machine learning for quantum many-body problems},\ }\href@noop {} {\bibfield  {journal} {\bibinfo  {journal} {Science}\ }\textbf {\bibinfo {volume} {377}},\ \bibinfo {pages} {eabk3333} (\bibinfo {year} {2022}{\natexlab{a}})}\BibitemShut {NoStop}%
\bibitem [{\citenamefont {Carleo}\ and\ \citenamefont {Troyer}(2017)}]{carleo2017solving}%
  \BibitemOpen
  \bibfield  {author} {\bibinfo {author} {\bibfnamefont {G.}~\bibnamefont {Carleo}}\ and\ \bibinfo {author} {\bibfnamefont {M.}~\bibnamefont {Troyer}},\ }\bibfield  {title} {\bibinfo {title} {Solving the quantum many-body problem with artificial neural networks},\ }\href@noop {} {\bibfield  {journal} {\bibinfo  {journal} {Science}\ }\textbf {\bibinfo {volume} {355}},\ \bibinfo {pages} {602} (\bibinfo {year} {2017})}\BibitemShut {NoStop}%
\bibitem [{\citenamefont {Torlai}\ \emph {et~al.}(2018)\citenamefont {Torlai}, \citenamefont {Mazzola}, \citenamefont {Carrasquilla}, \citenamefont {Troyer}, \citenamefont {Melko},\ and\ \citenamefont {Carleo}}]{torlai2018neural}%
  \BibitemOpen
  \bibfield  {author} {\bibinfo {author} {\bibfnamefont {G.}~\bibnamefont {Torlai}}, \bibinfo {author} {\bibfnamefont {G.}~\bibnamefont {Mazzola}}, \bibinfo {author} {\bibfnamefont {J.}~\bibnamefont {Carrasquilla}}, \bibinfo {author} {\bibfnamefont {M.}~\bibnamefont {Troyer}}, \bibinfo {author} {\bibfnamefont {R.}~\bibnamefont {Melko}},\ and\ \bibinfo {author} {\bibfnamefont {G.}~\bibnamefont {Carleo}},\ }\bibfield  {title} {\bibinfo {title} {Neural-network quantum state tomography},\ }\href@noop {} {\bibfield  {journal} {\bibinfo  {journal} {Nature physics}\ }\textbf {\bibinfo {volume} {14}},\ \bibinfo {pages} {447} (\bibinfo {year} {2018})}\BibitemShut {NoStop}%
\bibitem [{\citenamefont {Zhao}\ \emph {et~al.}(2024)\citenamefont {Zhao}, \citenamefont {Carleo},\ and\ \citenamefont {Vicentini}}]{zhao2024empirical}%
  \BibitemOpen
  \bibfield  {author} {\bibinfo {author} {\bibfnamefont {H.}~\bibnamefont {Zhao}}, \bibinfo {author} {\bibfnamefont {G.}~\bibnamefont {Carleo}},\ and\ \bibinfo {author} {\bibfnamefont {F.}~\bibnamefont {Vicentini}},\ }\bibfield  {title} {\bibinfo {title} {Empirical sample complexity of neural network mixed state reconstruction},\ }\href@noop {} {\bibfield  {journal} {\bibinfo  {journal} {Quantum}\ }\textbf {\bibinfo {volume} {8}},\ \bibinfo {pages} {1358} (\bibinfo {year} {2024})}\BibitemShut {NoStop}%
\bibitem [{\citenamefont {Zhao}\ \emph {et~al.}(2023)\citenamefont {Zhao}, \citenamefont {Lewis}, \citenamefont {Kannan}, \citenamefont {Quek}, \citenamefont {Huang},\ and\ \citenamefont {Caro}}]{zhao2023learning}%
  \BibitemOpen
  \bibfield  {author} {\bibinfo {author} {\bibfnamefont {H.}~\bibnamefont {Zhao}}, \bibinfo {author} {\bibfnamefont {L.}~\bibnamefont {Lewis}}, \bibinfo {author} {\bibfnamefont {I.}~\bibnamefont {Kannan}}, \bibinfo {author} {\bibfnamefont {Y.}~\bibnamefont {Quek}}, \bibinfo {author} {\bibfnamefont {H.-Y.}\ \bibnamefont {Huang}},\ and\ \bibinfo {author} {\bibfnamefont {M.~C.}\ \bibnamefont {Caro}},\ }\bibfield  {title} {\bibinfo {title} {Learning quantum states and unitaries of bounded gate complexity},\ }\href@noop {} {\bibfield  {journal} {\bibinfo  {journal} {arXiv preprint arXiv:2310.19882}\ } (\bibinfo {year} {2023})}\BibitemShut {NoStop}%
\bibitem [{\citenamefont {Hochreiter}\ and\ \citenamefont {Schmidhuber}(1997)}]{hochreiter1997long}%
  \BibitemOpen
  \bibfield  {author} {\bibinfo {author} {\bibfnamefont {S.}~\bibnamefont {Hochreiter}}\ and\ \bibinfo {author} {\bibfnamefont {J.}~\bibnamefont {Schmidhuber}},\ }\bibfield  {title} {\bibinfo {title} {Long short-term memory},\ }\href@noop {} {\bibfield  {journal} {\bibinfo  {journal} {Neural computation}\ }\textbf {\bibinfo {volume} {9}},\ \bibinfo {pages} {1735} (\bibinfo {year} {1997})}\BibitemShut {NoStop}%
\bibitem [{\citenamefont {Cho}\ \emph {et~al.}(2014)\citenamefont {Cho}, \citenamefont {Van~Merri{\"e}nboer}, \citenamefont {Gulcehre}, \citenamefont {Bahdanau}, \citenamefont {Bougares}, \citenamefont {Schwenk},\ and\ \citenamefont {Bengio}}]{cho2014learning}%
  \BibitemOpen
  \bibfield  {author} {\bibinfo {author} {\bibfnamefont {K.}~\bibnamefont {Cho}}, \bibinfo {author} {\bibfnamefont {B.}~\bibnamefont {Van~Merri{\"e}nboer}}, \bibinfo {author} {\bibfnamefont {C.}~\bibnamefont {Gulcehre}}, \bibinfo {author} {\bibfnamefont {D.}~\bibnamefont {Bahdanau}}, \bibinfo {author} {\bibfnamefont {F.}~\bibnamefont {Bougares}}, \bibinfo {author} {\bibfnamefont {H.}~\bibnamefont {Schwenk}},\ and\ \bibinfo {author} {\bibfnamefont {Y.}~\bibnamefont {Bengio}},\ }\bibfield  {title} {\bibinfo {title} {Learning phrase representations using rnn encoder--decoder for statistical machine translation},\ }in\ \href@noop {} {\emph {\bibinfo {booktitle} {Proceedings of the 2014 Conference on Empirical Methods in Natural Language Processing (EMNLP)}}}\ (\bibinfo {year} {2014})\ pp.\ \bibinfo {pages} {1724--1734}\BibitemShut {NoStop}%
\bibitem [{\citenamefont {Sutskever}\ \emph {et~al.}(2014)\citenamefont {Sutskever}, \citenamefont {Vinyals},\ and\ \citenamefont {Le}}]{sutskever2014sequence}%
  \BibitemOpen
  \bibfield  {author} {\bibinfo {author} {\bibfnamefont {I.}~\bibnamefont {Sutskever}}, \bibinfo {author} {\bibfnamefont {O.}~\bibnamefont {Vinyals}},\ and\ \bibinfo {author} {\bibfnamefont {Q.~V.}\ \bibnamefont {Le}},\ }\bibfield  {title} {\bibinfo {title} {Sequence to sequence learning with neural networks},\ }\href@noop {} {\bibfield  {journal} {\bibinfo  {journal} {Advances in neural information processing systems}\ }\textbf {\bibinfo {volume} {27}} (\bibinfo {year} {2014})}\BibitemShut {NoStop}%
\bibitem [{\citenamefont {Vaswani}\ \emph {et~al.}(2017)\citenamefont {Vaswani}, \citenamefont {Shazeer}, \citenamefont {Parmar}, \citenamefont {Uszkoreit}, \citenamefont {Jones}, \citenamefont {Gomez}, \citenamefont {Kaiser},\ and\ \citenamefont {Polosukhin}}]{vaswani2017attention}%
  \BibitemOpen
  \bibfield  {author} {\bibinfo {author} {\bibfnamefont {A.}~\bibnamefont {Vaswani}}, \bibinfo {author} {\bibfnamefont {N.}~\bibnamefont {Shazeer}}, \bibinfo {author} {\bibfnamefont {N.}~\bibnamefont {Parmar}}, \bibinfo {author} {\bibfnamefont {J.}~\bibnamefont {Uszkoreit}}, \bibinfo {author} {\bibfnamefont {L.}~\bibnamefont {Jones}}, \bibinfo {author} {\bibfnamefont {A.~N.}\ \bibnamefont {Gomez}}, \bibinfo {author} {\bibfnamefont {{\L}.}~\bibnamefont {Kaiser}},\ and\ \bibinfo {author} {\bibfnamefont {I.}~\bibnamefont {Polosukhin}},\ }\bibfield  {title} {\bibinfo {title} {Attention is all you need},\ }\href@noop {} {\bibfield  {journal} {\bibinfo  {journal} {Advances in neural information processing systems}\ }\textbf {\bibinfo {volume} {30}} (\bibinfo {year} {2017})}\BibitemShut {NoStop}%
\bibitem [{\citenamefont {Radford}\ \emph {et~al.}(2018)\citenamefont {Radford}, \citenamefont {Narasimhan}, \citenamefont {Salimans}, \citenamefont {Sutskever} \emph {et~al.}}]{radford2018improving}%
  \BibitemOpen
  \bibfield  {author} {\bibinfo {author} {\bibfnamefont {A.}~\bibnamefont {Radford}}, \bibinfo {author} {\bibfnamefont {K.}~\bibnamefont {Narasimhan}}, \bibinfo {author} {\bibfnamefont {T.}~\bibnamefont {Salimans}}, \bibinfo {author} {\bibfnamefont {I.}~\bibnamefont {Sutskever}}, \emph {et~al.},\ }\href {https://cdn.openai.com/research-covers/language-unsupervised/language_understanding_paper.pdf} {\bibinfo {title} {Improving language understanding by generative pre-training}} (\bibinfo {year} {2018})\BibitemShut {NoStop}%
\bibitem [{\citenamefont {Raz}(2018)}]{raz2018fast}%
  \BibitemOpen
  \bibfield  {author} {\bibinfo {author} {\bibfnamefont {R.}~\bibnamefont {Raz}},\ }\bibfield  {title} {\bibinfo {title} {Fast learning requires good memory: A time-space lower bound for parity learning},\ }\href@noop {} {\bibfield  {journal} {\bibinfo  {journal} {Journal of the ACM (JACM)}\ }\textbf {\bibinfo {volume} {66}},\ \bibinfo {pages} {1} (\bibinfo {year} {2018})}\BibitemShut {NoStop}%
\bibitem [{\citenamefont {Yao}(1993)}]{yao1993quantum}%
  \BibitemOpen
  \bibfield  {author} {\bibinfo {author} {\bibfnamefont {A.~C.-C.}\ \bibnamefont {Yao}},\ }\bibfield  {title} {\bibinfo {title} {Quantum circuit complexity},\ }in\ \href@noop {} {\emph {\bibinfo {booktitle} {Proceedings of 1993 IEEE 34th Annual Foundations of Computer Science}}}\ (\bibinfo {organization} {IEEE},\ \bibinfo {year} {1993})\ pp.\ \bibinfo {pages} {352--361}\BibitemShut {NoStop}%
\bibitem [{\citenamefont {Brassard}(2003)}]{brassard2003quantum}%
  \BibitemOpen
  \bibfield  {author} {\bibinfo {author} {\bibfnamefont {G.}~\bibnamefont {Brassard}},\ }\bibfield  {title} {\bibinfo {title} {Quantum communication complexity},\ }\href@noop {} {\bibfield  {journal} {\bibinfo  {journal} {Foundations of Physics}\ }\textbf {\bibinfo {volume} {33}},\ \bibinfo {pages} {1593} (\bibinfo {year} {2003})}\BibitemShut {NoStop}%
\bibitem [{\citenamefont {Cerezo}\ \emph {et~al.}(2021)\citenamefont {Cerezo}, \citenamefont {Arrasmith}, \citenamefont {Babbush}, \citenamefont {Benjamin}, \citenamefont {Endo}, \citenamefont {Fujii}, \citenamefont {McClean}, \citenamefont {Mitarai}, \citenamefont {Yuan}, \citenamefont {Cincio} \emph {et~al.}}]{cerezo2021variational}%
  \BibitemOpen
  \bibfield  {author} {\bibinfo {author} {\bibfnamefont {M.}~\bibnamefont {Cerezo}}, \bibinfo {author} {\bibfnamefont {A.}~\bibnamefont {Arrasmith}}, \bibinfo {author} {\bibfnamefont {R.}~\bibnamefont {Babbush}}, \bibinfo {author} {\bibfnamefont {S.~C.}\ \bibnamefont {Benjamin}}, \bibinfo {author} {\bibfnamefont {S.}~\bibnamefont {Endo}}, \bibinfo {author} {\bibfnamefont {K.}~\bibnamefont {Fujii}}, \bibinfo {author} {\bibfnamefont {J.~R.}\ \bibnamefont {McClean}}, \bibinfo {author} {\bibfnamefont {K.}~\bibnamefont {Mitarai}}, \bibinfo {author} {\bibfnamefont {X.}~\bibnamefont {Yuan}}, \bibinfo {author} {\bibfnamefont {L.}~\bibnamefont {Cincio}}, \emph {et~al.},\ }\bibfield  {title} {\bibinfo {title} {Variational quantum algorithms},\ }\href@noop {} {\bibfield  {journal} {\bibinfo  {journal} {Nature Reviews Physics}\ }\textbf {\bibinfo {volume} {3}},\ \bibinfo {pages} {625} (\bibinfo {year} {2021})}\BibitemShut {NoStop}%
\bibitem [{\citenamefont {Dalzell}\ \emph {et~al.}(2023)\citenamefont {Dalzell}, \citenamefont {McArdle}, \citenamefont {Berta}, \citenamefont {Bienias}, \citenamefont {Chen}, \citenamefont {Gily{\'e}n}, \citenamefont {Hann}, \citenamefont {Kastoryano}, \citenamefont {Khabiboulline}, \citenamefont {Kubica} \emph {et~al.}}]{dalzell2023quantum}%
  \BibitemOpen
  \bibfield  {author} {\bibinfo {author} {\bibfnamefont {A.~M.}\ \bibnamefont {Dalzell}}, \bibinfo {author} {\bibfnamefont {S.}~\bibnamefont {McArdle}}, \bibinfo {author} {\bibfnamefont {M.}~\bibnamefont {Berta}}, \bibinfo {author} {\bibfnamefont {P.}~\bibnamefont {Bienias}}, \bibinfo {author} {\bibfnamefont {C.-F.}\ \bibnamefont {Chen}}, \bibinfo {author} {\bibfnamefont {A.}~\bibnamefont {Gily{\'e}n}}, \bibinfo {author} {\bibfnamefont {C.~T.}\ \bibnamefont {Hann}}, \bibinfo {author} {\bibfnamefont {M.~J.}\ \bibnamefont {Kastoryano}}, \bibinfo {author} {\bibfnamefont {E.~T.}\ \bibnamefont {Khabiboulline}}, \bibinfo {author} {\bibfnamefont {A.}~\bibnamefont {Kubica}}, \emph {et~al.},\ }\bibfield  {title} {\bibinfo {title} {Quantum algorithms: A survey of applications and end-to-end complexities},\ }\href@noop {} {\bibfield  {journal} {\bibinfo  {journal} {arXiv preprint arXiv:2310.03011}\ } (\bibinfo {year} {2023})}\BibitemShut {NoStop}%
\bibitem [{\citenamefont {Lloyd}\ \emph {et~al.}(2014)\citenamefont {Lloyd}, \citenamefont {Mohseni},\ and\ \citenamefont {Rebentrost}}]{lloyd2014quantum}%
  \BibitemOpen
  \bibfield  {author} {\bibinfo {author} {\bibfnamefont {S.}~\bibnamefont {Lloyd}}, \bibinfo {author} {\bibfnamefont {M.}~\bibnamefont {Mohseni}},\ and\ \bibinfo {author} {\bibfnamefont {P.}~\bibnamefont {Rebentrost}},\ }\bibfield  {title} {\bibinfo {title} {Quantum principal component analysis},\ }\href@noop {} {\bibfield  {journal} {\bibinfo  {journal} {Nature physics}\ }\textbf {\bibinfo {volume} {10}},\ \bibinfo {pages} {631} (\bibinfo {year} {2014})}\BibitemShut {NoStop}%
\bibitem [{\citenamefont {Rebentrost}\ \emph {et~al.}(2014)\citenamefont {Rebentrost}, \citenamefont {Mohseni},\ and\ \citenamefont {Lloyd}}]{rebentrost2014quantum}%
  \BibitemOpen
  \bibfield  {author} {\bibinfo {author} {\bibfnamefont {P.}~\bibnamefont {Rebentrost}}, \bibinfo {author} {\bibfnamefont {M.}~\bibnamefont {Mohseni}},\ and\ \bibinfo {author} {\bibfnamefont {S.}~\bibnamefont {Lloyd}},\ }\bibfield  {title} {\bibinfo {title} {Quantum support vector machine for big data classification},\ }\href@noop {} {\bibfield  {journal} {\bibinfo  {journal} {Physical review letters}\ }\textbf {\bibinfo {volume} {113}},\ \bibinfo {pages} {130503} (\bibinfo {year} {2014})}\BibitemShut {NoStop}%
\bibitem [{\citenamefont {Dunjko}\ \emph {et~al.}(2016)\citenamefont {Dunjko}, \citenamefont {Taylor},\ and\ \citenamefont {Briegel}}]{dunjko2016quantum}%
  \BibitemOpen
  \bibfield  {author} {\bibinfo {author} {\bibfnamefont {V.}~\bibnamefont {Dunjko}}, \bibinfo {author} {\bibfnamefont {J.~M.}\ \bibnamefont {Taylor}},\ and\ \bibinfo {author} {\bibfnamefont {H.~J.}\ \bibnamefont {Briegel}},\ }\bibfield  {title} {\bibinfo {title} {Quantum-enhanced machine learning},\ }\href@noop {} {\bibfield  {journal} {\bibinfo  {journal} {Physical review letters}\ }\textbf {\bibinfo {volume} {117}},\ \bibinfo {pages} {130501} (\bibinfo {year} {2016})}\BibitemShut {NoStop}%
\bibitem [{\citenamefont {Gao}\ \emph {et~al.}(2018)\citenamefont {Gao}, \citenamefont {Zhang},\ and\ \citenamefont {Duan}}]{gao2018quantum}%
  \BibitemOpen
  \bibfield  {author} {\bibinfo {author} {\bibfnamefont {X.}~\bibnamefont {Gao}}, \bibinfo {author} {\bibfnamefont {Z.-Y.}\ \bibnamefont {Zhang}},\ and\ \bibinfo {author} {\bibfnamefont {L.-M.}\ \bibnamefont {Duan}},\ }\bibfield  {title} {\bibinfo {title} {A quantum machine learning algorithm based on generative models},\ }\href@noop {} {\bibfield  {journal} {\bibinfo  {journal} {Science advances}\ }\textbf {\bibinfo {volume} {4}},\ \bibinfo {pages} {eaat9004} (\bibinfo {year} {2018})}\BibitemShut {NoStop}%
\bibitem [{\citenamefont {Lloyd}\ and\ \citenamefont {Weedbrook}(2018)}]{lloyd2018quantum}%
  \BibitemOpen
  \bibfield  {author} {\bibinfo {author} {\bibfnamefont {S.}~\bibnamefont {Lloyd}}\ and\ \bibinfo {author} {\bibfnamefont {C.}~\bibnamefont {Weedbrook}},\ }\bibfield  {title} {\bibinfo {title} {Quantum generative adversarial learning},\ }\href@noop {} {\bibfield  {journal} {\bibinfo  {journal} {Physical review letters}\ }\textbf {\bibinfo {volume} {121}},\ \bibinfo {pages} {040502} (\bibinfo {year} {2018})}\BibitemShut {NoStop}%
\bibitem [{\citenamefont {Schuld}\ and\ \citenamefont {Killoran}(2019)}]{schuld2019quantum}%
  \BibitemOpen
  \bibfield  {author} {\bibinfo {author} {\bibfnamefont {M.}~\bibnamefont {Schuld}}\ and\ \bibinfo {author} {\bibfnamefont {N.}~\bibnamefont {Killoran}},\ }\bibfield  {title} {\bibinfo {title} {Quantum machine learning in feature hilbert spaces},\ }\href@noop {} {\bibfield  {journal} {\bibinfo  {journal} {Physical review letters}\ }\textbf {\bibinfo {volume} {122}},\ \bibinfo {pages} {040504} (\bibinfo {year} {2019})}\BibitemShut {NoStop}%
\bibitem [{\citenamefont {Liu}\ \emph {et~al.}(2021)\citenamefont {Liu}, \citenamefont {Arunachalam},\ and\ \citenamefont {Temme}}]{liu2021rigorous}%
  \BibitemOpen
  \bibfield  {author} {\bibinfo {author} {\bibfnamefont {Y.}~\bibnamefont {Liu}}, \bibinfo {author} {\bibfnamefont {S.}~\bibnamefont {Arunachalam}},\ and\ \bibinfo {author} {\bibfnamefont {K.}~\bibnamefont {Temme}},\ }\bibfield  {title} {\bibinfo {title} {A rigorous and robust quantum speed-up in supervised machine learning},\ }\href@noop {} {\bibfield  {journal} {\bibinfo  {journal} {Nature Physics}\ }\textbf {\bibinfo {volume} {17}},\ \bibinfo {pages} {1013} (\bibinfo {year} {2021})}\BibitemShut {NoStop}%
\bibitem [{\citenamefont {Huang}\ \emph {et~al.}(2022{\natexlab{b}})\citenamefont {Huang}, \citenamefont {Broughton}, \citenamefont {Cotler}, \citenamefont {Chen}, \citenamefont {Li}, \citenamefont {Mohseni}, \citenamefont {Neven}, \citenamefont {Babbush}, \citenamefont {Kueng}, \citenamefont {Preskill} \emph {et~al.}}]{huang2022quantum}%
  \BibitemOpen
  \bibfield  {author} {\bibinfo {author} {\bibfnamefont {H.-Y.}\ \bibnamefont {Huang}}, \bibinfo {author} {\bibfnamefont {M.}~\bibnamefont {Broughton}}, \bibinfo {author} {\bibfnamefont {J.}~\bibnamefont {Cotler}}, \bibinfo {author} {\bibfnamefont {S.}~\bibnamefont {Chen}}, \bibinfo {author} {\bibfnamefont {J.}~\bibnamefont {Li}}, \bibinfo {author} {\bibfnamefont {M.}~\bibnamefont {Mohseni}}, \bibinfo {author} {\bibfnamefont {H.}~\bibnamefont {Neven}}, \bibinfo {author} {\bibfnamefont {R.}~\bibnamefont {Babbush}}, \bibinfo {author} {\bibfnamefont {R.}~\bibnamefont {Kueng}}, \bibinfo {author} {\bibfnamefont {J.}~\bibnamefont {Preskill}}, \emph {et~al.},\ }\bibfield  {title} {\bibinfo {title} {Quantum advantage in learning from experiments},\ }\href@noop {} {\bibfield  {journal} {\bibinfo  {journal} {Science}\ }\textbf {\bibinfo {volume} {376}},\ \bibinfo {pages} {1182} (\bibinfo {year} {2022}{\natexlab{b}})}\BibitemShut {NoStop}%
\bibitem [{\citenamefont {Li}\ and\ \citenamefont {Deng}(2022)}]{li2022recent}%
  \BibitemOpen
  \bibfield  {author} {\bibinfo {author} {\bibfnamefont {W.}~\bibnamefont {Li}}\ and\ \bibinfo {author} {\bibfnamefont {D.-L.}\ \bibnamefont {Deng}},\ }\bibfield  {title} {\bibinfo {title} {Recent advances for quantum classifiers},\ }\href@noop {} {\bibfield  {journal} {\bibinfo  {journal} {Science China Physics, Mechanics \& Astronomy}\ }\textbf {\bibinfo {volume} {65}},\ \bibinfo {pages} {220301} (\bibinfo {year} {2022})}\BibitemShut {NoStop}%
\bibitem [{\citenamefont {Li}\ \emph {et~al.}(2021)\citenamefont {Li}, \citenamefont {Lu},\ and\ \citenamefont {Deng}}]{li2021quantum}%
  \BibitemOpen
  \bibfield  {author} {\bibinfo {author} {\bibfnamefont {W.}~\bibnamefont {Li}}, \bibinfo {author} {\bibfnamefont {S.}~\bibnamefont {Lu}},\ and\ \bibinfo {author} {\bibfnamefont {D.-L.}\ \bibnamefont {Deng}},\ }\bibfield  {title} {\bibinfo {title} {Quantum federated learning through blind quantum computing},\ }\href@noop {} {\bibfield  {journal} {\bibinfo  {journal} {Science China Physics, Mechanics \& Astronomy}\ }\textbf {\bibinfo {volume} {64}},\ \bibinfo {pages} {100312} (\bibinfo {year} {2021})}\BibitemShut {NoStop}%
\bibitem [{\citenamefont {Zhao}(2023)}]{zhao2023non}%
  \BibitemOpen
  \bibfield  {author} {\bibinfo {author} {\bibfnamefont {H.}~\bibnamefont {Zhao}},\ }\bibfield  {title} {\bibinfo {title} {Non-iid quantum federated learning with one-shot communication complexity},\ }\href@noop {} {\bibfield  {journal} {\bibinfo  {journal} {Quantum Machine Intelligence}\ }\textbf {\bibinfo {volume} {5}},\ \bibinfo {pages} {3} (\bibinfo {year} {2023})}\BibitemShut {NoStop}%
\bibitem [{\citenamefont {Caro}\ and\ \citenamefont {Datta}(2020)}]{caro2020pseudo}%
  \BibitemOpen
  \bibfield  {author} {\bibinfo {author} {\bibfnamefont {M.~C.}\ \bibnamefont {Caro}}\ and\ \bibinfo {author} {\bibfnamefont {I.}~\bibnamefont {Datta}},\ }\bibfield  {title} {\bibinfo {title} {Pseudo-dimension of quantum circuits},\ }\href@noop {} {\bibfield  {journal} {\bibinfo  {journal} {Quantum Machine Intelligence}\ }\textbf {\bibinfo {volume} {2}},\ \bibinfo {pages} {14} (\bibinfo {year} {2020})}\BibitemShut {NoStop}%
\bibitem [{\citenamefont {P{\'e}rez-Salinas}\ \emph {et~al.}(2021)\citenamefont {P{\'e}rez-Salinas}, \citenamefont {L{\'o}pez-N{\'u}{\~n}ez}, \citenamefont {Garc{\'\i}a-S{\'a}ez}, \citenamefont {Forn-D{\'\i}az},\ and\ \citenamefont {Latorre}}]{perez2021one}%
  \BibitemOpen
  \bibfield  {author} {\bibinfo {author} {\bibfnamefont {A.}~\bibnamefont {P{\'e}rez-Salinas}}, \bibinfo {author} {\bibfnamefont {D.}~\bibnamefont {L{\'o}pez-N{\'u}{\~n}ez}}, \bibinfo {author} {\bibfnamefont {A.}~\bibnamefont {Garc{\'\i}a-S{\'a}ez}}, \bibinfo {author} {\bibfnamefont {P.}~\bibnamefont {Forn-D{\'\i}az}},\ and\ \bibinfo {author} {\bibfnamefont {J.~I.}\ \bibnamefont {Latorre}},\ }\bibfield  {title} {\bibinfo {title} {One qubit as a universal approximant},\ }\href@noop {} {\bibfield  {journal} {\bibinfo  {journal} {Physical Review A}\ }\textbf {\bibinfo {volume} {104}},\ \bibinfo {pages} {012405} (\bibinfo {year} {2021})}\BibitemShut {NoStop}%
\bibitem [{\citenamefont {Schuld}\ \emph {et~al.}(2021)\citenamefont {Schuld}, \citenamefont {Sweke},\ and\ \citenamefont {Meyer}}]{schuld2021effect}%
  \BibitemOpen
  \bibfield  {author} {\bibinfo {author} {\bibfnamefont {M.}~\bibnamefont {Schuld}}, \bibinfo {author} {\bibfnamefont {R.}~\bibnamefont {Sweke}},\ and\ \bibinfo {author} {\bibfnamefont {J.~J.}\ \bibnamefont {Meyer}},\ }\bibfield  {title} {\bibinfo {title} {Effect of data encoding on the expressive power of variational quantum-machine-learning models},\ }\href@noop {} {\bibfield  {journal} {\bibinfo  {journal} {Physical Review A}\ }\textbf {\bibinfo {volume} {103}},\ \bibinfo {pages} {032430} (\bibinfo {year} {2021})}\BibitemShut {NoStop}%
\bibitem [{\citenamefont {Brunner}\ \emph {et~al.}(2014)\citenamefont {Brunner}, \citenamefont {Cavalcanti}, \citenamefont {Pironio}, \citenamefont {Scarani},\ and\ \citenamefont {Wehner}}]{brunner2014bell}%
  \BibitemOpen
  \bibfield  {author} {\bibinfo {author} {\bibfnamefont {N.}~\bibnamefont {Brunner}}, \bibinfo {author} {\bibfnamefont {D.}~\bibnamefont {Cavalcanti}}, \bibinfo {author} {\bibfnamefont {S.}~\bibnamefont {Pironio}}, \bibinfo {author} {\bibfnamefont {V.}~\bibnamefont {Scarani}},\ and\ \bibinfo {author} {\bibfnamefont {S.}~\bibnamefont {Wehner}},\ }\bibfield  {title} {\bibinfo {title} {Bell nonlocality},\ }\href@noop {} {\bibfield  {journal} {\bibinfo  {journal} {Reviews of modern physics}\ }\textbf {\bibinfo {volume} {86}},\ \bibinfo {pages} {419} (\bibinfo {year} {2014})}\BibitemShut {NoStop}%
\bibitem [{\citenamefont {Bell}(2004)}]{bell2004speakable}%
  \BibitemOpen
  \bibfield  {author} {\bibinfo {author} {\bibfnamefont {J.~S.}\ \bibnamefont {Bell}},\ }\href@noop {} {\emph {\bibinfo {title} {Speakable and unspeakable in quantum mechanics: Collected papers on quantum philosophy}}}\ (\bibinfo  {publisher} {Cambridge university press},\ \bibinfo {year} {2004})\BibitemShut {NoStop}%
\bibitem [{\citenamefont {Brassard}\ \emph {et~al.}(2005)\citenamefont {Brassard}, \citenamefont {Broadbent},\ and\ \citenamefont {Tapp}}]{brassard2005quantum}%
  \BibitemOpen
  \bibfield  {author} {\bibinfo {author} {\bibfnamefont {G.}~\bibnamefont {Brassard}}, \bibinfo {author} {\bibfnamefont {A.}~\bibnamefont {Broadbent}},\ and\ \bibinfo {author} {\bibfnamefont {A.}~\bibnamefont {Tapp}},\ }\bibfield  {title} {\bibinfo {title} {Quantum pseudo-telepathy},\ }\href@noop {} {\bibfield  {journal} {\bibinfo  {journal} {Foundations of Physics}\ }\textbf {\bibinfo {volume} {35}},\ \bibinfo {pages} {1877} (\bibinfo {year} {2005})}\BibitemShut {NoStop}%
\bibitem [{\citenamefont {Einstein}\ \emph {et~al.}(1935)\citenamefont {Einstein}, \citenamefont {Podolsky},\ and\ \citenamefont {Rosen}}]{einstein1935can}%
  \BibitemOpen
  \bibfield  {author} {\bibinfo {author} {\bibfnamefont {A.}~\bibnamefont {Einstein}}, \bibinfo {author} {\bibfnamefont {B.}~\bibnamefont {Podolsky}},\ and\ \bibinfo {author} {\bibfnamefont {N.}~\bibnamefont {Rosen}},\ }\bibfield  {title} {\bibinfo {title} {Can quantum-mechanical description of physical reality be considered complete?},\ }\href@noop {} {\bibfield  {journal} {\bibinfo  {journal} {Physical Review}\ }\textbf {\bibinfo {volume} {47}},\ \bibinfo {pages} {777} (\bibinfo {year} {1935})}\BibitemShut {NoStop}%
\bibitem [{\citenamefont {Mermin}(1990)}]{mermin1990simple}%
  \BibitemOpen
  \bibfield  {author} {\bibinfo {author} {\bibfnamefont {N.~D.}\ \bibnamefont {Mermin}},\ }\bibfield  {title} {\bibinfo {title} {Simple unified form for the major no-hidden-variables theorems},\ }\href@noop {} {\bibfield  {journal} {\bibinfo  {journal} {Physical review letters}\ }\textbf {\bibinfo {volume} {65}},\ \bibinfo {pages} {3373} (\bibinfo {year} {1990})}\BibitemShut {NoStop}%
\bibitem [{\citenamefont {Peres}(1990)}]{peres1990incompatible}%
  \BibitemOpen
  \bibfield  {author} {\bibinfo {author} {\bibfnamefont {A.}~\bibnamefont {Peres}},\ }\bibfield  {title} {\bibinfo {title} {Incompatible results of quantum measurements},\ }\href@noop {} {\bibfield  {journal} {\bibinfo  {journal} {Physics Letters A}\ }\textbf {\bibinfo {volume} {151}},\ \bibinfo {pages} {107} (\bibinfo {year} {1990})}\BibitemShut {NoStop}%
\bibitem [{\citenamefont {Cabello}(2001{\natexlab{a}})}]{cabello2001all}%
  \BibitemOpen
  \bibfield  {author} {\bibinfo {author} {\bibfnamefont {A.}~\bibnamefont {Cabello}},\ }\bibfield  {title} {\bibinfo {title} {“all versus nothing” inseparability for two observers},\ }\href@noop {} {\bibfield  {journal} {\bibinfo  {journal} {Physical Review Letters}\ }\textbf {\bibinfo {volume} {87}},\ \bibinfo {pages} {010403} (\bibinfo {year} {2001}{\natexlab{a}})}\BibitemShut {NoStop}%
\bibitem [{\citenamefont {Cabello}(2001{\natexlab{b}})}]{cabello2001bell}%
  \BibitemOpen
  \bibfield  {author} {\bibinfo {author} {\bibfnamefont {A.}~\bibnamefont {Cabello}},\ }\bibfield  {title} {\bibinfo {title} {Bell's theorem without inequalities and without probabilities for two observers},\ }\href@noop {} {\bibfield  {journal} {\bibinfo  {journal} {Physical review letters}\ }\textbf {\bibinfo {volume} {86}},\ \bibinfo {pages} {1911} (\bibinfo {year} {2001}{\natexlab{b}})}\BibitemShut {NoStop}%
\bibitem [{\citenamefont {Bravyi}\ \emph {et~al.}(2020)\citenamefont {Bravyi}, \citenamefont {Gosset}, \citenamefont {K{\"o}nig},\ and\ \citenamefont {Tomamichel}}]{bravyi2020quantum}%
  \BibitemOpen
  \bibfield  {author} {\bibinfo {author} {\bibfnamefont {S.}~\bibnamefont {Bravyi}}, \bibinfo {author} {\bibfnamefont {D.}~\bibnamefont {Gosset}}, \bibinfo {author} {\bibfnamefont {R.}~\bibnamefont {K{\"o}nig}},\ and\ \bibinfo {author} {\bibfnamefont {M.}~\bibnamefont {Tomamichel}},\ }\bibfield  {title} {\bibinfo {title} {Quantum advantage with noisy shallow circuits},\ }\href@noop {} {\bibfield  {journal} {\bibinfo  {journal} {Nature Physics}\ }\textbf {\bibinfo {volume} {16}},\ \bibinfo {pages} {1040} (\bibinfo {year} {2020})}\BibitemShut {NoStop}%
\bibitem [{\citenamefont {Rao}(2008)}]{rao2008parallel}%
  \BibitemOpen
  \bibfield  {author} {\bibinfo {author} {\bibfnamefont {A.}~\bibnamefont {Rao}},\ }\bibfield  {title} {\bibinfo {title} {Parallel repetition in projection games and a concentration bound},\ }in\ \href@noop {} {\emph {\bibinfo {booktitle} {Proceedings of the Fortieth Annual ACM Symposium on Theory of Computing}}}\ (\bibinfo {year} {2008})\ pp.\ \bibinfo {pages} {1--10}\BibitemShut {NoStop}%
\bibitem [{\citenamefont {Jain}\ and\ \citenamefont {Kundu}(2022)}]{jain2022direct}%
  \BibitemOpen
  \bibfield  {author} {\bibinfo {author} {\bibfnamefont {R.}~\bibnamefont {Jain}}\ and\ \bibinfo {author} {\bibfnamefont {S.}~\bibnamefont {Kundu}},\ }\bibfield  {title} {\bibinfo {title} {A direct product theorem for quantum communication complexity with applications to device-independent qkd},\ }in\ \href@noop {} {\emph {\bibinfo {booktitle} {2021 IEEE 62nd Annual Symposium on Foundations of Computer Science (FOCS)}}}\ (\bibinfo {organization} {IEEE},\ \bibinfo {year} {2022})\ pp.\ \bibinfo {pages} {1285--1295}\BibitemShut {NoStop}%
\bibitem [{\citenamefont {Bharti}\ and\ \citenamefont {Jain}(2023)}]{bharti2023power}%
  \BibitemOpen
  \bibfield  {author} {\bibinfo {author} {\bibfnamefont {K.}~\bibnamefont {Bharti}}\ and\ \bibinfo {author} {\bibfnamefont {R.}~\bibnamefont {Jain}},\ }\bibfield  {title} {\bibinfo {title} {On the power of geometrically-local classical and quantum circuits},\ }\href@noop {} {\bibfield  {journal} {\bibinfo  {journal} {arXiv preprint arXiv:2310.01540}\ } (\bibinfo {year} {2023})}\BibitemShut {NoStop}%
\bibitem [{\citenamefont {Vershynin}(2018)}]{vershynin2018high}%
  \BibitemOpen
  \bibfield  {author} {\bibinfo {author} {\bibfnamefont {R.}~\bibnamefont {Vershynin}},\ }\href@noop {} {\emph {\bibinfo {title} {High-Dimensional Probability: An Introduction with Applications in Data Science}}},\ Vol.~\bibinfo {volume} {47}\ (\bibinfo  {publisher} {Cambridge University Press},\ \bibinfo {year} {2018})\BibitemShut {NoStop}%
\bibitem [{\citenamefont {Chollet}\ \emph {et~al.}(2015)\citenamefont {Chollet} \emph {et~al.}}]{chollet2015keras}%
  \BibitemOpen
  \bibfield  {author} {\bibinfo {author} {\bibfnamefont {F.}~\bibnamefont {Chollet}} \emph {et~al.},\ }\href@noop {} {\bibinfo {title} {Keras}},\ \bibinfo {howpublished} {\url{https://keras.io}} (\bibinfo {year} {2015})\BibitemShut {NoStop}%
\end{thebibliography}%


\providecommand{\noopsort}[1]{}\providecommand{\singleletter}[1]{#1}%
\begin{thebibliography}{92}%
\makeatletter
\providecommand \@ifxundefined [1]{%
 \@ifx{#1\undefined}
}%
\providecommand \@ifnum [1]{%
 \ifnum #1\expandafter \@firstoftwo
 \else \expandafter \@secondoftwo
 \fi
}%
\providecommand \@ifx [1]{%
 \ifx #1\expandafter \@firstoftwo
 \else \expandafter \@secondoftwo
 \fi
}%
\providecommand \natexlab [1]{#1}%
\providecommand \enquote  [1]{``#1''}%
\providecommand \bibnamefont  [1]{#1}%
\providecommand \bibfnamefont [1]{#1}%
\providecommand \citenamefont [1]{#1}%
\providecommand \href@noop [0]{\@secondoftwo}%
\providecommand \href [0]{\begingroup \@sanitize@url \@href}%
\providecommand \@href[1]{\@@startlink{#1}\@@href}%
\providecommand \@@href[1]{\endgroup#1\@@endlink}%
\providecommand \@sanitize@url [0]{\catcode `\\12\catcode `\$12\catcode `\&12\catcode `\#12\catcode `\^12\catcode `\_12\catcode `\%12\relax}%
\providecommand \@@startlink[1]{}%
\providecommand \@@endlink[0]{}%
\providecommand \url  [0]{\begingroup\@sanitize@url \@url }%
\providecommand \@url [1]{\endgroup\@href {#1}{\urlprefix }}%
\providecommand \urlprefix  [0]{URL }%
\providecommand \Eprint [0]{\href }%
\providecommand \doibase [0]{https://doi.org/}%
\providecommand \selectlanguage [0]{\@gobble}%
\providecommand \bibinfo  [0]{\@secondoftwo}%
\providecommand \bibfield  [0]{\@secondoftwo}%
\providecommand \translation [1]{[#1]}%
\providecommand \BibitemOpen [0]{}%
\providecommand \bibitemStop [0]{}%
\providecommand \bibitemNoStop [0]{.\EOS\space}%
\providecommand \EOS [0]{\spacefactor3000\relax}%
\providecommand \BibitemShut  [1]{\csname bibitem#1\endcsname}%
\let\auto@bib@innerbib\@empty
\bibitem [{\citenamefont {Preskill}(2012)}]{preskill2012quantum}%
  \BibitemOpen
  \bibfield  {author} {\bibinfo {author} {\bibfnamefont {J.}~\bibnamefont {Preskill}},\ }\bibfield  {title} {\bibinfo {title} {Quantum computing and the entanglement frontier},\ }\href@noop {} {\bibfield  {journal} {\bibinfo  {journal} {arXiv preprint arXiv:1203.5813}\ } (\bibinfo {year} {2012})}\BibitemShut {NoStop}%
\bibitem [{\citenamefont {Preskill}(2018)}]{preskill2018quantum}%
  \BibitemOpen
  \bibfield  {author} {\bibinfo {author} {\bibfnamefont {J.}~\bibnamefont {Preskill}},\ }\bibfield  {title} {\bibinfo {title} {Quantum computing in the nisq era and beyond},\ }\href@noop {} {\bibfield  {journal} {\bibinfo  {journal} {Quantum}\ }\textbf {\bibinfo {volume} {2}},\ \bibinfo {pages} {79} (\bibinfo {year} {2018})}\BibitemShut {NoStop}%
\bibitem [{\citenamefont {Nielsen}\ and\ \citenamefont {Chuang}(2010)}]{nielsen2010quantum}%
  \BibitemOpen
  \bibfield  {author} {\bibinfo {author} {\bibfnamefont {M.~A.}\ \bibnamefont {Nielsen}}\ and\ \bibinfo {author} {\bibfnamefont {I.~L.}\ \bibnamefont {Chuang}},\ }\href@noop {} {\emph {\bibinfo {title} {Quantum computation and quantum information}}}\ (\bibinfo  {publisher} {Cambridge university press},\ \bibinfo {year} {2010})\BibitemShut {NoStop}%
\bibitem [{\citenamefont {Biamonte}\ \emph {et~al.}(2017)\citenamefont {Biamonte}, \citenamefont {Wittek}, \citenamefont {Pancotti}, \citenamefont {Rebentrost}, \citenamefont {Wiebe},\ and\ \citenamefont {Lloyd}}]{biamonte2017quantum}%
  \BibitemOpen
  \bibfield  {author} {\bibinfo {author} {\bibfnamefont {J.}~\bibnamefont {Biamonte}}, \bibinfo {author} {\bibfnamefont {P.}~\bibnamefont {Wittek}}, \bibinfo {author} {\bibfnamefont {N.}~\bibnamefont {Pancotti}}, \bibinfo {author} {\bibfnamefont {P.}~\bibnamefont {Rebentrost}}, \bibinfo {author} {\bibfnamefont {N.}~\bibnamefont {Wiebe}},\ and\ \bibinfo {author} {\bibfnamefont {S.}~\bibnamefont {Lloyd}},\ }\bibfield  {title} {\bibinfo {title} {Quantum machine learning},\ }\href@noop {} {\bibfield  {journal} {\bibinfo  {journal} {Nature}\ }\textbf {\bibinfo {volume} {549}},\ \bibinfo {pages} {195} (\bibinfo {year} {2017})}\BibitemShut {NoStop}%
\bibitem [{\citenamefont {Cerezo}\ \emph {et~al.}(2022)\citenamefont {Cerezo}, \citenamefont {Verdon}, \citenamefont {Huang}, \citenamefont {Cincio},\ and\ \citenamefont {Coles}}]{cerezo2022challenges}%
  \BibitemOpen
  \bibfield  {author} {\bibinfo {author} {\bibfnamefont {M.}~\bibnamefont {Cerezo}}, \bibinfo {author} {\bibfnamefont {G.}~\bibnamefont {Verdon}}, \bibinfo {author} {\bibfnamefont {H.-Y.}\ \bibnamefont {Huang}}, \bibinfo {author} {\bibfnamefont {L.}~\bibnamefont {Cincio}},\ and\ \bibinfo {author} {\bibfnamefont {P.~J.}\ \bibnamefont {Coles}},\ }\bibfield  {title} {\bibinfo {title} {Challenges and opportunities in quantum machine learning},\ }\href@noop {} {\bibfield  {journal} {\bibinfo  {journal} {Nature Computational Science}\ }\textbf {\bibinfo {volume} {2}},\ \bibinfo {pages} {567} (\bibinfo {year} {2022})}\BibitemShut {NoStop}%
\bibitem [{\citenamefont {Ciliberto}\ \emph {et~al.}(2018)\citenamefont {Ciliberto}, \citenamefont {Herbster}, \citenamefont {Ialongo}, \citenamefont {Pontil}, \citenamefont {Rocchetto}, \citenamefont {Severini},\ and\ \citenamefont {Wossnig}}]{ciliberto2018quantum}%
  \BibitemOpen
  \bibfield  {author} {\bibinfo {author} {\bibfnamefont {C.}~\bibnamefont {Ciliberto}}, \bibinfo {author} {\bibfnamefont {M.}~\bibnamefont {Herbster}}, \bibinfo {author} {\bibfnamefont {A.~D.}\ \bibnamefont {Ialongo}}, \bibinfo {author} {\bibfnamefont {M.}~\bibnamefont {Pontil}}, \bibinfo {author} {\bibfnamefont {A.}~\bibnamefont {Rocchetto}}, \bibinfo {author} {\bibfnamefont {S.}~\bibnamefont {Severini}},\ and\ \bibinfo {author} {\bibfnamefont {L.}~\bibnamefont {Wossnig}},\ }\bibfield  {title} {\bibinfo {title} {Quantum machine learning: a classical perspective},\ }\href@noop {} {\bibfield  {journal} {\bibinfo  {journal} {Proceedings of the Royal Society A: Mathematical, Physical and Engineering Sciences}\ }\textbf {\bibinfo {volume} {474}},\ \bibinfo {pages} {20170551} (\bibinfo {year} {2018})}\BibitemShut {NoStop}%
\bibitem [{\citenamefont {Carleo}\ \emph {et~al.}(2019)\citenamefont {Carleo}, \citenamefont {Cirac}, \citenamefont {Cranmer}, \citenamefont {Daudet}, \citenamefont {Schuld}, \citenamefont {Tishby}, \citenamefont {Vogt-Maranto},\ and\ \citenamefont {Zdeborov{\'a}}}]{carleo2019machine}%
  \BibitemOpen
  \bibfield  {author} {\bibinfo {author} {\bibfnamefont {G.}~\bibnamefont {Carleo}}, \bibinfo {author} {\bibfnamefont {I.}~\bibnamefont {Cirac}}, \bibinfo {author} {\bibfnamefont {K.}~\bibnamefont {Cranmer}}, \bibinfo {author} {\bibfnamefont {L.}~\bibnamefont {Daudet}}, \bibinfo {author} {\bibfnamefont {M.}~\bibnamefont {Schuld}}, \bibinfo {author} {\bibfnamefont {N.}~\bibnamefont {Tishby}}, \bibinfo {author} {\bibfnamefont {L.}~\bibnamefont {Vogt-Maranto}},\ and\ \bibinfo {author} {\bibfnamefont {L.}~\bibnamefont {Zdeborov{\'a}}},\ }\bibfield  {title} {\bibinfo {title} {Machine learning and the physical sciences},\ }\href@noop {} {\bibfield  {journal} {\bibinfo  {journal} {Reviews of Modern Physics}\ }\textbf {\bibinfo {volume} {91}},\ \bibinfo {pages} {045002} (\bibinfo {year} {2019})}\BibitemShut {NoStop}%
\bibitem [{\citenamefont {Goodfellow}\ \emph {et~al.}(2016)\citenamefont {Goodfellow}, \citenamefont {Bengio},\ and\ \citenamefont {Courville}}]{goodfellow2016deep}%
  \BibitemOpen
  \bibfield  {author} {\bibinfo {author} {\bibfnamefont {I.}~\bibnamefont {Goodfellow}}, \bibinfo {author} {\bibfnamefont {Y.}~\bibnamefont {Bengio}},\ and\ \bibinfo {author} {\bibfnamefont {A.}~\bibnamefont {Courville}},\ }\href@noop {} {\emph {\bibinfo {title} {Deep learning}}}\ (\bibinfo  {publisher} {MIT press},\ \bibinfo {year} {2016})\BibitemShut {NoStop}%
\bibitem [{\citenamefont {LeCun}\ \emph {et~al.}(2015)\citenamefont {LeCun}, \citenamefont {Bengio},\ and\ \citenamefont {Hinton}}]{lecun2015deep}%
  \BibitemOpen
  \bibfield  {author} {\bibinfo {author} {\bibfnamefont {Y.}~\bibnamefont {LeCun}}, \bibinfo {author} {\bibfnamefont {Y.}~\bibnamefont {Bengio}},\ and\ \bibinfo {author} {\bibfnamefont {G.}~\bibnamefont {Hinton}},\ }\bibfield  {title} {\bibinfo {title} {Deep learning},\ }\href@noop {} {\bibfield  {journal} {\bibinfo  {journal} {nature}\ }\textbf {\bibinfo {volume} {521}},\ \bibinfo {pages} {436} (\bibinfo {year} {2015})}\BibitemShut {NoStop}%
\bibitem [{\citenamefont {Jordan}\ and\ \citenamefont {Mitchell}(2015)}]{jordan2015machine}%
  \BibitemOpen
  \bibfield  {author} {\bibinfo {author} {\bibfnamefont {M.~I.}\ \bibnamefont {Jordan}}\ and\ \bibinfo {author} {\bibfnamefont {T.~M.}\ \bibnamefont {Mitchell}},\ }\bibfield  {title} {\bibinfo {title} {Machine learning: Trends, perspectives, and prospects},\ }\href@noop {} {\bibfield  {journal} {\bibinfo  {journal} {Science}\ }\textbf {\bibinfo {volume} {349}},\ \bibinfo {pages} {255} (\bibinfo {year} {2015})}\BibitemShut {NoStop}%
\bibitem [{\citenamefont {Dalzell}\ \emph {et~al.}(2023)\citenamefont {Dalzell}, \citenamefont {McArdle}, \citenamefont {Berta}, \citenamefont {Bienias}, \citenamefont {Chen}, \citenamefont {Gily{\'e}n}, \citenamefont {Hann}, \citenamefont {Kastoryano}, \citenamefont {Khabiboulline}, \citenamefont {Kubica} \emph {et~al.}}]{dalzell2023quantum}%
  \BibitemOpen
  \bibfield  {author} {\bibinfo {author} {\bibfnamefont {A.~M.}\ \bibnamefont {Dalzell}}, \bibinfo {author} {\bibfnamefont {S.}~\bibnamefont {McArdle}}, \bibinfo {author} {\bibfnamefont {M.}~\bibnamefont {Berta}}, \bibinfo {author} {\bibfnamefont {P.}~\bibnamefont {Bienias}}, \bibinfo {author} {\bibfnamefont {C.-F.}\ \bibnamefont {Chen}}, \bibinfo {author} {\bibfnamefont {A.}~\bibnamefont {Gily{\'e}n}}, \bibinfo {author} {\bibfnamefont {C.~T.}\ \bibnamefont {Hann}}, \bibinfo {author} {\bibfnamefont {M.~J.}\ \bibnamefont {Kastoryano}}, \bibinfo {author} {\bibfnamefont {E.~T.}\ \bibnamefont {Khabiboulline}}, \bibinfo {author} {\bibfnamefont {A.}~\bibnamefont {Kubica}}, \emph {et~al.},\ }\bibfield  {title} {\bibinfo {title} {Quantum algorithms: A survey of applications and end-to-end complexities},\ }\href@noop {} {\bibfield  {journal} {\bibinfo  {journal} {arXiv preprint arXiv:2310.03011}\ } (\bibinfo {year} {2023})}\BibitemShut {NoStop}%
\bibitem [{\citenamefont {Harrow}\ \emph {et~al.}(2009)\citenamefont {Harrow}, \citenamefont {Hassidim},\ and\ \citenamefont {Lloyd}}]{harrow2009quantum}%
  \BibitemOpen
  \bibfield  {author} {\bibinfo {author} {\bibfnamefont {A.~W.}\ \bibnamefont {Harrow}}, \bibinfo {author} {\bibfnamefont {A.}~\bibnamefont {Hassidim}},\ and\ \bibinfo {author} {\bibfnamefont {S.}~\bibnamefont {Lloyd}},\ }\bibfield  {title} {\bibinfo {title} {Quantum algorithm for linear systems of equations},\ }\href@noop {} {\bibfield  {journal} {\bibinfo  {journal} {Physical review letters}\ }\textbf {\bibinfo {volume} {103}},\ \bibinfo {pages} {150502} (\bibinfo {year} {2009})}\BibitemShut {NoStop}%
\bibitem [{\citenamefont {Lloyd}\ \emph {et~al.}(2014)\citenamefont {Lloyd}, \citenamefont {Mohseni},\ and\ \citenamefont {Rebentrost}}]{lloyd2014quantum}%
  \BibitemOpen
  \bibfield  {author} {\bibinfo {author} {\bibfnamefont {S.}~\bibnamefont {Lloyd}}, \bibinfo {author} {\bibfnamefont {M.}~\bibnamefont {Mohseni}},\ and\ \bibinfo {author} {\bibfnamefont {P.}~\bibnamefont {Rebentrost}},\ }\bibfield  {title} {\bibinfo {title} {Quantum principal component analysis},\ }\href@noop {} {\bibfield  {journal} {\bibinfo  {journal} {Nature physics}\ }\textbf {\bibinfo {volume} {10}},\ \bibinfo {pages} {631} (\bibinfo {year} {2014})}\BibitemShut {NoStop}%
\bibitem [{\citenamefont {Rebentrost}\ \emph {et~al.}(2014)\citenamefont {Rebentrost}, \citenamefont {Mohseni},\ and\ \citenamefont {Lloyd}}]{rebentrost2014quantum}%
  \BibitemOpen
  \bibfield  {author} {\bibinfo {author} {\bibfnamefont {P.}~\bibnamefont {Rebentrost}}, \bibinfo {author} {\bibfnamefont {M.}~\bibnamefont {Mohseni}},\ and\ \bibinfo {author} {\bibfnamefont {S.}~\bibnamefont {Lloyd}},\ }\bibfield  {title} {\bibinfo {title} {Quantum support vector machine for big data classification},\ }\href@noop {} {\bibfield  {journal} {\bibinfo  {journal} {Physical review letters}\ }\textbf {\bibinfo {volume} {113}},\ \bibinfo {pages} {130503} (\bibinfo {year} {2014})}\BibitemShut {NoStop}%
\bibitem [{\citenamefont {Dunjko}\ \emph {et~al.}(2016)\citenamefont {Dunjko}, \citenamefont {Taylor},\ and\ \citenamefont {Briegel}}]{dunjko2016quantum}%
  \BibitemOpen
  \bibfield  {author} {\bibinfo {author} {\bibfnamefont {V.}~\bibnamefont {Dunjko}}, \bibinfo {author} {\bibfnamefont {J.~M.}\ \bibnamefont {Taylor}},\ and\ \bibinfo {author} {\bibfnamefont {H.~J.}\ \bibnamefont {Briegel}},\ }\bibfield  {title} {\bibinfo {title} {Quantum-enhanced machine learning},\ }\href@noop {} {\bibfield  {journal} {\bibinfo  {journal} {Physical review letters}\ }\textbf {\bibinfo {volume} {117}},\ \bibinfo {pages} {130501} (\bibinfo {year} {2016})}\BibitemShut {NoStop}%
\bibitem [{\citenamefont {Gao}\ \emph {et~al.}(2018)\citenamefont {Gao}, \citenamefont {Zhang},\ and\ \citenamefont {Duan}}]{gao2018quantum}%
  \BibitemOpen
  \bibfield  {author} {\bibinfo {author} {\bibfnamefont {X.}~\bibnamefont {Gao}}, \bibinfo {author} {\bibfnamefont {Z.-Y.}\ \bibnamefont {Zhang}},\ and\ \bibinfo {author} {\bibfnamefont {L.-M.}\ \bibnamefont {Duan}},\ }\bibfield  {title} {\bibinfo {title} {A quantum machine learning algorithm based on generative models},\ }\href@noop {} {\bibfield  {journal} {\bibinfo  {journal} {Science advances}\ }\textbf {\bibinfo {volume} {4}},\ \bibinfo {pages} {eaat9004} (\bibinfo {year} {2018})}\BibitemShut {NoStop}%
\bibitem [{\citenamefont {Liu}\ \emph {et~al.}(2021)\citenamefont {Liu}, \citenamefont {Arunachalam},\ and\ \citenamefont {Temme}}]{liu2021rigorous}%
  \BibitemOpen
  \bibfield  {author} {\bibinfo {author} {\bibfnamefont {Y.}~\bibnamefont {Liu}}, \bibinfo {author} {\bibfnamefont {S.}~\bibnamefont {Arunachalam}},\ and\ \bibinfo {author} {\bibfnamefont {K.}~\bibnamefont {Temme}},\ }\bibfield  {title} {\bibinfo {title} {A rigorous and robust quantum speed-up in supervised machine learning},\ }\href@noop {} {\bibfield  {journal} {\bibinfo  {journal} {Nature Physics}\ }\textbf {\bibinfo {volume} {17}},\ \bibinfo {pages} {1013} (\bibinfo {year} {2021})}\BibitemShut {NoStop}%
\bibitem [{\citenamefont {Gyurik}\ and\ \citenamefont {Dunjko}(2023)}]{gyurik2023exponential}%
  \BibitemOpen
  \bibfield  {author} {\bibinfo {author} {\bibfnamefont {C.}~\bibnamefont {Gyurik}}\ and\ \bibinfo {author} {\bibfnamefont {V.}~\bibnamefont {Dunjko}},\ }\bibfield  {title} {\bibinfo {title} {Exponential separations between classical and quantum learners},\ }\href@noop {} {\bibfield  {journal} {\bibinfo  {journal} {arXiv preprint arXiv:2306.16028}\ } (\bibinfo {year} {2023})}\BibitemShut {NoStop}%
\bibitem [{\citenamefont {Lloyd}\ and\ \citenamefont {Weedbrook}(2018)}]{lloyd2018quantum}%
  \BibitemOpen
  \bibfield  {author} {\bibinfo {author} {\bibfnamefont {S.}~\bibnamefont {Lloyd}}\ and\ \bibinfo {author} {\bibfnamefont {C.}~\bibnamefont {Weedbrook}},\ }\bibfield  {title} {\bibinfo {title} {Quantum generative adversarial learning},\ }\href@noop {} {\bibfield  {journal} {\bibinfo  {journal} {Physical review letters}\ }\textbf {\bibinfo {volume} {121}},\ \bibinfo {pages} {040502} (\bibinfo {year} {2018})}\BibitemShut {NoStop}%
\bibitem [{\citenamefont {Schuld}\ and\ \citenamefont {Killoran}(2019)}]{schuld2019quantum}%
  \BibitemOpen
  \bibfield  {author} {\bibinfo {author} {\bibfnamefont {M.}~\bibnamefont {Schuld}}\ and\ \bibinfo {author} {\bibfnamefont {N.}~\bibnamefont {Killoran}},\ }\bibfield  {title} {\bibinfo {title} {Quantum machine learning in feature hilbert spaces},\ }\href@noop {} {\bibfield  {journal} {\bibinfo  {journal} {Physical review letters}\ }\textbf {\bibinfo {volume} {122}},\ \bibinfo {pages} {040504} (\bibinfo {year} {2019})}\BibitemShut {NoStop}%
\bibitem [{\citenamefont {Cerezo}\ \emph {et~al.}(2021)\citenamefont {Cerezo}, \citenamefont {Arrasmith}, \citenamefont {Babbush}, \citenamefont {Benjamin}, \citenamefont {Endo}, \citenamefont {Fujii}, \citenamefont {McClean}, \citenamefont {Mitarai}, \citenamefont {Yuan}, \citenamefont {Cincio} \emph {et~al.}}]{cerezo2021variational}%
  \BibitemOpen
  \bibfield  {author} {\bibinfo {author} {\bibfnamefont {M.}~\bibnamefont {Cerezo}}, \bibinfo {author} {\bibfnamefont {A.}~\bibnamefont {Arrasmith}}, \bibinfo {author} {\bibfnamefont {R.}~\bibnamefont {Babbush}}, \bibinfo {author} {\bibfnamefont {S.~C.}\ \bibnamefont {Benjamin}}, \bibinfo {author} {\bibfnamefont {S.}~\bibnamefont {Endo}}, \bibinfo {author} {\bibfnamefont {K.}~\bibnamefont {Fujii}}, \bibinfo {author} {\bibfnamefont {J.~R.}\ \bibnamefont {McClean}}, \bibinfo {author} {\bibfnamefont {K.}~\bibnamefont {Mitarai}}, \bibinfo {author} {\bibfnamefont {X.}~\bibnamefont {Yuan}}, \bibinfo {author} {\bibfnamefont {L.}~\bibnamefont {Cincio}}, \emph {et~al.},\ }\bibfield  {title} {\bibinfo {title} {Variational quantum algorithms},\ }\href@noop {} {\bibfield  {journal} {\bibinfo  {journal} {Nature Reviews Physics}\ }\textbf {\bibinfo {volume} {3}},\ \bibinfo {pages} {625} (\bibinfo {year} {2021})}\BibitemShut {NoStop}%
\bibitem [{\citenamefont {Huang}\ \emph {et~al.}(2022{\natexlab{a}})\citenamefont {Huang}, \citenamefont {Broughton}, \citenamefont {Cotler}, \citenamefont {Chen}, \citenamefont {Li}, \citenamefont {Mohseni}, \citenamefont {Neven}, \citenamefont {Babbush}, \citenamefont {Kueng}, \citenamefont {Preskill} \emph {et~al.}}]{huang2022quantum}%
  \BibitemOpen
  \bibfield  {author} {\bibinfo {author} {\bibfnamefont {H.-Y.}\ \bibnamefont {Huang}}, \bibinfo {author} {\bibfnamefont {M.}~\bibnamefont {Broughton}}, \bibinfo {author} {\bibfnamefont {J.}~\bibnamefont {Cotler}}, \bibinfo {author} {\bibfnamefont {S.}~\bibnamefont {Chen}}, \bibinfo {author} {\bibfnamefont {J.}~\bibnamefont {Li}}, \bibinfo {author} {\bibfnamefont {M.}~\bibnamefont {Mohseni}}, \bibinfo {author} {\bibfnamefont {H.}~\bibnamefont {Neven}}, \bibinfo {author} {\bibfnamefont {R.}~\bibnamefont {Babbush}}, \bibinfo {author} {\bibfnamefont {R.}~\bibnamefont {Kueng}}, \bibinfo {author} {\bibfnamefont {J.}~\bibnamefont {Preskill}}, \emph {et~al.},\ }\bibfield  {title} {\bibinfo {title} {Quantum advantage in learning from experiments},\ }\href@noop {} {\bibfield  {journal} {\bibinfo  {journal} {Science}\ }\textbf {\bibinfo {volume} {376}},\ \bibinfo {pages} {1182} (\bibinfo {year} {2022}{\natexlab{a}})}\BibitemShut {NoStop}%
\bibitem [{\citenamefont {Aharonov}\ \emph {et~al.}(2022)\citenamefont {Aharonov}, \citenamefont {Cotler},\ and\ \citenamefont {Qi}}]{aharonov2022quantum}%
  \BibitemOpen
  \bibfield  {author} {\bibinfo {author} {\bibfnamefont {D.}~\bibnamefont {Aharonov}}, \bibinfo {author} {\bibfnamefont {J.}~\bibnamefont {Cotler}},\ and\ \bibinfo {author} {\bibfnamefont {X.-L.}\ \bibnamefont {Qi}},\ }\bibfield  {title} {\bibinfo {title} {Quantum algorithmic measurement},\ }\href@noop {} {\bibfield  {journal} {\bibinfo  {journal} {Nature communications}\ }\textbf {\bibinfo {volume} {13}},\ \bibinfo {pages} {887} (\bibinfo {year} {2022})}\BibitemShut {NoStop}%
\bibitem [{\citenamefont {Oh}\ \emph {et~al.}(2024)\citenamefont {Oh}, \citenamefont {Chen}, \citenamefont {Wong}, \citenamefont {Zhou}, \citenamefont {Huang}, \citenamefont {Nielsen}, \citenamefont {Liu}, \citenamefont {Neergaard-Nielsen}, \citenamefont {Andersen}, \citenamefont {Jiang},\ and\ \citenamefont {Preskill}}]{oh2024entanglement}%
  \BibitemOpen
  \bibfield  {author} {\bibinfo {author} {\bibfnamefont {C.}~\bibnamefont {Oh}}, \bibinfo {author} {\bibfnamefont {S.}~\bibnamefont {Chen}}, \bibinfo {author} {\bibfnamefont {Y.}~\bibnamefont {Wong}}, \bibinfo {author} {\bibfnamefont {S.}~\bibnamefont {Zhou}}, \bibinfo {author} {\bibfnamefont {H.-Y.}\ \bibnamefont {Huang}}, \bibinfo {author} {\bibfnamefont {J.~A.}\ \bibnamefont {Nielsen}}, \bibinfo {author} {\bibfnamefont {Z.-H.}\ \bibnamefont {Liu}}, \bibinfo {author} {\bibfnamefont {J.~S.}\ \bibnamefont {Neergaard-Nielsen}}, \bibinfo {author} {\bibfnamefont {U.~L.}\ \bibnamefont {Andersen}}, \bibinfo {author} {\bibfnamefont {L.}~\bibnamefont {Jiang}},\ and\ \bibinfo {author} {\bibfnamefont {J.}~\bibnamefont {Preskill}},\ }\bibfield  {title} {\bibinfo {title} {Entanglement-enabled advantage for learning a bosonic random displacement channel},\ }\href@noop {} {\bibfield  {journal} {\bibinfo  {journal} {arXiv preprint arXiv:2402.18809}\ } (\bibinfo {year} {2024})}\BibitemShut {NoStop}%
\bibitem [{\citenamefont {Zhu}\ \emph {et~al.}(2022)\citenamefont {Zhu}, \citenamefont {Johri}, \citenamefont {Bacon}, \citenamefont {Esencan}, \citenamefont {Kim}, \citenamefont {Muir}, \citenamefont {Murgai}, \citenamefont {Nguyen}, \citenamefont {Pisenti}, \citenamefont {Schouela} \emph {et~al.}}]{zhu2022generative}%
  \BibitemOpen
  \bibfield  {author} {\bibinfo {author} {\bibfnamefont {E.~Y.}\ \bibnamefont {Zhu}}, \bibinfo {author} {\bibfnamefont {S.}~\bibnamefont {Johri}}, \bibinfo {author} {\bibfnamefont {D.}~\bibnamefont {Bacon}}, \bibinfo {author} {\bibfnamefont {M.}~\bibnamefont {Esencan}}, \bibinfo {author} {\bibfnamefont {J.}~\bibnamefont {Kim}}, \bibinfo {author} {\bibfnamefont {M.}~\bibnamefont {Muir}}, \bibinfo {author} {\bibfnamefont {N.}~\bibnamefont {Murgai}}, \bibinfo {author} {\bibfnamefont {J.}~\bibnamefont {Nguyen}}, \bibinfo {author} {\bibfnamefont {N.}~\bibnamefont {Pisenti}}, \bibinfo {author} {\bibfnamefont {A.}~\bibnamefont {Schouela}}, \emph {et~al.},\ }\bibfield  {title} {\bibinfo {title} {Generative quantum learning of joint probability distribution functions},\ }\href@noop {} {\bibfield  {journal} {\bibinfo  {journal} {Physical Review Research}\ }\textbf {\bibinfo {volume} {4}},\ \bibinfo {pages} {043092} (\bibinfo {year} {2022})}\BibitemShut {NoStop}%
\bibitem [{\citenamefont {Molteni}\ \emph {et~al.}(2024)\citenamefont {Molteni}, \citenamefont {Gyurik},\ and\ \citenamefont {Dunjko}}]{molteni2024exponential}%
  \BibitemOpen
  \bibfield  {author} {\bibinfo {author} {\bibfnamefont {R.}~\bibnamefont {Molteni}}, \bibinfo {author} {\bibfnamefont {C.}~\bibnamefont {Gyurik}},\ and\ \bibinfo {author} {\bibfnamefont {V.}~\bibnamefont {Dunjko}},\ }\bibfield  {title} {\bibinfo {title} {Exponential quantum advantages in learning quantum observables from classical data},\ }\href@noop {} {\bibfield  {journal} {\bibinfo  {journal} {arXiv preprint arXiv:2405.02027}\ } (\bibinfo {year} {2024})}\BibitemShut {NoStop}%
\bibitem [{\citenamefont {Aaronson}(2015)}]{aaronson2015read}%
  \BibitemOpen
  \bibfield  {author} {\bibinfo {author} {\bibfnamefont {S.}~\bibnamefont {Aaronson}},\ }\bibfield  {title} {\bibinfo {title} {Read the fine print},\ }\href@noop {} {\bibfield  {journal} {\bibinfo  {journal} {Nature Physics}\ }\textbf {\bibinfo {volume} {11}},\ \bibinfo {pages} {291} (\bibinfo {year} {2015})}\BibitemShut {NoStop}%
\bibitem [{\citenamefont {Tang}(2023)}]{tang2023quantum}%
  \BibitemOpen
  \bibfield  {author} {\bibinfo {author} {\bibfnamefont {E.}~\bibnamefont {Tang}},\ }\emph {\bibinfo {title} {Quantum Machine Learning Without Any Quantum}},\ \href@noop {} {Ph.D. thesis},\ \bibinfo  {school} {University of Washington} (\bibinfo {year} {2023})\BibitemShut {NoStop}%
\bibitem [{\citenamefont {Tang}(2022)}]{tang2022dequantizing}%
  \BibitemOpen
  \bibfield  {author} {\bibinfo {author} {\bibfnamefont {E.}~\bibnamefont {Tang}},\ }\bibfield  {title} {\bibinfo {title} {Dequantizing algorithms to understand quantum advantage in machine learning},\ }\href@noop {} {\bibfield  {journal} {\bibinfo  {journal} {Nature Reviews Physics}\ }\textbf {\bibinfo {volume} {4}},\ \bibinfo {pages} {692} (\bibinfo {year} {2022})}\BibitemShut {NoStop}%
\bibitem [{\citenamefont {Tang}(2019)}]{tang2019quantum}%
  \BibitemOpen
  \bibfield  {author} {\bibinfo {author} {\bibfnamefont {E.}~\bibnamefont {Tang}},\ }\bibfield  {title} {\bibinfo {title} {A quantum-inspired classical algorithm for recommendation systems},\ }in\ \href@noop {} {\emph {\bibinfo {booktitle} {Proceedings of the 51st annual ACM SIGACT symposium on theory of computing}}}\ (\bibinfo {year} {2019})\ pp.\ \bibinfo {pages} {217--228}\BibitemShut {NoStop}%
\bibitem [{\citenamefont {Tang}(2021)}]{tang2021quantum}%
  \BibitemOpen
  \bibfield  {author} {\bibinfo {author} {\bibfnamefont {E.}~\bibnamefont {Tang}},\ }\bibfield  {title} {\bibinfo {title} {Quantum principal component analysis only achieves an exponential speedup because of its state preparation assumptions},\ }\href@noop {} {\bibfield  {journal} {\bibinfo  {journal} {Physical Review Letters}\ }\textbf {\bibinfo {volume} {127}},\ \bibinfo {pages} {060503} (\bibinfo {year} {2021})}\BibitemShut {NoStop}%
\bibitem [{\citenamefont {Schuld}\ and\ \citenamefont {Killoran}(2022)}]{schuld2022quantum}%
  \BibitemOpen
  \bibfield  {author} {\bibinfo {author} {\bibfnamefont {M.}~\bibnamefont {Schuld}}\ and\ \bibinfo {author} {\bibfnamefont {N.}~\bibnamefont {Killoran}},\ }\bibfield  {title} {\bibinfo {title} {Is quantum advantage the right goal for quantum machine learning?},\ }\href@noop {} {\bibfield  {journal} {\bibinfo  {journal} {Prx Quantum}\ }\textbf {\bibinfo {volume} {3}},\ \bibinfo {pages} {030101} (\bibinfo {year} {2022})}\BibitemShut {NoStop}%
\bibitem [{\citenamefont {Cerezo}\ \emph {et~al.}(2023)\citenamefont {Cerezo}, \citenamefont {Larocca}, \citenamefont {Garc{\'\i}a-Mart{\'\i}n}, \citenamefont {Diaz}, \citenamefont {Braccia}, \citenamefont {Fontana}, \citenamefont {Rudolph}, \citenamefont {Bermejo}, \citenamefont {Ijaz}, \citenamefont {Thanasilp} \emph {et~al.}}]{cerezo2023does}%
  \BibitemOpen
  \bibfield  {author} {\bibinfo {author} {\bibfnamefont {M.}~\bibnamefont {Cerezo}}, \bibinfo {author} {\bibfnamefont {M.}~\bibnamefont {Larocca}}, \bibinfo {author} {\bibfnamefont {D.}~\bibnamefont {Garc{\'\i}a-Mart{\'\i}n}}, \bibinfo {author} {\bibfnamefont {N.~L.}\ \bibnamefont {Diaz}}, \bibinfo {author} {\bibfnamefont {P.}~\bibnamefont {Braccia}}, \bibinfo {author} {\bibfnamefont {E.}~\bibnamefont {Fontana}}, \bibinfo {author} {\bibfnamefont {M.~S.}\ \bibnamefont {Rudolph}}, \bibinfo {author} {\bibfnamefont {P.}~\bibnamefont {Bermejo}}, \bibinfo {author} {\bibfnamefont {A.}~\bibnamefont {Ijaz}}, \bibinfo {author} {\bibfnamefont {S.}~\bibnamefont {Thanasilp}}, \emph {et~al.},\ }\bibfield  {title} {\bibinfo {title} {Does provable absence of barren plateaus imply classical simulability? or, why we need to rethink variational quantum computing},\ }\href@noop {} {\bibfield  {journal} {\bibinfo  {journal} {arXiv preprint arXiv:2312.09121}\ } (\bibinfo {year} {2023})}\BibitemShut {NoStop}%
\bibitem [{\citenamefont {Gil-Fuster}\ \emph {et~al.}(2024)\citenamefont {Gil-Fuster}, \citenamefont {Gyurik}, \citenamefont {P{\'e}rez-Salinas},\ and\ \citenamefont {Dunjko}}]{gil2024relation}%
  \BibitemOpen
  \bibfield  {author} {\bibinfo {author} {\bibfnamefont {E.}~\bibnamefont {Gil-Fuster}}, \bibinfo {author} {\bibfnamefont {C.}~\bibnamefont {Gyurik}}, \bibinfo {author} {\bibfnamefont {A.}~\bibnamefont {P{\'e}rez-Salinas}},\ and\ \bibinfo {author} {\bibfnamefont {V.}~\bibnamefont {Dunjko}},\ }\bibfield  {title} {\bibinfo {title} {On the relation between trainability and dequantization of variational quantum learning models},\ }\href@noop {} {\bibfield  {journal} {\bibinfo  {journal} {arXiv preprint arXiv:2406.07072}\ } (\bibinfo {year} {2024})}\BibitemShut {NoStop}%
\bibitem [{\citenamefont {Bravyi}\ \emph {et~al.}(2018)\citenamefont {Bravyi}, \citenamefont {Gosset},\ and\ \citenamefont {K{\"o}nig}}]{bravyi2018quantum}%
  \BibitemOpen
  \bibfield  {author} {\bibinfo {author} {\bibfnamefont {S.}~\bibnamefont {Bravyi}}, \bibinfo {author} {\bibfnamefont {D.}~\bibnamefont {Gosset}},\ and\ \bibinfo {author} {\bibfnamefont {R.}~\bibnamefont {K{\"o}nig}},\ }\bibfield  {title} {\bibinfo {title} {Quantum advantage with shallow circuits},\ }\href@noop {} {\bibfield  {journal} {\bibinfo  {journal} {Science}\ }\textbf {\bibinfo {volume} {362}},\ \bibinfo {pages} {308} (\bibinfo {year} {2018})}\BibitemShut {NoStop}%
\bibitem [{\citenamefont {Bravyi}\ \emph {et~al.}(2020)\citenamefont {Bravyi}, \citenamefont {Gosset}, \citenamefont {K{\"o}nig},\ and\ \citenamefont {Tomamichel}}]{bravyi2020quantum}%
  \BibitemOpen
  \bibfield  {author} {\bibinfo {author} {\bibfnamefont {S.}~\bibnamefont {Bravyi}}, \bibinfo {author} {\bibfnamefont {D.}~\bibnamefont {Gosset}}, \bibinfo {author} {\bibfnamefont {R.}~\bibnamefont {K{\"o}nig}},\ and\ \bibinfo {author} {\bibfnamefont {M.}~\bibnamefont {Tomamichel}},\ }\bibfield  {title} {\bibinfo {title} {Quantum advantage with noisy shallow circuits},\ }\href@noop {} {\bibfield  {journal} {\bibinfo  {journal} {Nature Physics}\ }\textbf {\bibinfo {volume} {16}},\ \bibinfo {pages} {1040} (\bibinfo {year} {2020})}\BibitemShut {NoStop}%
\bibitem [{\citenamefont {Zhang}\ \emph {et~al.}(2024)\citenamefont {Zhang}, \citenamefont {Gong}, \citenamefont {Li},\ and\ \citenamefont {Deng}}]{zhang2024quantum}%
  \BibitemOpen
  \bibfield  {author} {\bibinfo {author} {\bibfnamefont {Z.}~\bibnamefont {Zhang}}, \bibinfo {author} {\bibfnamefont {W.}~\bibnamefont {Gong}}, \bibinfo {author} {\bibfnamefont {W.}~\bibnamefont {Li}},\ and\ \bibinfo {author} {\bibfnamefont {D.-L.}\ \bibnamefont {Deng}},\ }\bibfield  {title} {\bibinfo {title} {Quantum-classical separations in shallow-circuit-based learning with and without noises},\ }\href@noop {} {\bibfield  {journal} {\bibinfo  {journal} {Communications Physics}\ }\textbf {\bibinfo {volume} {7}},\ \bibinfo {pages} {290} (\bibinfo {year} {2024})}\BibitemShut {NoStop}%
\bibitem [{\citenamefont {Gao}\ \emph {et~al.}(2022)\citenamefont {Gao}, \citenamefont {Anschuetz}, \citenamefont {Wang}, \citenamefont {Cirac},\ and\ \citenamefont {Lukin}}]{gao2022enhancing}%
  \BibitemOpen
  \bibfield  {author} {\bibinfo {author} {\bibfnamefont {X.}~\bibnamefont {Gao}}, \bibinfo {author} {\bibfnamefont {E.~R.}\ \bibnamefont {Anschuetz}}, \bibinfo {author} {\bibfnamefont {S.-T.}\ \bibnamefont {Wang}}, \bibinfo {author} {\bibfnamefont {J.~I.}\ \bibnamefont {Cirac}},\ and\ \bibinfo {author} {\bibfnamefont {M.~D.}\ \bibnamefont {Lukin}},\ }\bibfield  {title} {\bibinfo {title} {Enhancing generative models via quantum correlations},\ }\href@noop {} {\bibfield  {journal} {\bibinfo  {journal} {Physical Review X}\ }\textbf {\bibinfo {volume} {12}},\ \bibinfo {pages} {021037} (\bibinfo {year} {2022})}\BibitemShut {NoStop}%
\bibitem [{\citenamefont {Anschuetz}\ \emph {et~al.}(2023)\citenamefont {Anschuetz}, \citenamefont {Hu}, \citenamefont {Huang},\ and\ \citenamefont {Gao}}]{anschuetz2023interpretable}%
  \BibitemOpen
  \bibfield  {author} {\bibinfo {author} {\bibfnamefont {E.~R.}\ \bibnamefont {Anschuetz}}, \bibinfo {author} {\bibfnamefont {H.-Y.}\ \bibnamefont {Hu}}, \bibinfo {author} {\bibfnamefont {J.-L.}\ \bibnamefont {Huang}},\ and\ \bibinfo {author} {\bibfnamefont {X.}~\bibnamefont {Gao}},\ }\bibfield  {title} {\bibinfo {title} {Interpretable quantum advantage in neural sequence learning},\ }\href@noop {} {\bibfield  {journal} {\bibinfo  {journal} {PRX Quantum}\ }\textbf {\bibinfo {volume} {4}},\ \bibinfo {pages} {020338} (\bibinfo {year} {2023})}\BibitemShut {NoStop}%
\bibitem [{\citenamefont {Anschuetz}\ and\ \citenamefont {Gao}(2024)}]{anschuetz2024arbitrary}%
  \BibitemOpen
  \bibfield  {author} {\bibinfo {author} {\bibfnamefont {E.~R.}\ \bibnamefont {Anschuetz}}\ and\ \bibinfo {author} {\bibfnamefont {X.}~\bibnamefont {Gao}},\ }\bibfield  {title} {\bibinfo {title} {Arbitrary polynomial separations in trainable quantum machine learning},\ }\href@noop {} {\bibfield  {journal} {\bibinfo  {journal} {arXiv preprint arXiv:2402.08606}\ } (\bibinfo {year} {2024})}\BibitemShut {NoStop}%
\bibitem [{\citenamefont {{IonQ}}(2024)}]{ionq}%
  \BibitemOpen
  \bibfield  {author} {\bibinfo {author} {\bibnamefont {{IonQ}}},\ }\href@noop {} {\bibinfo {title} {{IonQ Aria}}},\ \bibinfo {howpublished} {\url{https://ionq.com/quantum-systems/aria}} (\bibinfo {year} {2024}),\ \bibinfo {note} {accessed: 2024-10-03}\BibitemShut {NoStop}%
\bibitem [{\citenamefont {Yao}(1993)}]{yao1993quantum}%
  \BibitemOpen
  \bibfield  {author} {\bibinfo {author} {\bibfnamefont {A.~C.-C.}\ \bibnamefont {Yao}},\ }\bibfield  {title} {\bibinfo {title} {Quantum circuit complexity},\ }in\ \href@noop {} {\emph {\bibinfo {booktitle} {Proceedings of 1993 IEEE 34th Annual Foundations of Computer Science}}}\ (\bibinfo {organization} {IEEE},\ \bibinfo {year} {1993})\ pp.\ \bibinfo {pages} {352--361}\BibitemShut {NoStop}%
\bibitem [{\citenamefont {Brassard}(2003)}]{brassard2003quantum}%
  \BibitemOpen
  \bibfield  {author} {\bibinfo {author} {\bibfnamefont {G.}~\bibnamefont {Brassard}},\ }\bibfield  {title} {\bibinfo {title} {Quantum communication complexity},\ }\href@noop {} {\bibfield  {journal} {\bibinfo  {journal} {Foundations of Physics}\ }\textbf {\bibinfo {volume} {33}},\ \bibinfo {pages} {1593} (\bibinfo {year} {2003})}\BibitemShut {NoStop}%
\bibitem [{\citenamefont {Brassard}\ \emph {et~al.}(2005)\citenamefont {Brassard}, \citenamefont {Broadbent},\ and\ \citenamefont {Tapp}}]{brassard2005quantum}%
  \BibitemOpen
  \bibfield  {author} {\bibinfo {author} {\bibfnamefont {G.}~\bibnamefont {Brassard}}, \bibinfo {author} {\bibfnamefont {A.}~\bibnamefont {Broadbent}},\ and\ \bibinfo {author} {\bibfnamefont {A.}~\bibnamefont {Tapp}},\ }\bibfield  {title} {\bibinfo {title} {Quantum pseudo-telepathy},\ }\href@noop {} {\bibfield  {journal} {\bibinfo  {journal} {Foundations of Physics}\ }\textbf {\bibinfo {volume} {35}},\ \bibinfo {pages} {1877} (\bibinfo {year} {2005})}\BibitemShut {NoStop}%
\bibitem [{\citenamefont {Huggins}\ and\ \citenamefont {McClean}(2024)}]{huggins2024accelerating}%
  \BibitemOpen
  \bibfield  {author} {\bibinfo {author} {\bibfnamefont {W.~J.}\ \bibnamefont {Huggins}}\ and\ \bibinfo {author} {\bibfnamefont {J.~R.}\ \bibnamefont {McClean}},\ }\bibfield  {title} {\bibinfo {title} {Accelerating quantum algorithms with precomputation},\ }\href@noop {} {\bibfield  {journal} {\bibinfo  {journal} {Quantum}\ }\textbf {\bibinfo {volume} {8}},\ \bibinfo {pages} {1264} (\bibinfo {year} {2024})}\BibitemShut {NoStop}%
\bibitem [{\citenamefont {Manning}\ and\ \citenamefont {Schutze}(1999)}]{manning1999foundations}%
  \BibitemOpen
  \bibfield  {author} {\bibinfo {author} {\bibfnamefont {C.}~\bibnamefont {Manning}}\ and\ \bibinfo {author} {\bibfnamefont {H.}~\bibnamefont {Schutze}},\ }\href@noop {} {\emph {\bibinfo {title} {Foundations of statistical natural language processing}}}\ (\bibinfo  {publisher} {MIT press},\ \bibinfo {year} {1999})\BibitemShut {NoStop}%
\bibitem [{\citenamefont {Sutskever}\ \emph {et~al.}(2014)\citenamefont {Sutskever}, \citenamefont {Vinyals},\ and\ \citenamefont {Le}}]{sutskever2014sequence}%
  \BibitemOpen
  \bibfield  {author} {\bibinfo {author} {\bibfnamefont {I.}~\bibnamefont {Sutskever}}, \bibinfo {author} {\bibfnamefont {O.}~\bibnamefont {Vinyals}},\ and\ \bibinfo {author} {\bibfnamefont {Q.~V.}\ \bibnamefont {Le}},\ }\bibfield  {title} {\bibinfo {title} {Sequence to sequence learning with neural networks},\ }\href@noop {} {\bibfield  {journal} {\bibinfo  {journal} {Advances in neural information processing systems}\ }\textbf {\bibinfo {volume} {27}} (\bibinfo {year} {2014})}\BibitemShut {NoStop}%
\bibitem [{\citenamefont {Collins}\ \emph {et~al.}(2017)\citenamefont {Collins}, \citenamefont {Sohl-Dickstein},\ and\ \citenamefont {Sussillo}}]{collins2022capacity}%
  \BibitemOpen
  \bibfield  {author} {\bibinfo {author} {\bibfnamefont {J.}~\bibnamefont {Collins}}, \bibinfo {author} {\bibfnamefont {J.}~\bibnamefont {Sohl-Dickstein}},\ and\ \bibinfo {author} {\bibfnamefont {D.}~\bibnamefont {Sussillo}},\ }\bibfield  {title} {\bibinfo {title} {Capacity and trainability in recurrent neural networks},\ }in\ \href@noop {} {\emph {\bibinfo {booktitle} {International Conference on Learning Representations}}}\ (\bibinfo {year} {2017})\BibitemShut {NoStop}%
\bibitem [{\citenamefont {Vaswani}\ \emph {et~al.}(2017)\citenamefont {Vaswani}, \citenamefont {Shazeer}, \citenamefont {Parmar}, \citenamefont {Uszkoreit}, \citenamefont {Jones}, \citenamefont {Gomez}, \citenamefont {Kaiser},\ and\ \citenamefont {Polosukhin}}]{vaswani2017attention}%
  \BibitemOpen
  \bibfield  {author} {\bibinfo {author} {\bibfnamefont {A.}~\bibnamefont {Vaswani}}, \bibinfo {author} {\bibfnamefont {N.}~\bibnamefont {Shazeer}}, \bibinfo {author} {\bibfnamefont {N.}~\bibnamefont {Parmar}}, \bibinfo {author} {\bibfnamefont {J.}~\bibnamefont {Uszkoreit}}, \bibinfo {author} {\bibfnamefont {L.}~\bibnamefont {Jones}}, \bibinfo {author} {\bibfnamefont {A.~N.}\ \bibnamefont {Gomez}}, \bibinfo {author} {\bibfnamefont {{\L}.}~\bibnamefont {Kaiser}},\ and\ \bibinfo {author} {\bibfnamefont {I.}~\bibnamefont {Polosukhin}},\ }\bibfield  {title} {\bibinfo {title} {Attention is all you need},\ }\href@noop {} {\bibfield  {journal} {\bibinfo  {journal} {Advances in neural information processing systems}\ }\textbf {\bibinfo {volume} {30}} (\bibinfo {year} {2017})}\BibitemShut {NoStop}%
\bibitem [{\citenamefont {Mermin}(1990)}]{mermin1990simple}%
  \BibitemOpen
  \bibfield  {author} {\bibinfo {author} {\bibfnamefont {N.~D.}\ \bibnamefont {Mermin}},\ }\bibfield  {title} {\bibinfo {title} {Simple unified form for the major no-hidden-variables theorems},\ }\href@noop {} {\bibfield  {journal} {\bibinfo  {journal} {Physical review letters}\ }\textbf {\bibinfo {volume} {65}},\ \bibinfo {pages} {3373} (\bibinfo {year} {1990})}\BibitemShut {NoStop}%
\bibitem [{\citenamefont {Peres}(1990)}]{peres1990incompatible}%
  \BibitemOpen
  \bibfield  {author} {\bibinfo {author} {\bibfnamefont {A.}~\bibnamefont {Peres}},\ }\bibfield  {title} {\bibinfo {title} {Incompatible results of quantum measurements},\ }\href@noop {} {\bibfield  {journal} {\bibinfo  {journal} {Physics Letters A}\ }\textbf {\bibinfo {volume} {151}},\ \bibinfo {pages} {107} (\bibinfo {year} {1990})}\BibitemShut {NoStop}%
\bibitem [{\citenamefont {Cabello}(2001{\natexlab{a}})}]{cabello2001all}%
  \BibitemOpen
  \bibfield  {author} {\bibinfo {author} {\bibfnamefont {A.}~\bibnamefont {Cabello}},\ }\bibfield  {title} {\bibinfo {title} {“all versus nothing” inseparability for two observers},\ }\href@noop {} {\bibfield  {journal} {\bibinfo  {journal} {Physical Review Letters}\ }\textbf {\bibinfo {volume} {87}},\ \bibinfo {pages} {010403} (\bibinfo {year} {2001}{\natexlab{a}})}\BibitemShut {NoStop}%
\bibitem [{\citenamefont {Cabello}(2001{\natexlab{b}})}]{cabello2001bell}%
  \BibitemOpen
  \bibfield  {author} {\bibinfo {author} {\bibfnamefont {A.}~\bibnamefont {Cabello}},\ }\bibfield  {title} {\bibinfo {title} {Bell's theorem without inequalities and without probabilities for two observers},\ }\href@noop {} {\bibfield  {journal} {\bibinfo  {journal} {Physical review letters}\ }\textbf {\bibinfo {volume} {86}},\ \bibinfo {pages} {1911} (\bibinfo {year} {2001}{\natexlab{b}})}\BibitemShut {NoStop}%
\bibitem [{\citenamefont {Rao}(2008)}]{rao2008parallel}%
  \BibitemOpen
  \bibfield  {author} {\bibinfo {author} {\bibfnamefont {A.}~\bibnamefont {Rao}},\ }\bibfield  {title} {\bibinfo {title} {Parallel repetition in projection games and a concentration bound},\ }in\ \href@noop {} {\emph {\bibinfo {booktitle} {Proceedings of the Fortieth Annual ACM Symposium on Theory of Computing}}}\ (\bibinfo {year} {2008})\ pp.\ \bibinfo {pages} {1--10}\BibitemShut {NoStop}%
\bibitem [{\citenamefont {Jain}\ and\ \citenamefont {Kundu}(2022)}]{jain2022direct}%
  \BibitemOpen
  \bibfield  {author} {\bibinfo {author} {\bibfnamefont {R.}~\bibnamefont {Jain}}\ and\ \bibinfo {author} {\bibfnamefont {S.}~\bibnamefont {Kundu}},\ }\bibfield  {title} {\bibinfo {title} {A direct product theorem for quantum communication complexity with applications to device-independent qkd},\ }in\ \href@noop {} {\emph {\bibinfo {booktitle} {2021 IEEE 62nd Annual Symposium on Foundations of Computer Science (FOCS)}}}\ (\bibinfo {organization} {IEEE},\ \bibinfo {year} {2022})\ pp.\ \bibinfo {pages} {1285--1295}\BibitemShut {NoStop}%
\bibitem [{\citenamefont {Mohri}\ \emph {et~al.}(2018)\citenamefont {Mohri}, \citenamefont {Rostamizadeh},\ and\ \citenamefont {Talwalkar}}]{mohri2018foundations}%
  \BibitemOpen
  \bibfield  {author} {\bibinfo {author} {\bibfnamefont {M.}~\bibnamefont {Mohri}}, \bibinfo {author} {\bibfnamefont {A.}~\bibnamefont {Rostamizadeh}},\ and\ \bibinfo {author} {\bibfnamefont {A.}~\bibnamefont {Talwalkar}},\ }\href@noop {} {\emph {\bibinfo {title} {Foundations of machine learning}}}\ (\bibinfo  {publisher} {MIT press},\ \bibinfo {year} {2018})\BibitemShut {NoStop}%
\bibitem [{\citenamefont {Aaronson}\ and\ \citenamefont {Gottesman}(2004)}]{aaronson2004improved}%
  \BibitemOpen
  \bibfield  {author} {\bibinfo {author} {\bibfnamefont {S.}~\bibnamefont {Aaronson}}\ and\ \bibinfo {author} {\bibfnamefont {D.}~\bibnamefont {Gottesman}},\ }\bibfield  {title} {\bibinfo {title} {Improved simulation of stabilizer circuits},\ }\href@noop {} {\bibfield  {journal} {\bibinfo  {journal} {Physical Review A}\ }\textbf {\bibinfo {volume} {70}},\ \bibinfo {pages} {052328} (\bibinfo {year} {2004})}\BibitemShut {NoStop}%
\bibitem [{\citenamefont {Hu}\ \emph {et~al.}(2024)\citenamefont {Hu}, \citenamefont {Patti}, \citenamefont {Gu}, \citenamefont {Yi~Tan},\ and\ \citenamefont {You}}]{pyclifford}%
  \BibitemOpen
  \bibfield  {author} {\bibinfo {author} {\bibfnamefont {H.-Y.}\ \bibnamefont {Hu}}, \bibinfo {author} {\bibfnamefont {T.}~\bibnamefont {Patti}}, \bibinfo {author} {\bibfnamefont {A.}~\bibnamefont {Gu}}, \bibinfo {author} {\bibfnamefont {S.~Y.}\ \bibnamefont {Yi~Tan}},\ and\ \bibinfo {author} {\bibfnamefont {Y.-Z.}\ \bibnamefont {You}},\ }\href {https://github.com/hongyehu/PyClifford} {\bibinfo {title} {{PyClifford}: an python based simulator for clifford-dominated quantum circuits}} (\bibinfo {year} {2024})\BibitemShut {NoStop}%
\bibitem [{\citenamefont {Cho}\ \emph {et~al.}(2014)\citenamefont {Cho}, \citenamefont {Van~Merri{\"e}nboer}, \citenamefont {Gulcehre}, \citenamefont {Bahdanau}, \citenamefont {Bougares}, \citenamefont {Schwenk},\ and\ \citenamefont {Bengio}}]{cho2014learning}%
  \BibitemOpen
  \bibfield  {author} {\bibinfo {author} {\bibfnamefont {K.}~\bibnamefont {Cho}}, \bibinfo {author} {\bibfnamefont {B.}~\bibnamefont {Van~Merri{\"e}nboer}}, \bibinfo {author} {\bibfnamefont {C.}~\bibnamefont {Gulcehre}}, \bibinfo {author} {\bibfnamefont {D.}~\bibnamefont {Bahdanau}}, \bibinfo {author} {\bibfnamefont {F.}~\bibnamefont {Bougares}}, \bibinfo {author} {\bibfnamefont {H.}~\bibnamefont {Schwenk}},\ and\ \bibinfo {author} {\bibfnamefont {Y.}~\bibnamefont {Bengio}},\ }\bibfield  {title} {\bibinfo {title} {Learning phrase representations using rnn encoder--decoder for statistical machine translation},\ }in\ \href@noop {} {\emph {\bibinfo {booktitle} {Proceedings of the 2014 Conference on Empirical Methods in Natural Language Processing (EMNLP)}}}\ (\bibinfo {year} {2014})\ pp.\ \bibinfo {pages} {1724--1734}\BibitemShut {NoStop}%
\bibitem [{\citenamefont {Chollet}\ \emph {et~al.}(2015)\citenamefont {Chollet} \emph {et~al.}}]{chollet2015keras}%
  \BibitemOpen
  \bibfield  {author} {\bibinfo {author} {\bibfnamefont {F.}~\bibnamefont {Chollet}} \emph {et~al.},\ }\href@noop {} {\bibinfo {title} {Keras}},\ \bibinfo {howpublished} {\url{https://keras.io}} (\bibinfo {year} {2015})\BibitemShut {NoStop}%
\bibitem [{\citenamefont {Fattal}\ \emph {et~al.}(2004)\citenamefont {Fattal}, \citenamefont {Cubitt}, \citenamefont {Yamamoto}, \citenamefont {Bravyi},\ and\ \citenamefont {Chuang}}]{fattal2004entanglement}%
  \BibitemOpen
  \bibfield  {author} {\bibinfo {author} {\bibfnamefont {D.}~\bibnamefont {Fattal}}, \bibinfo {author} {\bibfnamefont {T.~S.}\ \bibnamefont {Cubitt}}, \bibinfo {author} {\bibfnamefont {Y.}~\bibnamefont {Yamamoto}}, \bibinfo {author} {\bibfnamefont {S.}~\bibnamefont {Bravyi}},\ and\ \bibinfo {author} {\bibfnamefont {I.~L.}\ \bibnamefont {Chuang}},\ }\bibfield  {title} {\bibinfo {title} {Entanglement in the stabilizer formalism},\ }\href@noop {} {\bibfield  {journal} {\bibinfo  {journal} {arXiv preprint quant-ph/0406168}\ } (\bibinfo {year} {2004})}\BibitemShut {NoStop}%
\bibitem [{\citenamefont {Tirrito}\ \emph {et~al.}(2024)\citenamefont {Tirrito}, \citenamefont {Tarabunga}, \citenamefont {Lami}, \citenamefont {Chanda}, \citenamefont {Leone}, \citenamefont {Oliviero}, \citenamefont {Dalmonte}, \citenamefont {Collura},\ and\ \citenamefont {Hamma}}]{tirrito2024quantifying}%
  \BibitemOpen
  \bibfield  {author} {\bibinfo {author} {\bibfnamefont {E.}~\bibnamefont {Tirrito}}, \bibinfo {author} {\bibfnamefont {P.~S.}\ \bibnamefont {Tarabunga}}, \bibinfo {author} {\bibfnamefont {G.}~\bibnamefont {Lami}}, \bibinfo {author} {\bibfnamefont {T.}~\bibnamefont {Chanda}}, \bibinfo {author} {\bibfnamefont {L.}~\bibnamefont {Leone}}, \bibinfo {author} {\bibfnamefont {S.~F.}\ \bibnamefont {Oliviero}}, \bibinfo {author} {\bibfnamefont {M.}~\bibnamefont {Dalmonte}}, \bibinfo {author} {\bibfnamefont {M.}~\bibnamefont {Collura}},\ and\ \bibinfo {author} {\bibfnamefont {A.}~\bibnamefont {Hamma}},\ }\bibfield  {title} {\bibinfo {title} {Quantifying nonstabilizerness through entanglement spectrum flatness},\ }\href@noop {} {\bibfield  {journal} {\bibinfo  {journal} {Physical Review A}\ }\textbf {\bibinfo {volume} {109}},\ \bibinfo {pages} {L040401} (\bibinfo {year} {2024})}\BibitemShut {NoStop}%
\bibitem [{\citenamefont {Gu}\ \emph {et~al.}(2024)\citenamefont {Gu}, \citenamefont {Oliviero},\ and\ \citenamefont {Leone}}]{gu2024magic}%
  \BibitemOpen
  \bibfield  {author} {\bibinfo {author} {\bibfnamefont {A.}~\bibnamefont {Gu}}, \bibinfo {author} {\bibfnamefont {S.~F.}\ \bibnamefont {Oliviero}},\ and\ \bibinfo {author} {\bibfnamefont {L.}~\bibnamefont {Leone}},\ }\bibfield  {title} {\bibinfo {title} {Magic-induced computational separation in entanglement theory},\ }\href@noop {} {\bibfield  {journal} {\bibinfo  {journal} {arXiv preprint arXiv:2403.19610}\ } (\bibinfo {year} {2024})}\BibitemShut {NoStop}%
\bibitem [{\citenamefont {Frau}\ \emph {et~al.}(2024)\citenamefont {Frau}, \citenamefont {Tarabunga}, \citenamefont {Collura}, \citenamefont {Dalmonte},\ and\ \citenamefont {Tirrito}}]{frau2024nonstabilizerness}%
  \BibitemOpen
  \bibfield  {author} {\bibinfo {author} {\bibfnamefont {M.}~\bibnamefont {Frau}}, \bibinfo {author} {\bibfnamefont {P.}~\bibnamefont {Tarabunga}}, \bibinfo {author} {\bibfnamefont {M.}~\bibnamefont {Collura}}, \bibinfo {author} {\bibfnamefont {M.}~\bibnamefont {Dalmonte}},\ and\ \bibinfo {author} {\bibfnamefont {E.}~\bibnamefont {Tirrito}},\ }\bibfield  {title} {\bibinfo {title} {Nonstabilizerness versus entanglement in matrix product states},\ }\href@noop {} {\bibfield  {journal} {\bibinfo  {journal} {Physical Review B}\ }\textbf {\bibinfo {volume} {110}},\ \bibinfo {pages} {045101} (\bibinfo {year} {2024})}\BibitemShut {NoStop}%
\bibitem [{\citenamefont {Wang}\ \emph {et~al.}(2024)\citenamefont {Wang}, \citenamefont {Li}, \citenamefont {Xu}, \citenamefont {Hu}, \citenamefont {Chen}, \citenamefont {Wu}, \citenamefont {Zhang}, \citenamefont {Jin}, \citenamefont {Zhu}, \citenamefont {Gao} \emph {et~al.}}]{wang2024probing}%
  \BibitemOpen
  \bibfield  {author} {\bibinfo {author} {\bibfnamefont {K.}~\bibnamefont {Wang}}, \bibinfo {author} {\bibfnamefont {W.}~\bibnamefont {Li}}, \bibinfo {author} {\bibfnamefont {S.}~\bibnamefont {Xu}}, \bibinfo {author} {\bibfnamefont {M.}~\bibnamefont {Hu}}, \bibinfo {author} {\bibfnamefont {J.}~\bibnamefont {Chen}}, \bibinfo {author} {\bibfnamefont {Y.}~\bibnamefont {Wu}}, \bibinfo {author} {\bibfnamefont {C.}~\bibnamefont {Zhang}}, \bibinfo {author} {\bibfnamefont {F.}~\bibnamefont {Jin}}, \bibinfo {author} {\bibfnamefont {X.}~\bibnamefont {Zhu}}, \bibinfo {author} {\bibfnamefont {Y.}~\bibnamefont {Gao}}, \emph {et~al.},\ }\bibfield  {title} {\bibinfo {title} {Probing many-body bell correlation depth with superconducting qubits},\ }\href@noop {} {\bibfield  {journal} {\bibinfo  {journal} {arXiv preprint arXiv:2406.17841}\ } (\bibinfo {year} {2024})}\BibitemShut {NoStop}%
\bibitem [{\citenamefont {Vidick}\ and\ \citenamefont {Wehner}(2023)}]{vidick2023introduction}%
  \BibitemOpen
  \bibfield  {author} {\bibinfo {author} {\bibfnamefont {T.}~\bibnamefont {Vidick}}\ and\ \bibinfo {author} {\bibfnamefont {S.}~\bibnamefont {Wehner}},\ }\href@noop {} {\emph {\bibinfo {title} {Introduction to quantum cryptography}}}\ (\bibinfo  {publisher} {Cambridge University Press},\ \bibinfo {year} {2023})\BibitemShut {NoStop}%
\bibitem [{\citenamefont {Watts}\ \emph {et~al.}(2024)\citenamefont {Watts}, \citenamefont {Gosset}, \citenamefont {Liu},\ and\ \citenamefont {Soleimanifar}}]{watts2024quantum}%
  \BibitemOpen
  \bibfield  {author} {\bibinfo {author} {\bibfnamefont {A.~B.}\ \bibnamefont {Watts}}, \bibinfo {author} {\bibfnamefont {D.}~\bibnamefont {Gosset}}, \bibinfo {author} {\bibfnamefont {Y.}~\bibnamefont {Liu}},\ and\ \bibinfo {author} {\bibfnamefont {M.}~\bibnamefont {Soleimanifar}},\ }\bibfield  {title} {\bibinfo {title} {Quantum advantage from measurement-induced entanglement in random shallow circuits},\ }\href@noop {} {\bibfield  {journal} {\bibinfo  {journal} {arXiv preprint arXiv:2407.21203}\ } (\bibinfo {year} {2024})}\BibitemShut {NoStop}%
\bibitem [{\citenamefont {Vershynin}(2018)}]{vershynin2018high}%
  \BibitemOpen
  \bibfield  {author} {\bibinfo {author} {\bibfnamefont {R.}~\bibnamefont {Vershynin}},\ }\href@noop {} {\emph {\bibinfo {title} {High-Dimensional Probability: An Introduction with Applications in Data Science}}},\ Vol.~\bibinfo {volume} {47}\ (\bibinfo  {publisher} {Cambridge University Press},\ \bibinfo {year} {2018})\BibitemShut {NoStop}%
\bibitem [{\citenamefont {Devlin}\ \emph {et~al.}(2019)\citenamefont {Devlin}, \citenamefont {Chang}, \citenamefont {Lee},\ and\ \citenamefont {Toutanova}}]{devlin2018bert}%
  \BibitemOpen
  \bibfield  {author} {\bibinfo {author} {\bibfnamefont {J.}~\bibnamefont {Devlin}}, \bibinfo {author} {\bibfnamefont {M.-W.}\ \bibnamefont {Chang}}, \bibinfo {author} {\bibfnamefont {K.}~\bibnamefont {Lee}},\ and\ \bibinfo {author} {\bibfnamefont {K.}~\bibnamefont {Toutanova}},\ }\bibfield  {title} {\bibinfo {title} {{BERT}: Pre-training of deep bidirectional transformers for language understanding},\ }in\ \href {https://doi.org/10.18653/v1/N19-1423} {\emph {\bibinfo {booktitle} {Proceedings of the 2019 Conference of the North {A}merican Chapter of the Association for Computational Linguistics: Human Language Technologies, Volume 1 (Long and Short Papers)}}},\ \bibinfo {editor} {edited by\ \bibinfo {editor} {\bibfnamefont {J.}~\bibnamefont {Burstein}}, \bibinfo {editor} {\bibfnamefont {C.}~\bibnamefont {Doran}},\ and\ \bibinfo {editor} {\bibfnamefont {T.}~\bibnamefont {Solorio}}}\ (\bibinfo  {publisher} {Association for Computational Linguistics},\ \bibinfo {address} {Minneapolis, Minnesota},\ \bibinfo
  {year} {2019})\ pp.\ \bibinfo {pages} {4171--4186}\BibitemShut {NoStop}%
\bibitem [{\citenamefont {Brown}\ \emph {et~al.}(2020)\citenamefont {Brown}, \citenamefont {Mann}, \citenamefont {Ryder}, \citenamefont {Subbiah}, \citenamefont {Kaplan}, \citenamefont {Dhariwal}, \citenamefont {Neelakantan}, \citenamefont {Shyam}, \citenamefont {Sastry}, \citenamefont {Askell} \emph {et~al.}}]{brown2020language}%
  \BibitemOpen
  \bibfield  {author} {\bibinfo {author} {\bibfnamefont {T.}~\bibnamefont {Brown}}, \bibinfo {author} {\bibfnamefont {B.}~\bibnamefont {Mann}}, \bibinfo {author} {\bibfnamefont {N.}~\bibnamefont {Ryder}}, \bibinfo {author} {\bibfnamefont {M.}~\bibnamefont {Subbiah}}, \bibinfo {author} {\bibfnamefont {J.~D.}\ \bibnamefont {Kaplan}}, \bibinfo {author} {\bibfnamefont {P.}~\bibnamefont {Dhariwal}}, \bibinfo {author} {\bibfnamefont {A.}~\bibnamefont {Neelakantan}}, \bibinfo {author} {\bibfnamefont {P.}~\bibnamefont {Shyam}}, \bibinfo {author} {\bibfnamefont {G.}~\bibnamefont {Sastry}}, \bibinfo {author} {\bibfnamefont {A.}~\bibnamefont {Askell}}, \emph {et~al.},\ }\bibfield  {title} {\bibinfo {title} {Language models are few-shot learners},\ }\href@noop {} {\bibfield  {journal} {\bibinfo  {journal} {Advances in neural information processing systems}\ }\textbf {\bibinfo {volume} {33}},\ \bibinfo {pages} {1877} (\bibinfo {year} {2020})}\BibitemShut {NoStop}%
\bibitem [{\citenamefont {Achiam}\ \emph {et~al.}(2023)\citenamefont {Achiam}, \citenamefont {Adler}, \citenamefont {Agarwal}, \citenamefont {Ahmad}, \citenamefont {Akkaya}, \citenamefont {Aleman}, \citenamefont {Almeida}, \citenamefont {Altenschmidt}, \citenamefont {Altman}, \citenamefont {Anadkat} \emph {et~al.}}]{achiam2023gpt}%
  \BibitemOpen
  \bibfield  {author} {\bibinfo {author} {\bibfnamefont {J.}~\bibnamefont {Achiam}}, \bibinfo {author} {\bibfnamefont {S.}~\bibnamefont {Adler}}, \bibinfo {author} {\bibfnamefont {S.}~\bibnamefont {Agarwal}}, \bibinfo {author} {\bibfnamefont {L.}~\bibnamefont {Ahmad}}, \bibinfo {author} {\bibfnamefont {I.}~\bibnamefont {Akkaya}}, \bibinfo {author} {\bibfnamefont {F.~L.}\ \bibnamefont {Aleman}}, \bibinfo {author} {\bibfnamefont {D.}~\bibnamefont {Almeida}}, \bibinfo {author} {\bibfnamefont {J.}~\bibnamefont {Altenschmidt}}, \bibinfo {author} {\bibfnamefont {S.}~\bibnamefont {Altman}}, \bibinfo {author} {\bibfnamefont {S.}~\bibnamefont {Anadkat}}, \emph {et~al.},\ }\bibfield  {title} {\bibinfo {title} {Gpt-4 technical report},\ }\href@noop {} {\bibfield  {journal} {\bibinfo  {journal} {arXiv preprint arXiv:2303.08774}\ } (\bibinfo {year} {2023})}\BibitemShut {NoStop}%
\bibitem [{\citenamefont {Jumper}\ \emph {et~al.}(2021)\citenamefont {Jumper}, \citenamefont {Evans}, \citenamefont {Pritzel}, \citenamefont {Green}, \citenamefont {Figurnov}, \citenamefont {Ronneberger}, \citenamefont {Tunyasuvunakool}, \citenamefont {Bates}, \citenamefont {{\v{Z}}{\'\i}dek}, \citenamefont {Potapenko} \emph {et~al.}}]{jumper2021highly}%
  \BibitemOpen
  \bibfield  {author} {\bibinfo {author} {\bibfnamefont {J.}~\bibnamefont {Jumper}}, \bibinfo {author} {\bibfnamefont {R.}~\bibnamefont {Evans}}, \bibinfo {author} {\bibfnamefont {A.}~\bibnamefont {Pritzel}}, \bibinfo {author} {\bibfnamefont {T.}~\bibnamefont {Green}}, \bibinfo {author} {\bibfnamefont {M.}~\bibnamefont {Figurnov}}, \bibinfo {author} {\bibfnamefont {O.}~\bibnamefont {Ronneberger}}, \bibinfo {author} {\bibfnamefont {K.}~\bibnamefont {Tunyasuvunakool}}, \bibinfo {author} {\bibfnamefont {R.}~\bibnamefont {Bates}}, \bibinfo {author} {\bibfnamefont {A.}~\bibnamefont {{\v{Z}}{\'\i}dek}}, \bibinfo {author} {\bibfnamefont {A.}~\bibnamefont {Potapenko}}, \emph {et~al.},\ }\bibfield  {title} {\bibinfo {title} {Highly accurate protein structure prediction with alphafold},\ }\href@noop {} {\bibfield  {journal} {\bibinfo  {journal} {Nature}\ }\textbf {\bibinfo {volume} {596}},\ \bibinfo {pages} {583} (\bibinfo {year} {2021})}\BibitemShut {NoStop}%
\bibitem [{\citenamefont {Merchant}\ \emph {et~al.}(2023)\citenamefont {Merchant}, \citenamefont {Batzner}, \citenamefont {Schoenholz}, \citenamefont {Aykol}, \citenamefont {Cheon},\ and\ \citenamefont {Cubuk}}]{merchant2023scaling}%
  \BibitemOpen
  \bibfield  {author} {\bibinfo {author} {\bibfnamefont {A.}~\bibnamefont {Merchant}}, \bibinfo {author} {\bibfnamefont {S.}~\bibnamefont {Batzner}}, \bibinfo {author} {\bibfnamefont {S.~S.}\ \bibnamefont {Schoenholz}}, \bibinfo {author} {\bibfnamefont {M.}~\bibnamefont {Aykol}}, \bibinfo {author} {\bibfnamefont {G.}~\bibnamefont {Cheon}},\ and\ \bibinfo {author} {\bibfnamefont {E.~D.}\ \bibnamefont {Cubuk}},\ }\bibfield  {title} {\bibinfo {title} {Scaling deep learning for materials discovery},\ }\href@noop {} {\bibfield  {journal} {\bibinfo  {journal} {Nature}\ }\textbf {\bibinfo {volume} {624}},\ \bibinfo {pages} {80} (\bibinfo {year} {2023})}\BibitemShut {NoStop}%
\bibitem [{\citenamefont {Akiyama}\ \emph {et~al.}(2019)\citenamefont {Akiyama}, \citenamefont {Alberdi}, \citenamefont {Alef}, \citenamefont {Asada}, \citenamefont {Azulay}, \citenamefont {Baczko}, \citenamefont {Ball}, \citenamefont {Balokovi{\'c}}, \citenamefont {Barrett}, \citenamefont {Bintley} \emph {et~al.}}]{akiyama2019first}%
  \BibitemOpen
  \bibfield  {author} {\bibinfo {author} {\bibfnamefont {K.}~\bibnamefont {Akiyama}}, \bibinfo {author} {\bibfnamefont {A.}~\bibnamefont {Alberdi}}, \bibinfo {author} {\bibfnamefont {W.}~\bibnamefont {Alef}}, \bibinfo {author} {\bibfnamefont {K.}~\bibnamefont {Asada}}, \bibinfo {author} {\bibfnamefont {R.}~\bibnamefont {Azulay}}, \bibinfo {author} {\bibfnamefont {A.-K.}\ \bibnamefont {Baczko}}, \bibinfo {author} {\bibfnamefont {D.}~\bibnamefont {Ball}}, \bibinfo {author} {\bibfnamefont {M.}~\bibnamefont {Balokovi{\'c}}}, \bibinfo {author} {\bibfnamefont {J.}~\bibnamefont {Barrett}}, \bibinfo {author} {\bibfnamefont {D.}~\bibnamefont {Bintley}}, \emph {et~al.},\ }\bibfield  {title} {\bibinfo {title} {First m87 event horizon telescope results. iv. imaging the central supermassive black hole},\ }\href@noop {} {\bibfield  {journal} {\bibinfo  {journal} {The Astrophysical Journal Letters}\ }\textbf {\bibinfo {volume} {875}},\ \bibinfo {pages} {L4} (\bibinfo {year} {2019})}\BibitemShut {NoStop}%
\bibitem [{\citenamefont {Zhao}\ and\ \citenamefont {Zhu}(2022)}]{zhao2022magic}%
  \BibitemOpen
  \bibfield  {author} {\bibinfo {author} {\bibfnamefont {H.}~\bibnamefont {Zhao}}\ and\ \bibinfo {author} {\bibfnamefont {W.}~\bibnamefont {Zhu}},\ }\bibfield  {title} {\bibinfo {title} {Magic: Microlensing analysis guided by intelligent computation},\ }\href@noop {} {\bibfield  {journal} {\bibinfo  {journal} {The Astronomical Journal}\ }\textbf {\bibinfo {volume} {164}},\ \bibinfo {pages} {192} (\bibinfo {year} {2022})}\BibitemShut {NoStop}%
\bibitem [{\citenamefont {Huang}\ \emph {et~al.}(2022{\natexlab{b}})\citenamefont {Huang}, \citenamefont {Kueng}, \citenamefont {Torlai}, \citenamefont {Albert},\ and\ \citenamefont {Preskill}}]{huang2022provably}%
  \BibitemOpen
  \bibfield  {author} {\bibinfo {author} {\bibfnamefont {H.-Y.}\ \bibnamefont {Huang}}, \bibinfo {author} {\bibfnamefont {R.}~\bibnamefont {Kueng}}, \bibinfo {author} {\bibfnamefont {G.}~\bibnamefont {Torlai}}, \bibinfo {author} {\bibfnamefont {V.~V.}\ \bibnamefont {Albert}},\ and\ \bibinfo {author} {\bibfnamefont {J.}~\bibnamefont {Preskill}},\ }\bibfield  {title} {\bibinfo {title} {Provably efficient machine learning for quantum many-body problems},\ }\href@noop {} {\bibfield  {journal} {\bibinfo  {journal} {Science}\ }\textbf {\bibinfo {volume} {377}},\ \bibinfo {pages} {eabk3333} (\bibinfo {year} {2022}{\natexlab{b}})}\BibitemShut {NoStop}%
\bibitem [{\citenamefont {Carleo}\ and\ \citenamefont {Troyer}(2017)}]{carleo2017solving}%
  \BibitemOpen
  \bibfield  {author} {\bibinfo {author} {\bibfnamefont {G.}~\bibnamefont {Carleo}}\ and\ \bibinfo {author} {\bibfnamefont {M.}~\bibnamefont {Troyer}},\ }\bibfield  {title} {\bibinfo {title} {Solving the quantum many-body problem with artificial neural networks},\ }\href@noop {} {\bibfield  {journal} {\bibinfo  {journal} {Science}\ }\textbf {\bibinfo {volume} {355}},\ \bibinfo {pages} {602} (\bibinfo {year} {2017})}\BibitemShut {NoStop}%
\bibitem [{\citenamefont {Torlai}\ \emph {et~al.}(2018)\citenamefont {Torlai}, \citenamefont {Mazzola}, \citenamefont {Carrasquilla}, \citenamefont {Troyer}, \citenamefont {Melko},\ and\ \citenamefont {Carleo}}]{torlai2018neural}%
  \BibitemOpen
  \bibfield  {author} {\bibinfo {author} {\bibfnamefont {G.}~\bibnamefont {Torlai}}, \bibinfo {author} {\bibfnamefont {G.}~\bibnamefont {Mazzola}}, \bibinfo {author} {\bibfnamefont {J.}~\bibnamefont {Carrasquilla}}, \bibinfo {author} {\bibfnamefont {M.}~\bibnamefont {Troyer}}, \bibinfo {author} {\bibfnamefont {R.}~\bibnamefont {Melko}},\ and\ \bibinfo {author} {\bibfnamefont {G.}~\bibnamefont {Carleo}},\ }\bibfield  {title} {\bibinfo {title} {Neural-network quantum state tomography},\ }\href@noop {} {\bibfield  {journal} {\bibinfo  {journal} {Nature physics}\ }\textbf {\bibinfo {volume} {14}},\ \bibinfo {pages} {447} (\bibinfo {year} {2018})}\BibitemShut {NoStop}%
\bibitem [{\citenamefont {Zhao}\ \emph {et~al.}(2024)\citenamefont {Zhao}, \citenamefont {Carleo},\ and\ \citenamefont {Vicentini}}]{zhao2024empirical}%
  \BibitemOpen
  \bibfield  {author} {\bibinfo {author} {\bibfnamefont {H.}~\bibnamefont {Zhao}}, \bibinfo {author} {\bibfnamefont {G.}~\bibnamefont {Carleo}},\ and\ \bibinfo {author} {\bibfnamefont {F.}~\bibnamefont {Vicentini}},\ }\bibfield  {title} {\bibinfo {title} {Empirical sample complexity of neural network mixed state reconstruction},\ }\href@noop {} {\bibfield  {journal} {\bibinfo  {journal} {Quantum}\ }\textbf {\bibinfo {volume} {8}},\ \bibinfo {pages} {1358} (\bibinfo {year} {2024})}\BibitemShut {NoStop}%
\bibitem [{\citenamefont {Zhao}\ \emph {et~al.}(2023)\citenamefont {Zhao}, \citenamefont {Lewis}, \citenamefont {Kannan}, \citenamefont {Quek}, \citenamefont {Huang},\ and\ \citenamefont {Caro}}]{zhao2023learning}%
  \BibitemOpen
  \bibfield  {author} {\bibinfo {author} {\bibfnamefont {H.}~\bibnamefont {Zhao}}, \bibinfo {author} {\bibfnamefont {L.}~\bibnamefont {Lewis}}, \bibinfo {author} {\bibfnamefont {I.}~\bibnamefont {Kannan}}, \bibinfo {author} {\bibfnamefont {Y.}~\bibnamefont {Quek}}, \bibinfo {author} {\bibfnamefont {H.-Y.}\ \bibnamefont {Huang}},\ and\ \bibinfo {author} {\bibfnamefont {M.~C.}\ \bibnamefont {Caro}},\ }\bibfield  {title} {\bibinfo {title} {Learning quantum states and unitaries of bounded gate complexity},\ }\href@noop {} {\bibfield  {journal} {\bibinfo  {journal} {arXiv preprint arXiv:2310.19882}\ } (\bibinfo {year} {2023})}\BibitemShut {NoStop}%
\bibitem [{\citenamefont {Hochreiter}\ and\ \citenamefont {Schmidhuber}(1997)}]{hochreiter1997long}%
  \BibitemOpen
  \bibfield  {author} {\bibinfo {author} {\bibfnamefont {S.}~\bibnamefont {Hochreiter}}\ and\ \bibinfo {author} {\bibfnamefont {J.}~\bibnamefont {Schmidhuber}},\ }\bibfield  {title} {\bibinfo {title} {Long short-term memory},\ }\href@noop {} {\bibfield  {journal} {\bibinfo  {journal} {Neural computation}\ }\textbf {\bibinfo {volume} {9}},\ \bibinfo {pages} {1735} (\bibinfo {year} {1997})}\BibitemShut {NoStop}%
\bibitem [{\citenamefont {Radford}\ \emph {et~al.}(2018)\citenamefont {Radford}, \citenamefont {Narasimhan}, \citenamefont {Salimans}, \citenamefont {Sutskever} \emph {et~al.}}]{radford2018improving}%
  \BibitemOpen
  \bibfield  {author} {\bibinfo {author} {\bibfnamefont {A.}~\bibnamefont {Radford}}, \bibinfo {author} {\bibfnamefont {K.}~\bibnamefont {Narasimhan}}, \bibinfo {author} {\bibfnamefont {T.}~\bibnamefont {Salimans}}, \bibinfo {author} {\bibfnamefont {I.}~\bibnamefont {Sutskever}}, \emph {et~al.},\ }\href {https://cdn.openai.com/research-covers/language-unsupervised/language_understanding_paper.pdf} {\bibinfo {title} {Improving language understanding by generative pre-training}} (\bibinfo {year} {2018})\BibitemShut {NoStop}%
\bibitem [{\citenamefont {Li}\ and\ \citenamefont {Deng}(2022)}]{li2022recent}%
  \BibitemOpen
  \bibfield  {author} {\bibinfo {author} {\bibfnamefont {W.}~\bibnamefont {Li}}\ and\ \bibinfo {author} {\bibfnamefont {D.-L.}\ \bibnamefont {Deng}},\ }\bibfield  {title} {\bibinfo {title} {Recent advances for quantum classifiers},\ }\href@noop {} {\bibfield  {journal} {\bibinfo  {journal} {Science China Physics, Mechanics \& Astronomy}\ }\textbf {\bibinfo {volume} {65}},\ \bibinfo {pages} {220301} (\bibinfo {year} {2022})}\BibitemShut {NoStop}%
\bibitem [{\citenamefont {Li}\ \emph {et~al.}(2021)\citenamefont {Li}, \citenamefont {Lu},\ and\ \citenamefont {Deng}}]{li2021quantum}%
  \BibitemOpen
  \bibfield  {author} {\bibinfo {author} {\bibfnamefont {W.}~\bibnamefont {Li}}, \bibinfo {author} {\bibfnamefont {S.}~\bibnamefont {Lu}},\ and\ \bibinfo {author} {\bibfnamefont {D.-L.}\ \bibnamefont {Deng}},\ }\bibfield  {title} {\bibinfo {title} {Quantum federated learning through blind quantum computing},\ }\href@noop {} {\bibfield  {journal} {\bibinfo  {journal} {Science China Physics, Mechanics \& Astronomy}\ }\textbf {\bibinfo {volume} {64}},\ \bibinfo {pages} {100312} (\bibinfo {year} {2021})}\BibitemShut {NoStop}%
\bibitem [{\citenamefont {Zhao}(2023)}]{zhao2023non}%
  \BibitemOpen
  \bibfield  {author} {\bibinfo {author} {\bibfnamefont {H.}~\bibnamefont {Zhao}},\ }\bibfield  {title} {\bibinfo {title} {Non-iid quantum federated learning with one-shot communication complexity},\ }\href@noop {} {\bibfield  {journal} {\bibinfo  {journal} {Quantum Machine Intelligence}\ }\textbf {\bibinfo {volume} {5}},\ \bibinfo {pages} {3} (\bibinfo {year} {2023})}\BibitemShut {NoStop}%
\bibitem [{\citenamefont {Caro}\ and\ \citenamefont {Datta}(2020)}]{caro2020pseudo}%
  \BibitemOpen
  \bibfield  {author} {\bibinfo {author} {\bibfnamefont {M.~C.}\ \bibnamefont {Caro}}\ and\ \bibinfo {author} {\bibfnamefont {I.}~\bibnamefont {Datta}},\ }\bibfield  {title} {\bibinfo {title} {Pseudo-dimension of quantum circuits},\ }\href@noop {} {\bibfield  {journal} {\bibinfo  {journal} {Quantum Machine Intelligence}\ }\textbf {\bibinfo {volume} {2}},\ \bibinfo {pages} {14} (\bibinfo {year} {2020})}\BibitemShut {NoStop}%
\bibitem [{\citenamefont {P{\'e}rez-Salinas}\ \emph {et~al.}(2021)\citenamefont {P{\'e}rez-Salinas}, \citenamefont {L{\'o}pez-N{\'u}{\~n}ez}, \citenamefont {Garc{\'\i}a-S{\'a}ez}, \citenamefont {Forn-D{\'\i}az},\ and\ \citenamefont {Latorre}}]{perez2021one}%
  \BibitemOpen
  \bibfield  {author} {\bibinfo {author} {\bibfnamefont {A.}~\bibnamefont {P{\'e}rez-Salinas}}, \bibinfo {author} {\bibfnamefont {D.}~\bibnamefont {L{\'o}pez-N{\'u}{\~n}ez}}, \bibinfo {author} {\bibfnamefont {A.}~\bibnamefont {Garc{\'\i}a-S{\'a}ez}}, \bibinfo {author} {\bibfnamefont {P.}~\bibnamefont {Forn-D{\'\i}az}},\ and\ \bibinfo {author} {\bibfnamefont {J.~I.}\ \bibnamefont {Latorre}},\ }\bibfield  {title} {\bibinfo {title} {One qubit as a universal approximant},\ }\href@noop {} {\bibfield  {journal} {\bibinfo  {journal} {Physical Review A}\ }\textbf {\bibinfo {volume} {104}},\ \bibinfo {pages} {012405} (\bibinfo {year} {2021})}\BibitemShut {NoStop}%
\bibitem [{\citenamefont {Schuld}\ \emph {et~al.}(2021)\citenamefont {Schuld}, \citenamefont {Sweke},\ and\ \citenamefont {Meyer}}]{schuld2021effect}%
  \BibitemOpen
  \bibfield  {author} {\bibinfo {author} {\bibfnamefont {M.}~\bibnamefont {Schuld}}, \bibinfo {author} {\bibfnamefont {R.}~\bibnamefont {Sweke}},\ and\ \bibinfo {author} {\bibfnamefont {J.~J.}\ \bibnamefont {Meyer}},\ }\bibfield  {title} {\bibinfo {title} {Effect of data encoding on the expressive power of variational quantum-machine-learning models},\ }\href@noop {} {\bibfield  {journal} {\bibinfo  {journal} {Physical Review A}\ }\textbf {\bibinfo {volume} {103}},\ \bibinfo {pages} {032430} (\bibinfo {year} {2021})}\BibitemShut {NoStop}%
\bibitem [{\citenamefont {Brunner}\ \emph {et~al.}(2014)\citenamefont {Brunner}, \citenamefont {Cavalcanti}, \citenamefont {Pironio}, \citenamefont {Scarani},\ and\ \citenamefont {Wehner}}]{brunner2014bell}%
  \BibitemOpen
  \bibfield  {author} {\bibinfo {author} {\bibfnamefont {N.}~\bibnamefont {Brunner}}, \bibinfo {author} {\bibfnamefont {D.}~\bibnamefont {Cavalcanti}}, \bibinfo {author} {\bibfnamefont {S.}~\bibnamefont {Pironio}}, \bibinfo {author} {\bibfnamefont {V.}~\bibnamefont {Scarani}},\ and\ \bibinfo {author} {\bibfnamefont {S.}~\bibnamefont {Wehner}},\ }\bibfield  {title} {\bibinfo {title} {Bell nonlocality},\ }\href@noop {} {\bibfield  {journal} {\bibinfo  {journal} {Reviews of modern physics}\ }\textbf {\bibinfo {volume} {86}},\ \bibinfo {pages} {419} (\bibinfo {year} {2014})}\BibitemShut {NoStop}%
\bibitem [{\citenamefont {Bell}(2004)}]{bell2004speakable}%
  \BibitemOpen
  \bibfield  {author} {\bibinfo {author} {\bibfnamefont {J.~S.}\ \bibnamefont {Bell}},\ }\href@noop {} {\emph {\bibinfo {title} {Speakable and unspeakable in quantum mechanics: Collected papers on quantum philosophy}}}\ (\bibinfo  {publisher} {Cambridge university press},\ \bibinfo {year} {2004})\BibitemShut {NoStop}%
\bibitem [{\citenamefont {Einstein}\ \emph {et~al.}(1935)\citenamefont {Einstein}, \citenamefont {Podolsky},\ and\ \citenamefont {Rosen}}]{einstein1935can}%
  \BibitemOpen
  \bibfield  {author} {\bibinfo {author} {\bibfnamefont {A.}~\bibnamefont {Einstein}}, \bibinfo {author} {\bibfnamefont {B.}~\bibnamefont {Podolsky}},\ and\ \bibinfo {author} {\bibfnamefont {N.}~\bibnamefont {Rosen}},\ }\bibfield  {title} {\bibinfo {title} {Can quantum-mechanical description of physical reality be considered complete?},\ }\href@noop {} {\bibfield  {journal} {\bibinfo  {journal} {Physical Review}\ }\textbf {\bibinfo {volume} {47}},\ \bibinfo {pages} {777} (\bibinfo {year} {1935})}\BibitemShut {NoStop}%
\bibitem [{\citenamefont {Bharti}\ and\ \citenamefont {Jain}(2023)}]{bharti2023power}%
  \BibitemOpen
  \bibfield  {author} {\bibinfo {author} {\bibfnamefont {K.}~\bibnamefont {Bharti}}\ and\ \bibinfo {author} {\bibfnamefont {R.}~\bibnamefont {Jain}},\ }\bibfield  {title} {\bibinfo {title} {On the power of geometrically-local classical and quantum circuits},\ }\href@noop {} {\bibfield  {journal} {\bibinfo  {journal} {arXiv preprint arXiv:2310.01540}\ } (\bibinfo {year} {2023})}\BibitemShut {NoStop}%
\end{thebibliography}%

\clearpage{}

\end{document}


\title{Supplementary Materials for Entanglement-induced provable and robust quantum learning advantages}

\author{Haimeng Zhao}
\email{haimengzhao@icloud.com}
\affiliation{Center for Quantum Information, IIIS, Tsinghua University, Beijing 100084, China}
\affiliation{Institute for Quantum Information and Matter, California Institute of Technology, Pasadena, CA 91125, USA}

 \author{Dong-Ling Deng}
\email[]{dldeng@tsinghua.edu.cn}
\affiliation{Center for Quantum Information, IIIS, Tsinghua University, Beijing 100084, China}
\affiliation{Hefei National Laboratory, Hefei 230088, PR China}
\affiliation{Shanghai Qi Zhi Institute, Shanghai 200232, China}

\maketitle

\tableofcontents

\section{Preliminaries}
\label{app:prelim}

In this section, we provide a brief review of classical machine learning (\Cref{app:cmodel}), quantum machine learning (\Cref{app:qmodel}), and quantum non-local games (\Cref{app:game}).
These basic ingredients will be extensively used in proving our main results (\Cref{app:inf,app:train}).
Throughout this paper, we use $x_i$ to denote the $i$-th bit of a bitstring $x\in\{0, 1\}^\star$. 
For any two bitstrings $x\in \{0, 1\}^n, y\in \{0, 1\}^m$, we use $xy\in \{0, 1\}^{n+m}$ to denote the concatenation of these two strings.

\subsection{Classical machine learning}
\label{app:cmodel}

We begin by reviewing the classical ML models for sequence translation tasks $R: \{0, 1\}^{4n}\times \{0, 1\}^{4n}\to \{0, 1\}$, where $R(x, y)=1$ indicates that $y$ is a valid translation of $x$.
In standard ML literature \cite{goodfellow2016deep}, the translation rule $R$ is often given implicitly as the support of the conditional probability distribution $p(y|x)$ that governs the data generation of $y$ given $x$.
Then the goal of the ML task is to model the distribution $p(y|x)$ using a parameterized classical model $p_\theta(y|x)$ based on a set of training data $\{x^{(i)}, y^{(i)}\}_{i=1}^N, x\sim p(x), y\sim p(y|x)$ drawn from the distribution.
This is achieved by tuning the parameters $\theta$ to minimize certain loss function $\mathcal{L}(\theta)$ that represents how far away $p_\theta$ is from $p$.
A typical loss function is the Kullback-Leibler (KL) divergence (or relative entropy)
\begin{equation}
    \mathcal{L}(\theta) = D(p\|p_\theta) = \E_{x, y \sim p}\left[\log\frac{p(y|x)}{p_\theta(y|x)}\right],
\end{equation}
where $D(p\|p_\theta)=0$ if and only if $p$ and $p_\theta$ are the same (i.e., perfect training).
Since only its $\theta$ dependence matters in optimization, we can ignore the constant terms in $\mathcal{L}(\theta)$ that do not depend on $\theta$ and approximate the expectation value with a sample mean calculated from the training data:
\begin{equation}
\begin{split}
    \mathcal{L}(\theta) &= \mathrm{const.} + \E_{x, y \sim p}\left[-\log{p_\theta(y|x)}\right] \\
    &\approx\mathrm{const.} + \frac{1}{N}\sum_{i=1}^N\left[-\log{p_\theta(y^{(i)}|x^{(i)})}\right]. \\
\end{split}
\end{equation}
This allow us to train the model by minimizing $\mathcal{L}(\theta)$ (often referred to minimizing negative log-likelihood or maximum likelihood estimation) based on the training data and the parameterized model $p_\theta$.
After training, new translation results can be obtained by sampling from the distribution $p_\theta(y|x)\approx p(y|x)$.
This ML method for sequence learning has found wide application in many fields, including natural language processing \cite{devlin2018bert,brown2020language,achiam2023gpt}, scientific discoveries \cite{jumper2021highly,merchant2023scaling,akiyama2019first,zhao2022magic}, and modeling physical systems \cite{huang2022provably,carleo2017solving,torlai2018neural,zhao2024empirical,zhao2023learning}.

The performance and efficiency of such ML models depend crucially on the structure of the parameterized models $p_\theta(y|x)$.
To avoid direct parameterization of a distribution over an exponentially large space $y\in \{0, 1\}^{4n}$, one can decompose $p_\theta(y|x)$ sequentially into a product of conditional probability
\begin{equation}
    p_\theta(y|x) = \prod_{i=1}^{4n}p_\theta (y_i|y_{<i}, x) \approx \prod_{i=1}^{4n}p_\theta (y_i|y_{<i}, x_{\leq i}),
\end{equation}
where $y_{<i}=y_1\ldots y_{i-1}, x_{\leq i}=x_i\ldots x_i$, and we approximate each conditional probability using the causal order of the sequences (i.e., $y_i$ only depends on $y_{<i}$ and $x_{i\leq i}$).
Then we only need to parameterize each $p_\theta (y_i|y_{<i}, x_{\leq i})$ which only lives on a constant size space $y_i\in \{0, 1\}$.
Such models are called \emph{autoregressive} models.
For example, one can use a recurrent neural network (RNN), or its variants long short term memory (LSTM) \cite{hochreiter1997long} and gated recurrent unit (GRU) \cite{cho2014learning}, to parameterize $p_\theta(y_i|y_{<i}, x_{\leq i})$.
As shown in Figure 1(d) in the main text, we define a hidden state $h_i\in \mathbb{R}^d$ to record the information extracted from previous elements of the sequences, and recurrently update its value:
\begin{align}
    h_i &= f_\theta(h_{i-1}, x_i, y_{i-1}),\\
    p_i &= g_\theta(h_i),\\
    y_i &\sim \mathrm{Bern}(p_i),
\end{align}
where $\mathrm{Bern}(p)$ denotes the Bernoulli distribution with probability $p$ of taking $1$, and the details of $f_\theta, g_\theta$ depends on the choice of RNN.
Here, the update rules $f_\theta, g_\theta$ for each $i$ are kept the same, to avoid unlimited growth of parameter size when the sequence length $4n$ increases.
This reflects the \emph{weight sharing} property described in the main text.

Another mainstream way to parameterize $p_\theta(y|x)$ is to introduce a latent space $\lambda \in \mathbb{R}^d$ to store the features of the input sequence $x$, and then use $\lambda$ to generate the output $y$.
Such models are called \emph{encoder-decoder} models, where an encoder $\mathcal{E}$ encodes the input $x$ into the latent space $\lambda$ and a decoder $\mathcal{D}$ maps $\lambda$ to the output $y$, as shown in Figure 1(d) in the main text.
Typically, the encoder $\mathcal{E}$ is chosen to be deterministic $p_\theta(\lambda|x) = \delta(\lambda-\mathcal{E}(x))$, since it is easier to parameterize a function than a distribution.
But the decoder $\mathcal{D}$ necessarily parameterizes the distribution $p_\theta(y|\lambda)$ to accommodate the probabilistic nature of the sequence translation task.
Therefore, $\mathcal{D}$ alone is often constructed as an autoregressive model.
Notable examples of encoder-decoder models include the sequence-to-sequence models (seq2seq) \cite{sutskever2014sequence} used by Google Translate and Transformers \cite{vaswani2017attention} and their descendant generative pre-trained transformers (GPTs) \cite{radford2018improving} that empower the latest development of large language models (LLMs) \cite{achiam2023gpt}.

In all the above models, the size of the model $d$ controls the amount of communication $c$ between different parts of the sequence allowed by the model.
For example, in autoregressive models, the communication capacity $c$ is bounded by the size of the hidden state $h_i$, since $h_i$ is the only thing that carries information.
Its information capacity $c$ is bounded by the sum of the capacity provided by each entry of the $d$ dimensional hidden state. 
Therefore, taking machine precision into account, we have
\begin{equation}
\label{eq:com-size}
    c\leq \sum_{i=1}^{d} \bigo(1)=\bigo(d).
\end{equation}
The same holds for encoder-decoder model, as the only thing that carries information is the latent variable $\lambda\in \mathbb{R}^d$.

\ins{
In the most general case, we follow the standard definition of space complexity (see e.g., \cite{raz2018fast}) and define the model size $d$ as the total number of bits used to store the intermediate state of the model at any time step during the forward computation of the model.
Since parameters are stored using a constant number of bits in computers, this general definition of model size $d$ reduces (up to a constant factor) to counting the number of parameters in the special cases discussed above for autoregressive and encoder-decoder models.
As for communication complexity, we follow the standard definition \cite{yao1993quantum,brassard2003quantum} and define it in terms of two-player games.
In particular, we assume that two players each hold a segment of the input sequence and they want to solve a task.
The communication complexity is defined as the minimal number of input-dependent bits that they need to send to each other to solve the task.
Meanwhile, the communication capacity of a specific strategy is defined as the number of bits that are actually sent in this strategy.
The corresponding quantum notions are defined similarly with bits replaced by qubits.
We note that \Cref{eq:com-size} still holds in this most general case, because each bit of communication in $c$ must have been carried by certain internal variables in the model that are counted in $d$.
}

\ins{
We remark that our classical lower bounds (Theorems 1 and 2) apply to more general classes of classical ML models besides autoregressive and encoder-decoder models, as long as their communication capacity is bounded by their size (i.e., \Cref{eq:com-size} holds).
We also note that we allow the classical models to have shared randomness distributed across different parts of the model (as the classical analog of shared entanglement), and our lower bounds still hold.
}

\subsection{Quantum machine learning}
\label{app:qmodel}

Motivated by the success of classical ML, it is natural to study its quantum counterpart and regard quantum ML as a promising route towards practical quantum advantage.
The basic idea of quantum ML is to parameterize the model $p_\theta(y|x)$ with a parameterized quantum circuit (also known as a quantum neural network).
Such models are also called variational quantum algorithms \cite{cerezo2021variational} as they can be viewed as quantum algorithms with variable parameters.
Proposals have been made showing potential advantage in speed, memory cost, data efficiency, representation power, privacy, etc \cite{dalzell2023quantum,lloyd2014quantum,rebentrost2014quantum,dunjko2016quantum,gao2018quantum,lloyd2018quantum,schuld2019quantum,liu2021rigorous,huang2022quantum,cerezo2021variational,li2022recent,li2021quantum,zhao2023non}. 

In particular, a general way to represent a function with parameterized quantum circuit is to encode the data into some initial state $\ket{x}$, evolve it with a parameterized quantum circuit $U_\theta(x)$ that can also depend on $x$, and readout the expectation value of some observable $O$ as the function output.
If we absorb the state preparation procedure into the circuit $U_\theta(x)$, we can express the function as
\begin{equation}
    f_\theta(x) = \bra{0}U_\theta(x)^\dagger OU_\theta(x)\ket{0}.
\end{equation}
Extensive studies have been done to understand the expressivity and structure of such functions \cite{zhao2023learning,caro2020pseudo,perez2021one,schuld2021effect}.

More generally, one can parameterize conditional distributions with a set of measurement $\{P_y\}$ (e.g., computational basis measurement) after the action of a parameterized quantum circuit $U_\theta(x)$ on an initial state $\rho_x$:
\begin{equation}
    p_\theta(y|x) = \Tr(P_y U_\theta(x) \rho_x U_\theta(x)^\dagger).
\end{equation}
The output $y$ can be sampled by running the circuit and making the measurement $\{P_y\}$.
This gives a quantum model for sequence translation.
In this work, we consider a simple quantum model (see Figure 1(c) in the main text), where the initial state is set to $2n$ entangled pairs $\ket{\mathrm{EPR}}^{\otimes 2n}$, the unitary $U_\theta(x)$ is realized with a single layer of two qubit gates $\{U_i(x)\}_{i=1}^{2n}$, and the measurement is the computational basis measurement.

\begin{figure}
    \centering
    \includegraphics[width=0.5\linewidth]{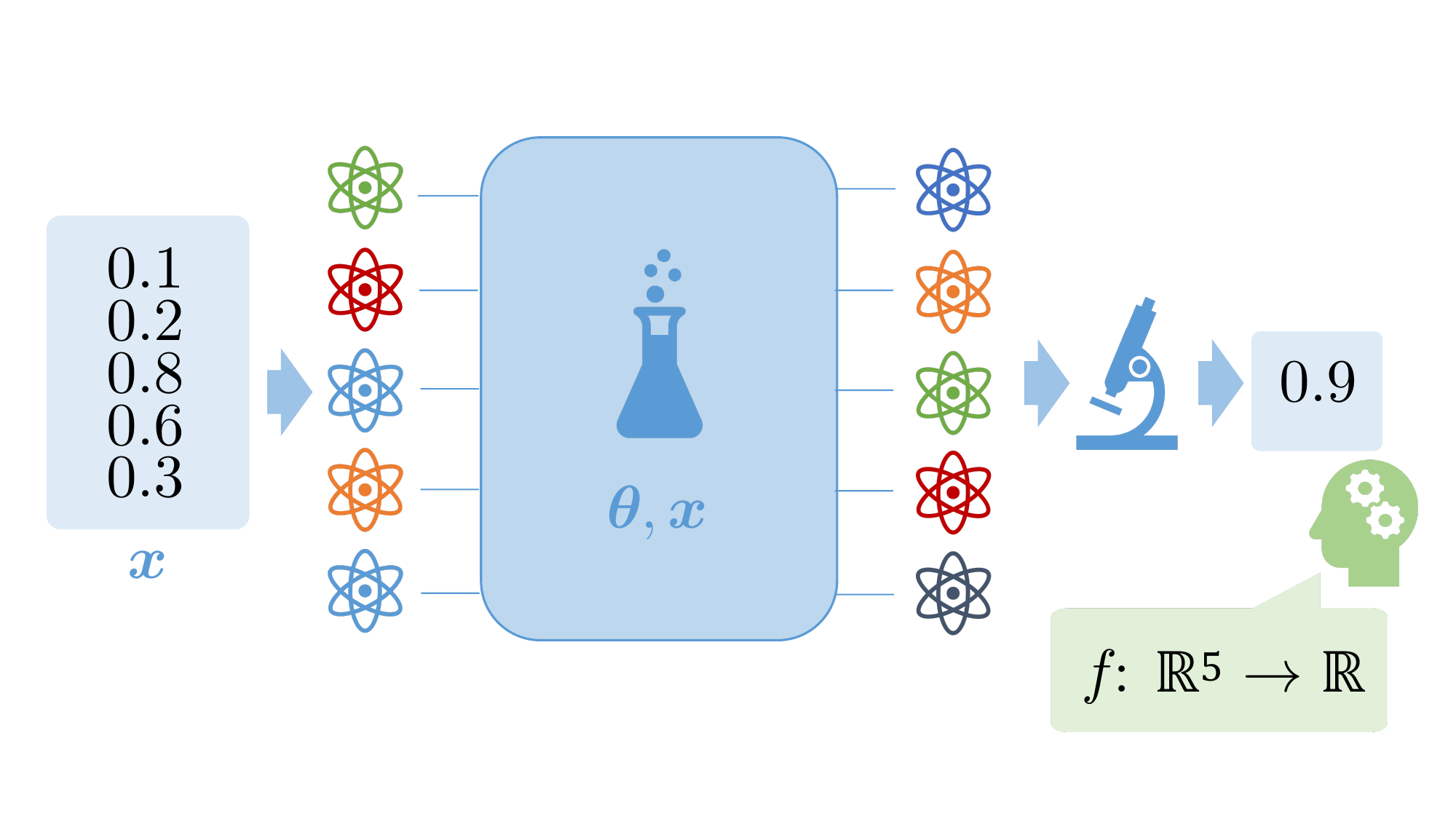}
    \caption{Illustration of a quantum machine learning model $f_\theta(x)=\bra{0}U_\theta(x)^\dagger OU_\theta(x)\ket{0}$.}
    \label{fig:qml}
\end{figure}

\subsection{Quantum non-local games}
\label{app:game}

A non-local game $\mathcal{G}$ is an interactive protocol between some players and a verifier \cite{brunner2014bell}.
We focus on the case where there are only two players $A$ and $B$.
The verifier gives each player $i\in \{A, B\}$ a query $x^i\in \mathcal{X}$, and require it to output an answer $y^i\in \mathcal{Y}$.
We assume that the queries are independently sampled from the uniform distribution over $\mathcal{X}$.
The rule (or winning condition) of the game is specified by a predicate function $V: \mathcal{X}\times \mathcal{X}\times \mathcal{Y} \times \mathcal{Y}\to \{0, 1\}$, where $V(x^A, x^B, y^A, y^B)=1$ indicates winning.

The players can coordinate with each other to set up a strategy for producing the answers, but they cannot communicate anymore after receiving the queries (i.e., a player does not know what the queries of the other players are).
A classical strategy is a strategy in which players can access a shared classical source of randomness and prior information (i.e., \emph{hidden variables} \cite{bell2004speakable}) to make the plan.
In a quantum strategy, players can access a shared (possibly entangled) state and perform quantum operations on it to output the answers.
The maximal winning probability of classical (quantum) strategies is called the \emph{classical (quantum) value} of the game.
We use $\omega(\mathcal{G})$ and $\omega^\star(\mathcal{G})$ to denote the classical and quantum values of the game $\mathcal{G}$, respectively.

A game is a \emph{quantum pseudo-telepathy} game if it has quantum value one but classical value strictly smaller than one \cite{brassard2005quantum}.
In other words, a quantum pseudo-telepathy game can be won by a quantum strategy with certainty, but requires a certain amount of communication to be won perfectly by any classical strategy.
Such games have found applications in proving that entanglement can be used to reduce the classical communication burden of certain distributed computational tasks \cite{yao1993quantum,brassard2003quantum}.
We remark that this reduction of classical communication is not via transmitting information with entanglement, which would have violated special relativity \cite{einstein1935can}.
Instead, it is achieved by allowing quantum protocols which are more general than classical strategies.

In this work, we extensively use a quantum pseudo-telepathy game called the Mermin-Peres magic square game \cite{mermin1990simple,peres1990incompatible,cabello2001all,cabello2001bell}.
It is a two-player game where the queries $x^A, x^B\in \mathcal{X}=\{1, 2, 3\}$ and the answers $y^A, y^B\in \{0, 1\}^3$.
The game is won if all of the following three conditions are satisfied:
\begin{equation}
    \bigoplus_{j=1}^3 y^A_j = 1, \quad 
    \bigoplus_{j=1}^3 y^B_j = 0, \quad
    y^A_{x^B} = y^B_{x^A},
\end{equation}
where $\oplus$ is the addition over $\mathbb{Z}_2$.
These conditions force the answers of the players upon receiving each possible queries (i.e., their strategy) to form a square, as shown in the left panel of \Cref{tab:msg}.
In the square, the $x^A$-th column represents the answer $y^A$ given by $A$ upon receiving the query $x^A$, and the $x^B$-th row represents the answer $y^B$ given by $B$ upon receiving the query $x^B$.
Each column must sum to $1$, and each row must sum to $0$.
But this is classically impossible, as it means that the sum of all numbers in the square must be odd and even simultaneously.
Therefore, for any classical strategy, there must be at least one case in all the $3\times 3=9$ cases where the game is not won.
This means that the classical value of this game is $\omega=8/9$.

\begingroup
\setlength{\tabcolsep}{6pt} 
\renewcommand{\arraystretch}{1.5} 
\begin{table}
  \centering
  \begin{minipage}{0.45\linewidth}
    \centering
    \begin{tabular}{ccccc}
                       & 1                      & 2                      & 3                      & $x^A$ \\ \cline{2-4}
\multicolumn{1}{c|}{1} & \multicolumn{1}{c|}{0} & \multicolumn{1}{c|}{0} & \multicolumn{1}{c|}{0} &       \\ \cline{2-4}
\multicolumn{1}{c|}{2} & \multicolumn{1}{c|}{0} & \multicolumn{1}{c|}{1} & \multicolumn{1}{c|}{1} &       \\ \cline{2-4}
\multicolumn{1}{c|}{3} & \multicolumn{1}{c|}{1} & \multicolumn{1}{c|}{0} & \multicolumn{1}{c|}{?} &       \\ \cline{2-4}
$x^B$                  &                        &                        &                        &      
\end{tabular}
  \end{minipage}\hfill
  \begin{minipage}{0.45\linewidth}
    \centering
    \begin{tabular}{ccccc}
                       & 1                              & 2                              & 3                             & $x^A$ \\ \cline{2-4}
\multicolumn{1}{c|}{1} & \multicolumn{1}{c|}{$X_1$}     & \multicolumn{1}{c|}{$X_2$}     & \multicolumn{1}{c|}{$X_1X_2$} &       \\ \cline{2-4}
\multicolumn{1}{c|}{2} & \multicolumn{1}{c|}{$Z_2$}     & \multicolumn{1}{c|}{$Z_1$}     & \multicolumn{1}{c|}{$Z_1Z_2$} &       \\ \cline{2-4}
\multicolumn{1}{c|}{3} & \multicolumn{1}{c|}{$-X_1Z_2$} & \multicolumn{1}{c|}{$-Z_1X_2$} & \multicolumn{1}{c|}{$Y_1Y_2$} &       \\ \cline{2-4}
$x^B$                  &                                &                                &                               &      
\end{tabular}
  \end{minipage}
  \caption{A classical strategy (left) and a quantum winning strategy (right) of the magic square game.}
  \label{tab:msg}
\end{table}
\endgroup

On the other hand, there is a quantum strategy that wins the game with certainty, as shown in the right panel of \Cref{tab:msg}.
The two players $A, B$ share an entangled state
\begin{equation}
    \ket{\mathrm{EPR}}^{\otimes 2} = \frac{1}{\sqrt{2}}(\ket{0}_a\ket{0}_b+\ket{1}_a\ket{1}_b)\frac{1}{\sqrt{2}}(\ket{0}_c\ket{0}_d+\ket{1}_c\ket{1}_d),
\end{equation}
where $A$ holds register $a$ and $c$, and $B$ holds the rest.
Upon receiving queries $x^A, x^B$, they produce answers $y^A, y^B$ by measuring the observables in the $x^A$-th column and $x^B$-th row and output the measured bitstrings.
Note that in any column or row, the observables commute with each other.
Therefore, they can indeed be measured simultaneously and produce length-$3$ bitstrings $y^A, y^B$.
One can easily verify that all the observables in each column multiplies to $-1$ and those in each row multiplies to $1$, thus satisfying the first two winning conditions of the game.
The last condition is guaranteed by the fact that measuring the same observable on the two parts of an EPR pair yields the same result, since the reduced density matrices are the same.
This means that the quantum strategy in \Cref{tab:msg} can win the game with certainty using two entangled pairs and the quantum value of the game is one.

For our purposes, we adopt a variant of the magic square game \cite{bravyi2020quantum} where the input and output formats are regularized.
Specifically, let $x^A, x^B\in \mathcal{X}=\{0, 1\}^2$ and $y^A, y^B\in \mathcal{Y}=\{0, 1\}^2$.
Define $I(x)=2x_1+x_2\in \{0, 1, 2, 3\}$ and
\begin{align}
    y^A_3 &= y^A_1\oplus y^A_2\oplus 1, \\
    y^B_3 &= y^B_1\oplus y^B_2.
\end{align}
The game is won if and only if at least one of the following two conditions are satisfied: (1) either $x^A$ or $x^B$ is $00$; (2) $y^B_{I(x^A)}=y^A_{I(x^B)}$.
We immediately see that the non-trivial part of the game (i.e., when neither of $x^A, x^B$ is $00$) is equivalent to the magic square game.
Therefore, it has quantum value $\omega^\star=1$ and classical value $\omega = 1-(3/4)^2\times (1/9)=15/16$.
In the rest of the appendices, we use $\mathcal{G}^0$ to denote this game.

\section{Quantum advantage in inference}
\label{app:inf}

In this section, we give detailed proofs of our results on quantum advantage in noiseless inference (Theorem 1) and noisy inference (Theorem 2).
We refer the readers to the Methods section in the main text for an overview of the proof.

\subsection{The magic square translation task}

We begin by formally defining the magic square translation task $R: \{0, 1\}^{4n}\times \{0, 1\}^{4n}\to \{0, 1\}$.
Recall that a translation $(x, y)$ is deemed valid if $R(x, y)=1$.
We first construct a sub-task $R^0: \{0, 1\}^4\times \{0, 1\}^4 \to \{0, 1\}$ based on the modified magic square game $\mathcal{G}$ introduced in \Cref{app:game}.
For clarity, we include a transparent illustration of this translation rule $R^0$ in \Cref{tab:r0}.

\begin{table}[]
    \centering
    \begin{tabular}{c|c}
    \hline
         input $x$ & valid translation outputs $y$ (i.e., $R^0(x, y)=1$)  \\
    \hline
         $00{*}{*}$ & ${*}{*}{*}{*}$ \\
         ${*}{*}00$ & ${*}{*}{*}{*}$ \\
         $0101$ & $0{*}0{*}$, $1{*}1{*}$ \\
         $1001$ & $0{*}{*}0$, $1{*}{*}1$ \\
         $1101$ & $0{*}01$, $0{*}10$, $1{*}00$, $1{*}11$ \\
         $0110$ & ${*}00{*}$, ${*}11{*}$ \\
         $1010$ & ${*}0{*}0$, ${*}1{*}1$ \\
         $1110$ & ${*}001$, ${*}010$, ${*}100$, ${*}111$ \\
         $0111$ & $010{*}$, $100{*}$, $001{*}$, $111{*}$ \\
         $1011$ & $01{*}0$, $10{*}0$, $00{*}1$, $11{*}1$ \\
         $1111$ & $0101$, $1010$, $0000$, $1111$ \\
    \hline
    \end{tabular}
    \caption{Illustration of the magic square translation sub-task rule $R^0: \{0, 1\}^4\times \{0, 1\}^{4}\to \{0, 1\}$.
    For each possible input sequence $x\in\{0, 1\}^4$. we list all the valid translation outputs $y\in \{0, 1\}^4$ such that $R^0(x, y)=1$. 
    Here, $*$ can be any single bit.}
    \label{tab:r0}
\end{table}

\begin{definition}[The magic square translation sub-task]
    For any $x, y\in \{0, 1\}^4$, let $x^A=x_1x_2, x^B=x_3x_4\in \{0, 1\}^2$ and $y^A=y_1y_2(y_1\oplus y_2\oplus 1), y^B=y_3y_4(y_3\oplus y_4)\in \{0, 1\}^3$.
    Define $I(a)=2a_1+a_2\in \{1, 2, 3\}$ for any $a\in \{0, 1\}^2\setminus \{00\}$.
    The magic square translation sub-task is a relation $R^0: \{0, 1\}^4\times \{0, 1\}^4\to \{0, 1\}$ such that for any $x, y\in \{0, 1\}^4$, $R^0(x, y)=1$ if and only if at least one of the following conditions are satisfied:
    (1) either $x^A$ or $x^B$ is $00$;
    (2) $y^B_{I(x^A)}=y^A_{I(x^B)}$.
\end{definition}

The score of any model $y=\mathcal{M}(x)$ on this magic square translation sub-task $R^0$, defined as $S(\mathcal{M})=\E[R^0(x, y)]=\mathbb{P}[R^0(x, y)=1]$, is the same as the winning probability of the underlying magic square game $\mathcal{G}^0$.
From the discussion in \Cref{app:game}, we know that for any classical non-communicating model (i.e., the generation of $y^A$ does not depend on $x^B$ and vise versa), the score of the model cannot exceed the classical value of the game $\omega(\mathcal{G}^0)=15/16$.
On the other hand, since the game has quantum value $\omega^\star(\mathcal{G}^0)=1$, there is a quantum (non-communicating) model that can achieve a score of one, as shown in \Cref{tab:msg}.

To boost the quantum-classical separation in score (or equivalently, the value of the underlying game), we embed an $n$-fold parallel repetition of the sub-task $R^0$ to into the magic square translation task.

\begin{definition}[The magic square translation task]
    For any positive integer $n$, the magic square translation task is a relation $R: \{0, 1\}^{4n}\times \{0, 1\}^{4n}\to \{0, 1\}$ such that for any $x, y\in \{0, 1\}^{4n}$, $R(x, y)=1$ if and only if 
    \begin{equation}
        R^0(x_{2i-1}x_{2i}x_{2n+2i-1}x_{2n+2i}, y_{2i-1}y_{2i}y_{2n+2i-1}y_{2n+2i}) = 1
    \end{equation}
    for any $1\leq i\leq n$.
\end{definition}

Intuitively, as shown in Figure 1(b) in the main text, the magic square translation task $R$ is composed of $n$ sub-tasks $R^0$ each locating at the bits $2i-1, 2i$ and $2n+2i-1, 2n+2i$.
It can be viewed as a two-player game $\mathcal{G}$ where the player $A$ ($B$) inputs and outputs the first (last) $2n$ bits of $x$ and $y$, respectively.

Since $\mathcal{G}$ is an $n$-fold parallel repetition of $\mathcal{G}$, it can be won with certainty by repeating the quantum winning strategy of $\mathcal{G}^0$.
Therefore, it also has quantum value $\omega^\star(\mathcal{G})=1$.
To derive its classical value $\omega(\mathcal{G})$, we use the following lemma \cite{rao2008parallel} on the parallel repetition of games.

\begin{lemma}[Concentration inequality of parallel repeated games, {\cite[Theorem 7]{rao2008parallel}}]
\label{lem:parallel-rep}
    There is a universal constant $\alpha>0$ such that for any $\delta>0$ and any game with classical value $1-\epsilon \in [0, 1]$ and output length $c$ (in bits), the probability that, out of $n$ parallel repetition of this game, at least $(1-\epsilon+\delta)n$ can be won is less than
    \begin{equation}
        2\left(1-\frac{\delta/2}{1-\epsilon+3\delta/4}\right)^r,
    \end{equation}
    where
    \begin{equation}
        r=\frac{c\delta^2n}{c-\log(1-\epsilon+\delta/4)}.
    \end{equation}
\end{lemma}

This allows as to show that the score of any non-communicating classical model is exponentially small on the magic square translation task $R$.
By non-communicating, we always mean that the two players of the underlying game cannot communicate information about their inputs.

\begin{corollary}[Hardness of $R$ for classical non-communicating models]
\label{cor:classical-non-com-hardness}
    The score of any classical non-communicating model $\mathcal{M}$ on the magic square translation task $R: \{0, 1\}^{4n}\times \{0, 1\}^{4n}\to \{0, 1\}$ cannot exceed
    \begin{equation}
        S(\mathcal{M})\leq 2^{-\Omega(n)}.
    \end{equation}
    In other words, the classical value of the underlying game $\mathcal{G}$ is bounded by $\omega(\mathcal{G})\leq 2^{-\Omega(n)}$.
\end{corollary}

\begin{proof}
    Recall that for a translation $(x, y)$ to be valid under $R$ (i.e., $R(x, y)=1$), the $n$ sub-tasks $R^0$ must be simultaneously solved.
    In the language of non-local games, the winning condition of the game $\mathcal{G}$ is that all the $n$ parallel repetition of $\mathcal{G}^0$ are won.
    By \Cref{lem:parallel-rep} with $\epsilon=\delta = 1-\omega(\mathcal{G}^0)=1/16$ and $c=2$, we have $r=\Theta(n)$ and
    \begin{equation}
        \mathbb{P}[R(x, y)=1]\leq 2^{-\Omega(n)}
    \end{equation}
    for any classical non-communicating model $y=\mathcal{M}(x)$.
    The desired bounds on score and classical value follows.
\end{proof}

\subsection{Communication complexity of the task}

In \Cref{cor:classical-non-com-hardness}, we have shown that a classical model lacking communication cannot solve $R$ with exponentially small score.
However, in real classical models, communication does happen, and the communication capacity $c$ is generally bounded by the size of the model $d$ (\Cref{eq:com-size}): $c = \bigo(d)$.
Therefore, to prove quantum advantage, we need to characterize the communication complexity of the magic square translation task $R$.
In particular, we derive the relationship between the score of a classical model and its communication capacity.

\begin{prop}[Score of classical communicating models]
\label{prop:score-com}
    For any classical model $\mathcal{M}$ with communication capacity $c$, the score it achieves on the magic square translation task $R: \{0, 1\}^{4n}\to \{0, 1\}^{4n} \to \{0, 1\}$ cannot exceed
    \begin{equation}
        S(\mathcal{M})\leq 2^{-\Omega(n)+c}.
    \end{equation}
    That is, the communication complexity of the task $R$ is at least $\Omega(n)$.
\end{prop}

\begin{proof}
    We prove \Cref{prop:score-com} via a reduction from communicating models to non-communicating models, similar to the proof idea of \cite{jain2022direct,bharti2023power}.
    Recall that in the task $R$, the underlying game $\mathcal{G}$ has two players $A, B$, each taking the first and last $2n$ bits of the sequences $x, y$.
    Let $\mathcal{M}_c$ be a classical model with communication capacity $c$ between $A$ and $B$ and maximal score $\omega_c$.
    It can be decomposed into two parts: $\mathcal{A}^1(x^1, r^1, m(x^1, x^2, r^m))$ and $\mathcal{A}^2(x^2, r^2, m(x^1, x^2, r^m))$, where $\mathcal{A}^1, \mathcal{A}^2$ are the strategies used by $A$ and $B$, $x^1=x_{\leq 2n}, x^2=x_{>2n}$ are the inputs to $A$ and $B$, $m\in \{0, 1\}^c$ is the message transmitted between the players, and $r^1, r^2, r^m$ are the randomness used by $\mathcal{A}^1, \mathcal{A}^2, m$.
    That is, the output of the model reads
    \begin{equation}
        y=\mathcal{M}(x)=\mathcal{A}^1(x^1, r^1, m(x^1, x^2, r^m))\mathcal{A}^2(x^2, r^2, m(x^1, x^2, r^m)).
    \end{equation}

    \begin{figure}
        \centering
        $\mathcal{M}=$
        \begin{quantikz}[align equals at=4]
            \lstick{$x^1$}&\ctrl{2}&&\gate[2][1.5cm]{\mathcal{A}^1}&\rstick[2]{$y^1$}\\
            \lstick{$r^1$}&&&&\\
            &\wireoverride{n}&\gate[3]{m}\\
            \lstick{$r^m$}&&&\ctrl{-2}\wire[d][2]{q}\\
            &\wireoverride{n}&\\
            \lstick{$r^2$}&&&\gate[2][1.5cm]{\mathcal{A}^2}&\rstick[2]{$y^2$}\\
            \lstick{$x^2$}&\ctrl{-2}&&&\\
        \end{quantikz}
        $\quad \mathcal{M}_{\mathrm{non-com}}=$
        \begin{quantikz}[align equals at=2.5]
            \lstick{$x^1$}&&&\gate[2][1.5cm]{\mathcal{A}^1}&\rstick[2]{$y^1$}\\
            \lstick{$r^1, r$}&&&&\\
            \lstick{$r^2, r$}&&&\gate[2][1.5cm]{\mathcal{A}^2}&\rstick[2]{$y^2$}\\
            \lstick{$x^2$}&&&&\\
        \end{quantikz}
        \caption{Simulating a communicating model $\mathrm{M}$ with a non-communicating model $\mathcal{M}_\mathrm{non-com}$.}
        \label{fig:com-noncom}
    \end{figure}
    
    Now we construct a non-communicating model $\mathcal{M}_\mathrm{non-com}$ from $\mathcal{M}$, as shown in \Cref{fig:com-noncom}.
    We do so by first sampling a random bitstring $r\in \{0, 1\}^c$ from the shared random source, use it as the message, and feed it into $\mathcal{A}^1, \mathcal{A}^2$.
    In other words, we generate the output by executing $\mathcal{A}^1(x^1, r^1, r)$ and $\mathcal{A}^2(x^2, r^2, r)$:
    \begin{equation}
        y=\mathcal{M}_\mathrm{non-com}(x)=\mathcal{A}^1(x^1, r^1, r)\mathcal{A}^2(x^2, r^2, r).
    \end{equation}
    The probability of guessing $m$ successfully is
    \begin{equation}
        \mathbb{P}[r=m] = \frac{1}{2^c}.
    \end{equation}
    Therefore, the score of the non-communicating model $\mathcal{M}_\mathrm{non-com}$ reads
    \begin{equation}
    \begin{split}
        \omega_\mathrm{non-com}&=\mathbb{P}[R(x, \mathcal{M}_\mathrm{non-com}(x)=1]\\
        &=\mathbb{P}[r=m]\mathbb{P}[R(x, \mathcal{A}^1(x^1, r^1, r)\mathcal{A}^2(x^2, r^2, r))=1|r=m] \\
        &+ \mathbb{P}[r\neq m]\mathbb{P}[R(x, \mathcal{A}^1(x^1, r^1, r)\mathcal{A}^2(x^2, r^2, r))=1|r\neq m] \\
        &\geq \frac{1}{2^c}\omega_c.
    \end{split}
    \end{equation}
    Meanwhile, from \Cref{cor:classical-non-com-hardness}, we have
    \begin{equation}
        \omega_\mathrm{non-com}\leq 2^{-\Omega(n)}.
    \end{equation}
    Therefore, we arrive at the desired result
    \begin{equation}
        \omega_c\leq 2^{-\Omega(n)+c}.
    \end{equation}
    In other words, any classical model must have communication capacity at least $\Omega(n)$ to achieve $2^{-o(n)}$ score.
\end{proof}

\subsection{An optimal quantum model}

Now that we have obtained a lower bound on the classical communication complexity of the magic square translation task $R$, we describe a \emph{quantum pseudo-telepathy} model $\mathcal{M}_Q$ that can solve the task perfectly in $\bigo(1)$ depth and $\bigo(1)$ parameters without any communication.

\begin{figure}
    \centering
    \includegraphics[width=0.8\linewidth]{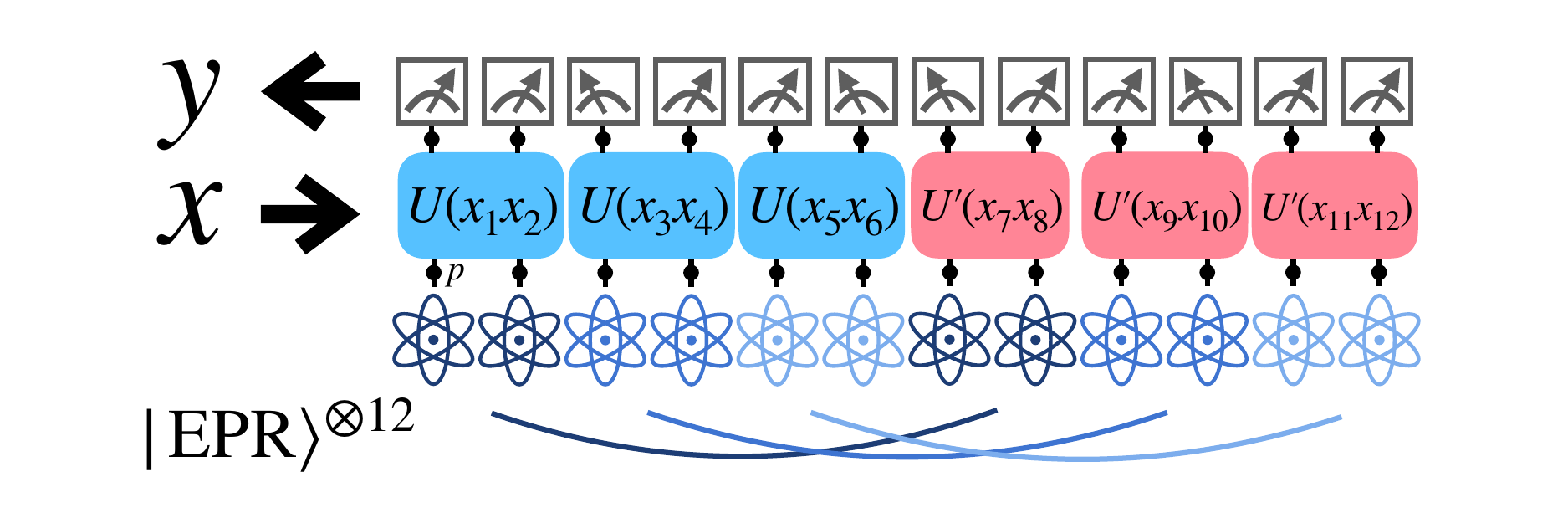}
    \caption{Illustration of an optimal quantum model $\mathcal{M}_Q$ of the magic square translation task $R$ when $n=3$.
    $U$ and $U'$ are two parameterized unitary gates that depend on the input $x$, and gates with the same color indicate weight sharing.
    Measurements are conducted in the computational basis to generate the output $y$.
    Each four initial qubits with the same color form two Bell states $\ket{\mathrm{EPR}}^{\otimes 2}$.}
    \label{fig:qmodel}
\end{figure}

\begin{prop}[Quantum model for $R$]
\label{prop:qmodel}
    There is a quantum model $\mathcal{M}_Q$ in the form of \Cref{fig:qmodel} that achieves perfect score $S(\mathcal{M}_Q)=1$ on the magic square translation task $R:\{0, 1\}^{4n}\times \{0, 1\}^{4n}\to \{0, 1\}$ using $\bigo(1)$ depth, $\bigo(1)$ parameters, and $2n$ Bell pairs as the initial state.
\end{prop}

\begin{proof}
    We construct the quantum model $\mathcal{M}_Q$ based on the winning strategy of the magic square game $\mathcal{G}$.
    An illustration of the model is shown in \Cref{fig:qmodel}.
    Recall that the task $R$ is an $n$-fold parallel repetition of $R^0$.
    Thus we only need to construct the model for each sub-task $R^0$ (or equivalently, the corresponding game $\mathcal{G}^0$), and copy it to other places via weight sharing.

    In the right panel of \Cref{tab:msg}, we already illustrated the observables that can be measured to win the game $\mathcal{G}^0$ perfectly.
    They can be converted into computational basis measurements following a set of unitary gates summarized in \Cref{tab:msg-gate} \cite{bravyi2020quantum}.

    \begin{table}
    \centering
    \begin{tabular}{|c|c|c|c|c|}
    \hline
    $a\in \{0, 1\}^2$    & 00  & 01       & 10                       & 11                                    \\ \hline
    $U(a)$ & $I$ & $H_1$    & $H_1\cdot \mathrm{SWAP}$ & $H_1\cdot \mathrm{CNOT}$              \\ \hline
    $U'(a)$ & $I$ & $H_1H_2$ & $\mathrm{SWAP}$          & $H_1H_2\cdot \mathrm{CZ}\cdot Z_1Z_2$ \\ \hline
    \end{tabular}
    \caption{The unitary gates that implements the quantum winning strategy of the magic square game $\mathcal{G}^0$ \cite{bravyi2020quantum}.
    The $U$ and $U'$ gates are acted on the first and last $2n$ qubits, respectively.}
    \label{tab:msg-gate}
    \end{table}
    
    This allows us to construct the quantum model $\mathcal{M}_Q$ as follows.
    For each $1\leq i\leq n$, we prepare two Bell states 
    \begin{equation}
        \ket{\mathrm{EPR}}^{\otimes 2} = \frac{1}{\sqrt{2}}(\ket{0}_{2i-1}\ket{0}_{2n+2i-1}+\ket{1}_{2i-1}\ket{1}_{2n+2i-1})\frac{1}{\sqrt{2}}(\ket{0}_{2i}\ket{0}_{2n+2i}+\ket{1}_{2i}\ket{1}_{2n+2i}).
    \end{equation}
    Then we act the two-qubit gates $U(x_{2i-1}x_{2i})$ and $U'(x_{2n+2i-1}x_{2n+2i})$ on registers $(2i-1, 2i)$ and $(2n+2i-1, 2n+2i)$, respectively.
    After a single layer of gates, we measure the qubits in computational basis and output the outcome bitstring as the translation result $y$.
    From the discussion of $\mathcal{G}^0$ in \Cref{app:game}, this procedure produces a valid translation $(x, y)$ weith certainty.
    In other words, this model $\mathcal{M}_Q$ can achieve score
    \begin{equation}
        S(\mathcal{M}_Q)=\mathbb{P}[R(x, \mathcal{M}_Q(x))=1]=1.
    \end{equation}
    Note that due to weight sharing in this model, we only need to parameterize the two-qubit gates $U(a), U'(a)\in \mathrm{U}(4)$ under different inputs $a\in \{0, 1\}^2$, this amounts to 
    \begin{equation}
        2 \cdot \mathrm{dim}(U(4)) \cdot |\{0, 1\}^2| = 128 = \bigo(1)
    \end{equation}
    real parameters in total.
    Moreover, the model uses $2n$ Bell pairs as the initial state and only has $\bigo(1)$ depth.
    This completes the proof of \Cref{prop:qmodel}.
\end{proof}

\subsection{Proof of Theorem 1}

We are now ready to prove Theorem 1.
\begin{proof}[Proof of Theorem 1]
    For the quantum model, we know from \Cref{prop:qmodel} that there exists a quantum model $\mathcal{M}_Q$ achieving perfect score $S(\mathcal{M}_Q)=1$ on the magic square translation task $R$ using $\bigo(1)$ depth, $\bigo(1)$ parameters, and $2n$ Bell pairs as the initial state.

    For the classical hardness, \Cref{prop:score-com} asserts that any classical model $\mathcal{M}_C$ with communication capacity $c$ cannot have a score higher than 
    \begin{equation}
        S(\mathcal{M}_C)\leq 2^{-\Omega(n)+c}.
    \end{equation}
    In other words, to achieve a score higher than $2^{-o(n)}$, the classical model must have communication capacity at least
    \begin{equation}
        c\geq \Omega(n).
    \end{equation}
    Meanwhile, from the discussion in \Cref{app:cmodel}, we have 
    \begin{equation}
        c\leq \bigo(d),
    \end{equation}
    where $d$ is the size of the classical model.
    Therefore, we must have 
    \begin{equation}
        d\geq \Omega(n)
    \end{equation}
    to achieve the desired score.
    This concludes the proof of Theorem 1.
\end{proof}

\subsection{Adding noise}
\label{app:noise}

\begin{figure}
    \centering
    \begin{quantikz}[wire types={c,c,q,q,c,c,q,q}]
        \lstick{$x_1$}&&&&&&\ctrl[vertical
wire=c]{2}\\
        \lstick{$x_2$}&&&&&&\control{}\\
        \lstick{$\ket{0}$}&\gate{H}&\phase{p}&\ctrl{4}&&\phase{p}&\gate[2]{U(x_1x_2)}&\phase{p}&\meter{}&\setwiretype{c}\rstick{$y_1$}\\
        \lstick{$\ket{0}$}&\gate{H}&\phase{p}&&\ctrl{4}&\phase{p}&&\phase{p}&\meter{}&\setwiretype{c}\rstick{$y_2$}\\
        \lstick{$x_3$}&&&&&&\ctrl[vertical
wire=c]{2}\\
        \lstick{$x_4$}&&&&&&\control{}\\
        \lstick{$\ket{0}$}&&&\targ{}&&\phase{p}&\gate[2]{U'(x_3x_4)}&\phase{p}&\meter{}&\setwiretype{c}\rstick{$y_3$}\\
        \lstick{$\ket{0}$}&&&&\targ{}&\phase{p}&&\phase{p}&\meter{}&\setwiretype{c}\rstick{$y_4$}\\
    \end{quantikz}
    \caption{Illustration of the quantum model for $R^0$ under depolarization noise of strength $p$.}
    \label{fig:noisy-msg-subcircuit}
\end{figure}

When depolarization noise of strength $p$ corrupts a state, there is a probability of at least $1-p$ that the state stays the same.
If there are $t$ places where noise could happen during a computation, the probability that the output stays the same decays as $(1-p)^t$ (at worst).
In \Cref{fig:noisy-msg-subcircuit}, we draw the circuit of the optimal quantum model for one sub-task $R^0$, where we include the preparation procedure of the initial state $\ket{\mathrm{EPR}}^{\otimes 2}$ to mimic the scenarios in real experiments.
We note that there are in total $10$ places where the noise can occur, therefore the success probability (i.e., quantum value) of the quantum model satisfies 
\begin{equation}
    \omega_p^\star \geq (1-p)^{10}.
\end{equation}
We note that here we're assuming a specific noise model, where two-qubit gate noise is a simple tensor product of two single-qubit noise.
We also assume that the gates are implemented as \Cref{fig:noisy-msg-subcircuit}.
In general, the number $10$ may change, but it remains a constant since there are only constant number of gates.

As long as $\omega_p^\star > \omega$, we still have a quantum-classical separation of winning probability.
This condition translate into
\begin{equation}
    (1-p)^{10}>\frac{15}{16},
\end{equation}
which gives
\begin{equation}
    p<p^\star=1-\left(\frac{15}{16}\right)^{1/10}\approx 0.0064.
\end{equation}

When $p<p^\star=\bigo(1)$, we can still boost the quantum-classical separation by a fractional version of the $n$-fold parallel repetition.
Specifically, we define the noisy magic square translation task as follows.

\begin{definition}[The noisy magic square translation task]
    For any positive integer $n$ and any $0<p<p^\star\approx 0.0064$, let $\eta_p=((1-p)^{10}+15/16)/2$.
    The noisy magic square translation task is a relation $R_p: \{0, 1\}^{4n}\times \{0, 1\}^{4n}\to \{0, 1\}$ such that for any $x, y\in \{0, 1\}^{4n}$, $R_p(x, y)=1$ if and only if there exists a set of indices $I\subseteq \{1, \ldots, n\}$ with $|I|\geq \eta_p n$ such that
    \begin{equation}
    \label{eq:win-cond}
        R^0(x_{2i-1}x_{2i}x_{2n+2i-1}x_{2n+2i}, y_{2i-1}y_{2i}y_{2n+2i-1}y_{2n+2i}) = 1
    \end{equation}
    for any $i \in I$.
\end{definition}

The fraction $\eta_p = ((1-p)^{10}+15/16)/2$ is chosen such that $\omega<\eta_p<(1-p)^{10}\leq \omega^\star_p$ when $p < p^\star$.

\subsection{Proof of Theorem 2}

By standard concentration inequalities \cite{vershynin2018high}, we obtain basically the same exponential quantum-classical separation on $R_p$ as in \Cref{cor:classical-non-com-hardness}. 
Combined with the arguments in \Cref{prop:score-com} and \Cref{prop:qmodel}, this allows us to prove Theorem 2.

\begin{proof}[Proof of Theorem 2]
    Recall that when $p<p^\star$, we have $\omega<\eta_p<(1-p)^{10}\leq \omega^\star_p$.
    We begin by proving the part of quantum model in Theorem 2.
    From \Cref{app:noise}, we know that the optimal quantum model $\mathcal{M}_Q$ in \Cref{prop:qmodel} has the desired properties, but can only solve the sub-task $R^0$ with probability $\omega^\star_p\geq (1-p)^{10}$ under depolarization noise of strength $p$.
    For any $1\leq i\leq n$, let $s_i\in \{0, 1\}$ denotes the success of the $i$-th sub-game $R^0$ (i.e., $s_i=1$ when \Cref{eq:win-cond} is satisfied).
    Then we have $s_i\sim \mathrm{Bern}(\omega_p^\star)$, where each $\mathrm{Bern}(\omega_p^\star)$ is an independent Bernoulli distribution with probability $\omega_p^\star$ of taking one.
    The expectation value of $s_i$ is $\E[s_i]=\omega_p^\star$.
    For any $\delta>0$, the probability that the model $\mathcal{M}_Q$ solving at least $(\omega_p^\star-\delta)n$ faction of the $n$ sub-tasks
    reads
    \begin{equation}
    \begin{split}
        \mathbb{P}\left[\sum_{i=1}^n s_i\geq (\omega_p^\star-\delta)n\right] &= 1-\Pr\left[\sum_{i=1}^n s_i - \E\left[\sum_{i=1}^n s_i\right]< -\delta n\right]\\
        &\geq 1-\exp\left(-\frac{2\delta^2 n^2}{\sum_{i=1}^n 1}\right)\\
        &=1 - \exp(-2\delta^2n),
    \end{split}
    \end{equation}
    where we have used Hoeffding's inequality \cite{vershynin2018high}.
    Since $\omega_p^\star>(1-p)^{10}>\eta_p$, we can take a positive $\delta = \omega_p^\star-\eta_p=\bigo(1)$.
    Therefore, we have
    \begin{equation}
        S(\mathcal{M}_Q) = \mathbb{P}\left[\sum_{i=1}^n s_i\geq \eta_p n\right] \geq 1-2^{-\Omega(n)}.
    \end{equation}

    Next, we prove the part of classical model in Theorem 2.
    We only need to prove the fact that any classical non-communicating model $\mathcal{M}_C$ cannot solve $R_p$ with a score higher than $2^{-\Omega(n)}$.
    Then by the same reduction argument from communicating models to non-communicating ones in \Cref{prop:score-com}, we arrive at the desired hardness result for classical models.
    On the other hand, this fact follows directly from \Cref{lem:parallel-rep} by setting $\epsilon=1-\omega=1/16$, $\delta = \eta_p-\omega>0$.
    This completes the proof of Theorem 2.
\end{proof}

\section{Quantum advantage in training}
\label{app:train}

In this section, we provide detailed proofs of our results on quantum advantage in training (Theorem 3).
We first introduce the training algorithm based on maximum likelihood estimation, and then prove that it has the desired efficiency.

\subsection{The training algorithm}
\label{app:train-alg}

We begin by introducing the training algorithm for the quantum model $\mathcal{M}_Q$.
We use $\theta$ to denote all the parameters of $U, U'$ in the model (\Cref{fig:qmodel}).
Since the optimal model (\Cref{tab:msg-gate}) is composed of two-qubit Clifford gates $\mathcal{C}(2)$ only, we restrict the parameter space to Clifford gates.
In other words, the parameter space is $\Theta = \mathcal{C}(2)^{2\cdot |\{0, 1\}^2|}$ with size $|\Theta|=\bigo(1)$.
We note that the difference between any pairs of $\theta, \theta'\in \Theta$ (measured in the difference in empirical log-likelihood) is a constant $\epsilon=\Theta(1)$, since the definition of the finite-size set $\Theta$ is independent of $n$.
The parameters of the optimal model is denoted as $\theta^\star\in \Theta$.
We assume that the training data $\{(x^{(i)}, y^{(i)})\}_{i=1}^N$ are drawn from $p(x), p(y|x)$ where $p(y|x)=p_{\theta^\star}(y|x)$.
The training algorithm is based on the maximum likelihood estimation as stated in \Cref{app:cmodel}:
\begin{equation}
    \max_{\theta\in \Theta} \E_{x, y\sim p}\left[\log p_\theta(y|x)\right]
    \approx\max_{\theta\in \Theta}\frac{1}{N}\sum_{i=1}^N\left[\log p_\theta(y^{(i)}|x^{(i)})\right].
\end{equation}

Specifically, the training algorithm works as follows.
It conducts an exhaustive search over all $\theta\in \Theta$.
For each $\theta$, we run the quantum model $\mathcal{M}_Q$ with parameters $\theta$ and calculate the probability $p_\theta(y^{(i)}|x^{(i)})$ for every data pair $(x^{(i)}, y^{(i)})$.
To calculate $p_\theta(y^{(i)}|x^{(i)})$, it measures the expectation values of a family of single-qubit observables
\begin{equation}
    \mathcal{O}(a) = (1-a)\ket{0}\bra{0}+a\ket{1}\bra{1} = \frac{1}{2}I + \left(\frac{1}{2}-a\right)Z,
\end{equation}
for $a\in \{0, 1\}$, where $\mathcal{O}(0)=\ket{0}\bra{0}$ and $\mathcal{O}(1)=\ket{1}\bra{1}$ measures the probability of sampling $0$ and $1$.
This amounts to measuring $\expval{Z}$ and classically calculating $\expval{\mathcal{O}(a)}=1/2+(1/2-a)\expval{Z}$.
In particular, the training algorithm measures $\mathcal{O}(y^{(i)}_j)$ on the $j$-th qubit and calculate the product
\begin{equation}
    \expval{\prod_{j=1}^{4n}\mathcal{O}_j\left(y^{(i)}_j\right)} = p_\theta(y^{(i)}|x^{(i)})
\end{equation}
to obtain the empirical likelihood $p_\theta(y^{(i)}|x^{(i)})$.
For later purposes, we decompose it into the product of $n$ sub-distributions $p_\theta^0$ over $\{0, 1\}^4\times \{0, 1\}^4$ based on the fact that the model is an independent parallel repetition of the sub-models for the sub-tasks $R^0$.
The algorithm then calculates and stores 
\begin{equation}
    l(\theta)=\frac{1}{nN}\sum_{i=1}^{nN} \log (p_\theta^0(y^{[i]}|x^{[i]})+\xi),
\end{equation}
where $\xi=\Theta(1)$ is a small constant to avoid divergence (e.g., $\xi$ represents the machine precision of say $\xi=10^{-8}$), and $\{(x^{[i]}, y^{[i]})\}_{i=1}^{nN}$ are independent samples derived from $\{x^{(i)}, y^{(i)}\}_{i=1}^N$ and effectively drawn from $p_{\theta^\star}^0$.
After repeating the above procedure for every $\theta\in \Theta$, the algorithm finds the $\hat{\theta}\in \Theta$ that has the largest likelihood
\begin{equation}
    \hat{\theta} = \argmax_{\theta\in \Theta} l(\theta)
\end{equation}
and output $\hat{\theta}$ as the learned model.

\subsection{Proof of Theorem 3}

We are now ready to analyze the resource consumption of the training algorithm and prove Theorem 3.

\begin{proof}[Proof of Theorem 3]
    The training algorithm in \Cref{app:train-alg} proceeds with an exhaustive search over all possible parameters $\theta\in \Theta$.
    For each $\theta\in \Theta$, it runs the circuit to calculate $\log(p_\theta^0(y^{[i]}|x^{[i]})+\xi)$.
    Since the random variable $\log(p_\theta^0(y^{[i]}|x^{[i]})+\xi)\in [-\log(1/\xi), \log(1+\xi)]$ is bounded, by Hoeffding's inequality \cite{vershynin2018high}, we have
    \begin{equation}
        \Pr\left[\left|\frac{1}{nN}\sum_{i=1}^{nN}\log(p_\theta^0(y^{[i]}|x^{[i]})+\xi)-\E\left[\log(p_\theta^0(y^{[i]}|x^{[i]})+\xi)\right]\right|>\epsilon\right]\leq 2\exp(-\frac{2nN\epsilon^2}{\log^2(1/\xi)}).
    \end{equation}
    Setting the right hand side to $\delta\in (0, 1)$, we have that, with success probability at least $1-\delta$, a sample size of 
    \begin{equation}
        N=\bigo\left(\frac{\log^2(1/\xi)\log(1/\delta)}{n \epsilon^2}\right) = \bigo\left(\frac{\log(1/\delta)}{n\epsilon^2}\right)
    \end{equation}
    suffice to estimate the empirical log-likelihood $l(\theta)$ to $\epsilon$ error.
    To estimate $l(\theta)$ to $\epsilon$ error for all $\theta\in \Theta$ simultaneously, we note that the union bound yields
    \begin{equation}
        \Pr[\text{success on all $\theta$}] = 1- \Pr[\text{fail on one $\theta$}] \geq 1- \sum_{\theta\in \Theta} \Pr[\text{fail on $\theta$}]\geq 1-|\Theta|\delta.
    \end{equation}
    Thus, we set $\delta'=|\Theta|\delta=1/3$ (i.e., $\delta=\frac{1}{3|\Theta|}=\Theta(1)$) and we can estimate $l(\theta)$ to $\epsilon$ error for all $\theta \in \Theta$ at once with success probability at least $2/3$.
    Meanwhile, a constant error $\epsilon=\Theta(1)$ is enough to distinguish between different $\theta\in \Theta$.
    Therefore, the algorithm can find the $\hat{\theta}=\argmax_{\theta\in \Theta}l(\theta)$ correctly (i.e., $\hat{\theta}=\theta^\star$) with probability at least $2/3$ using 
    \begin{equation}
        N=\bigo\left(\frac{\log(3|\Theta|)}{n \epsilon^2}\right)=\bigo\left(\frac{1}{n}\right)
    \end{equation}
    samples.
    Each sample $(x^{(i)}, y^{(i)})$ is $\Theta(n)$ bits long, so the total memory cost for storing the data is $M=\Theta(nN)=\bigo(1)$.
    The total number of entangled pairs needed is also $\bigo(1)$.
    The running time of the training algorithm is
    \begin{equation}
        T=|\Theta|\cdot N \cdot \bigo\left(\frac{n}{\epsilon^2}\right) + \bigo(|\Theta|) = \bigo(1),
    \end{equation}
    where the first term is the time needed to estimate can calculate the log-likelihood ($\bigo(n/\epsilon^2)$) for each data pair ($N$) and each parameter ($|\Theta|$), and the second term is the time for finding the maximum.
    This concludes the proof of Theorem 3.
\end{proof}

\section{Implementation details of the experiments}
\label{app:numerics}

In this section, we summarize the implementation details of the classical ML models used in the numerical experiments.
Code and data can be accessed at \url{https://github.com/haimengzhao/qml-advantage}.

The autoregressive model is composed of the standard Gated Recurrent Unit (GRU) network from Keras \cite{chollet2015keras} with varying latent dimension $d$ from $4$ to $512$ with log-equal spacing.
The biases are initialized to be all $-1$.
The rest of the settings are by default in Keras.
The output sequences of the GRU are fed into a fully connected layer with $3$ outputs, acted by the softmax function to give a probability distribution over $\{0, 1, 2\}$.
Here, $0, 1$ are the usual bits and $2$ is the placeholder for starting the sequence.
For each problem size $n$ from $4$ to $40$, we train the model by minimizing the negative log-likelihood using the Adam optimizer with learning rate $10^{-2}$ and batch size $1000$ for $10^4$ epochs.
When the loss stops to decrease for $500$ epochs, we also terminate the training.
For testing, we set the threshold $\eta_p=0.95$ to accommodate the inaccuracy of the model.

In the encoder-decoder model, the encoder has the same structure of the GRU used in the autoregressive model, and we use its final hidden state as the latent variable.
The decoder is another GRU of the same structure, whose initial hidden state is set to be the latent variable and its output sequences are fed into a fully connected layer with $4$ outputs acted by the softmax function to give a probability distribution over $\{0, 1, 2, 3\}$.
Again, $0, 1$ are the usual bits and $2, 3$ is the placeholder for starting and ending the sequence, respectively.
The training procedure is the same as that of autoregressive models, except that the learning rate is tuned to $5\times 10^{-3}$.

For the trapped-ion experiment, the IonQ 25-qubit trapped-ion device Aria has average single-qubit fidelity $99.980\%$, two-qubit fidelity $98.870\%$, and readout fidelity $99.480\%$.
It has decoherence time $T_1=10^8\text{\textmu s}$, $T_2=10^6\text{\textmu s}$, and the average duration for single-qubit gates, two-qubit gates, and readout are $135\text{\textmu s}$, $600\text{\textmu s}$, and $300\text{\textmu s}$, respectively.
To generate translation results with moderate cost, we randomly sample $10$ inputs $x$, and execute the quantum model $10^2$ times to generate $y$ for each $x$.
This results in a total of $10^3$ samples, with which we calculate the score.

\bibliography{ref}